\documentclass{article}

\usepackage{arxiv}

\usepackage[utf8]{inputenc} % allow utf-8 input
\usepackage[T1]{fontenc}    % use 8-bit T1 fonts
\usepackage{hyperref}       % hyperlinks
\usepackage{url}            % simple URL typesetting
\usepackage{booktabs}       % professional-quality tables
\usepackage{amsfonts}       % blackboard math symbols
\usepackage{nicefrac}       % compact symbols for 1/2, etc.
\usepackage{microtype}      % microtypography
\usepackage{lipsum}

\usepackage{amssymb}
\usepackage{amsmath}
\usepackage{mathtools}
\usepackage{amsthm}

\usepackage{booktabs} % For formal tables
\usepackage[ruled]{algorithm2e} % For algorithms

\SetAlFnt{\small}
\SetAlCapFnt{\small}
\SetAlCapNameFnt{\small}
\SetAlCapHSkip{0pt}
\IncMargin{-\parindent}

\usepackage{caption}
\usepackage{subcaption}
\usepackage{color}
\usepackage{algorithmic}
\usepackage{enumerate}
\usepackage{graphicx}
\newcommand{\Nn}{{\mathbb{N}}}

\newcommand{\Gg}{{\mathcal{G}}}

\newcommand{\lsb}{\left[}
\newcommand{\rsb}{\right]}
\newcommand{\lb}{\left(}
\newcommand{\rb}{\right)}

\newcommand{\Mm}{\mathcal{M}}
\newcommand{\Bb}{\mathcal{B}} 
 \newtheorem{example}{Example}
 \newtheorem{theorem}{Theorem}
 \newtheorem{lemma}{Lemma}
 \newtheorem{prop}{Proposition}
 \newtheorem{remark}{Remark}

\title{A Robust Reputation-based Group Ranking System and its Resistance to Bribery}

\author{
  Jo\~ao Sa\'ude \\
  Department of Electrical and Computer Engineering\\
  Carnegie-Melon University\\
  Pittsburgh, USA\\
  \texttt{jsaude@alumni.cmu.edu} \\
   \And
   Guilherme Ramos\\
   Department of Electrical and Computer Engineering\\
   Faculty of Engineering, University of Porto\\
   Porto, Portugal\\
   \texttt{guilhermeramos21@gmail.com}\\
   \And
   Ludovico Boratto\\
   Data Science and Big Data Analytics, EURECAT\\
   Barcelona, Spain\\
   \texttt{ludovico.boratto@acm.org}\\
   \And
   Carlos Caleiro\\
   SQIG - Instituto de Telecomunica\c{c}\~{o}es,\\
   Dept. of Mathematics,\\
   Instituto Superior T\'{e}cnico, University of Lisbon\\
   Lisbon, Portugal
}

\begin{document}
\maketitle

\begin{abstract}
    The spread of online reviews and opinions and its growing influence on people's behavior and decisions, boosted the interest to extract meaningful information from this data deluge. 
Hence, crowdsourced ratings of products and services gained a critical role in business and governments.
Current state-of-the-art solutions rank the items with an average of the ratings expressed for an item, with a consequent lack of personalization for the users, and the exposure to attacks and spamming/spurious users. 
Using these ratings to group users with similar preferences might be useful to present users with items that reflect their preferences and overcome those vulnerabilities. 
In this paper, we propose a new reputation-based ranking system, utilizing multipartite rating subnetworks, which clusters users by their similarities using three measures, two of them based on Kolmogorov complexity. 
We also study its resistance to bribery and how to design optimal bribing strategies.
Our system is novel in that it reflects the diversity of preferences by (possibly) assigning distinct rankings to the same item, for different groups of users. 
We prove the convergence and efficiency of the system. 
By testing it on synthetic and real data, we see that it copes better with spamming/spurious users, being more robust to attacks than state-of-the-art approaches.
Also, by clustering users, the effect of bribery in the proposed multipartite ranking system is dimmed, comparing to the bipartite case.
\end{abstract}

% keywords can be removed

\section{Introduction} % (fold)
\label{sec:introduction} 
    In our daily life, electronic commerce, streaming media, and collaborative economy are ubiquitous.
    Moreover, people's opinions can be as effective as an advertisement.
    These facts inspired the development of crowd-sourced ratings/reviews.
    Consumers started to use, and rely on this information to decide whether or not to buy a product/service, have a meal at a restaurant, or attend an event~\cite{sparks2011impact}.
    The sellers, aware of how the ratings of products/services impact sales~\cite{chevalier2006effect}, started to rely on the ratings and reviews of their products to assess their commercial viability as well as to predict sales~\cite{dellarocas2007exploring}.
    
    A domain in which ratings and reviews can be employed effectively is the systems that rank the items for the users (e.g., Netflix and IMDB provide to the logged-in users a ranking of the items).
    Given the relevance that ratings and reviews have for both users and companies, it is of primary importance to detect and, automatically, correct rating manipulations through fake users' ratings. 
    
    A simple way to collect and process ratings is to compute their \emph{arithmetic average} (AA).
    The main drawback of the AA is the indistinguishability of users, as it treats, in the same way, the most relevant raters and spam. 
    Therefore AA is prone to manipulation of ratings through malicious attacks or spamming.
    Further, AA might be misleading, because it does not capture the possible multimodal behavior of ratings~\cite{hu2006can}.
    For instance, in a bimodal ratings' distribution on the opposite extremes, the average is in the middle where the density of votes is low.
    By using weighted average algorithms, we can attribute different importance to the users.
    The authors in~\cite{li2012robust} proposed to weigh the importance of the users through a novel formulation of {\em reputation}, which takes into account the distance between the rating of the user for an item and the ratings of the other users for the same item (the higher is the distance, the lower is the reputation).

    \textbf{Open issues.} Considering the existing work in the literature, two main open issues arise. The first is that the similarity between the users is ignored. Indeed, by using AA or weighted average to rank the items, we are not taking into account any explicit relations between users or users' preferences. Hence, a ranking system does not make the most out of the efforts made by the users to rate the items. Indeed, the current solutions do not offer any form of personalization to the users. At the same time, it would be desirable that, if they belong to a segment with specific preferences, these should be reflected in the ranking (i.e., the users should be presented first with items they might be interested in).
The second open issue is related to the existing formulation of reputation. While we acknowledge the work done by the authors in~\cite{li2012robust} to introduce a notion of reputation in ranking systems, their formulation is such that the final ranking does not accurately reflect the actual preferences of the users. This inaccuracy happens because they weigh ratings by users' reputations but do not normalize with the sum of weights (users' reputations); indeed, they divide the weighted ratings' sum by the number of raters. Hence, when all users rate an item with the same value, the ranking is below that value and can further be smaller than the minimum allowed rating. Section~\ref{sub:undesirable} will provide details of this open issue.
    
\begin{figure}
  		\centering
  		\includegraphics[width=0.37\textwidth]{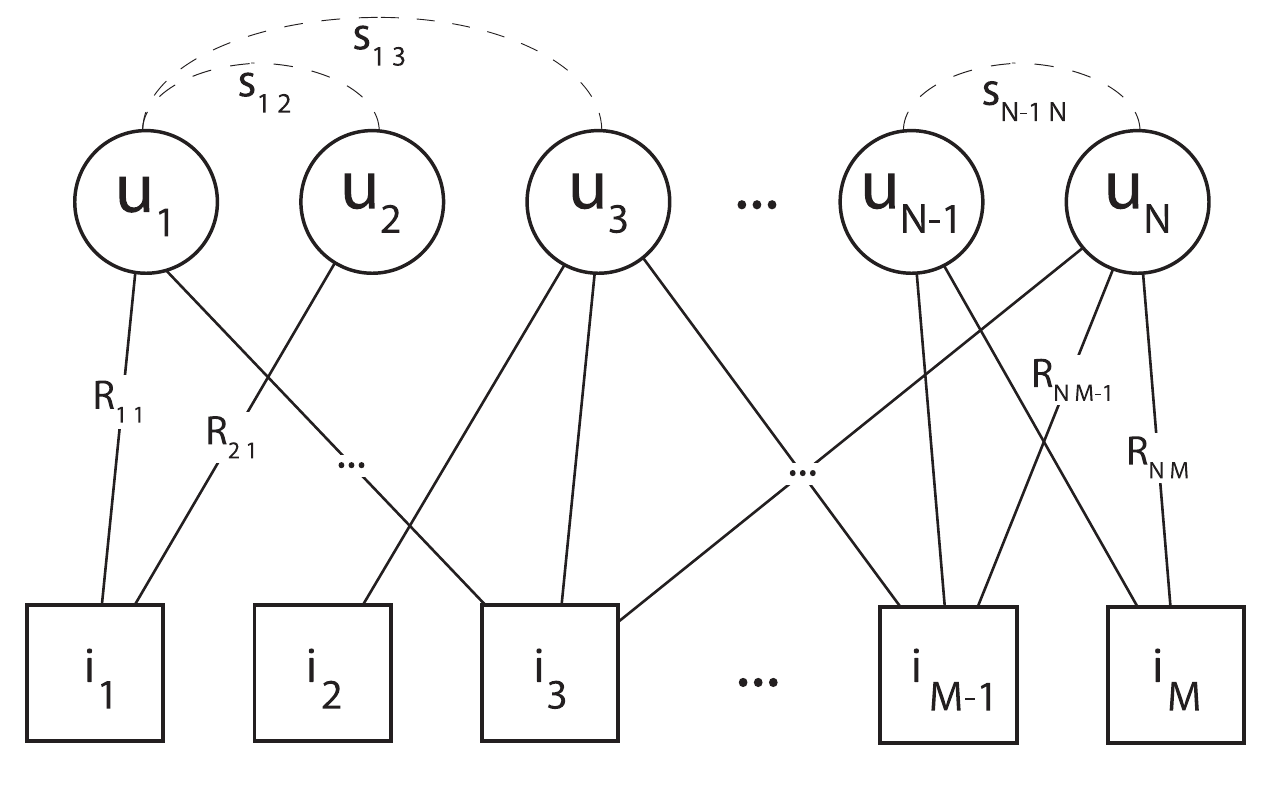}
 	 		 	\caption{Multipartite graph representing $N$ users, $M$ items, and the ratings given by user $u$ to item $i$, $R_{ui}$. 
				 The lines represent the connection, through ratings, from the users to the items.
				 The dashed lines represent links between users, through their similarities, $s_{mn}$.}
	 	\label{fig:multipartite}
	\end{figure}

    \textbf{Our contributions.} 
    In this paper, we describe a generic class of iterative reputation-based ranking systems. 
    Furthermore, we provide conditions such that the algorithms in that class converge and are efficient.
We present a new reputation-based ranking system in this class to improve the useful properties of the system proposed in~\cite{li2012robust}.
   Our approach also moves from a bipartite to a multipartite graph of the preferences. This new aspect helps to improve robustness and personalization perspectives. 
    Indeed, to design the system, we use similarities between users.
The similarities allow us to cluster users based, solely, on their ratings (see \figurename~\ref{fig:multipartite}, where two subnetworks of users are depicted in dashed lines, i.e., $\{u_1, u_2, u_3\}$ and $\{u_{N-1}, u_N\}$). To cluster users, we propose two novel similarity measures, the linear similarity (LS) and the Kolmogorov similarity (KS), and we test them against the normalized compression similarity (CS), derived from the distance metric proposed in~\cite{li2004similarity}. 

	After, we compute for the different subnetworks/clusters (possible) different rankings for the same item.
    Therefore, our method enables us to present, custom-built, items' rankings to each cluster. 	
    
    Our approach adapts better to the preferences of similar users and also improves robustness against both spurious users and spamming/malicious attacks.
    Further, it embeds the multimodal behavior of ratings' distribution.
    The existing approaches, instead, neglect the smaller subgroups that do not identify with the majority because they are averaged out.  
    We overcome this issue, since we rank the items on the base of a user clustering, while the other approaches consider the whole set of users.
    
    %Our proposed similarities not only perform better but also have smaller computational complexity than using CS.  
    %When comparing LS with KS, the former responds better to noisy spam, and the latter is more robust to targeted attacks to a set of items.
    %Both carry the same order of computational complexity, although in our implementation KS is slightly faster than LS.
    %Finally, by using LS, we obtain better robustness results than state-of-the-art approaches.
    
    Lastly, we study the resistance to bribery of the proposed bipartite and multipartite reputation-based ranking systems.
    We show that, in the bipartite scenario, users are bribable if their reputation is above the average reputation of users that rated the item.
    Whereas, in the multipartite scenario we propose, the ranking system is much more robust to bribing, and a user is bribable if s/he has a reputation above the average of the reputations of users that rated the item in the cluster s/he belongs.
    The model we propose may also be used to evaluate marketing campaigns, where a company wants to invest money to either boost the number of sales or to improve their reviews.
    
    Let us point out that this is not a recommender system\footnote{A problem similar to ours, applied to a pure recommendation setting was tackled in~\cite{RamosBC20}.}, but a ranking system that combines the existing preferences of the users, so that items are proposed to them according to the preferences they previously expressed. In our work, we do not predict the missing ratings, thus saving much computational effort. At the same time, by producing group rankings for each cluster, the rankings will be closer to the individual preferences than a unique ranking. Therefore, our work stays in the middle between classic ranking systems and the personalization provided by a recommender system. %To place our work in one of the two previously mentioned application scenarios, our approach could be applied to rank the items in given category in Netflix, will the ``Recommended for you'' sections would be produced by a classic recommender system.
    %\note{The ICDM reference in the next sentence will disappear in the first revision of the paper. It is only a way to show the reviewers that they're dealing with stuff that was published in a top-tier conference :)}
    This work extends our IEEE ICDM conference paper~\cite{SaudeRCK17} in the following ways: 
($i$) we improve the bipartite ranking system proposed in~\cite{li2012robust} to eliminate some unintuitive properties that characterize it (as presented in Section~\ref{sub:undesirable}); 
($ii$) we propose a method to cluster users based solely on their rating patterns, proposing three similarity metrics. 
With this, we can present rankings that are more tailored to users according to group preferences. The method is independent of the underlying bipartite ranking system to apply for each group; 
($iii$) we generalize the results of bribing in bipartite reputation-based ranking systems for three new cases where: 
(a) $N$ raters are bribed, 
(b) $M$ non-raters are bribed, and  
(c) $N$ raters and $M$ non-raters are bribed.

Henceforth, to the best of our knowledge, this is the first time that Kolmogorov-based measures are used in the scope of ranking systems. Also, it is the first time that a detailed and theoretical bribing analysis of reputation-based ranking systems (or ranking systems that calculate rankings as weighted average or ratings) is performed. 
%{\color{red}Shall we say that the multipartite approach id independent of the underlying ranking system? 
%Something like: }

%\todo{Does this help avoiding criticism to not compare with more literature? }
    
    %\todo{Complete, once we know the extension w.r.t. the ICDM paper. Let's try to be detailed, to show the we did a lot (theory, algorithms, experiments, ...) and where the extensions can be concretely found in this paper.}
    
    \textbf{Paper structure.} 
    In Section~\ref{related}, we present an overview of the related work. In Section~\ref{notation}, we provide the notation used in this work. 
    In Section~\ref{sec:bipartite}, we introduce a generic class of reputation-based ranking iterative algorithms, prove their convergence and efficiency, and show limitations of the existing reputation-based ranking system. 
    In Section~\ref{sec:multipartite}, we design a new reputation-based ranking system, prove its convergence, and explain its implementation.
    The experimental setup is described in Section~\ref{sec:experimental_setup} and we discuss our results in Section~\ref{sec:experimental_results}.
    In Section~\ref{sec:conclusions}, we conclude the paper.
    To improve the readability of the paper, we collected all the proofs in Appendix A.  

\section{Related Work}\label{related}
    As mentioned in the Introduction, previous ranking systems have explored a weighted average to combine the individual ratings; examples of works in this direction are~\cite{yu2006decoding,de2010iterative}. 
    In~\cite{mizzaro2003quality}, Mizzaro used an additional time-dependent quantity to weigh the ratings of users.
    Li et al.~\cite{li2012robust} introduced the concept of {\em reputation}, which measures how close are the preferences of a user to those of the others.
    Reputation is used to generate a unique ranking of the items.
    The ratings of the users who rated a specific item are weighted by the reputation of the respective user. 
    In \cite{li2015topic}, the authors use the same method to compute user reputation based on the topics associated with items (e.g., by considering Epinions and Amazon's product categories) and the score given by the user. This method allows them to build a topic-biased model (TBM), which leads to six algorithms evaluated on both real-world and synthetic datasets.  Results show that considering item categories leads to more robust item scores concerning existing approaches. Our proposal is more general and can apply to any domain, even though items may be associated with metadata, such as topics. 
    
\begin{figure}
  		\centering
  		\includegraphics[width=0.37\textwidth]{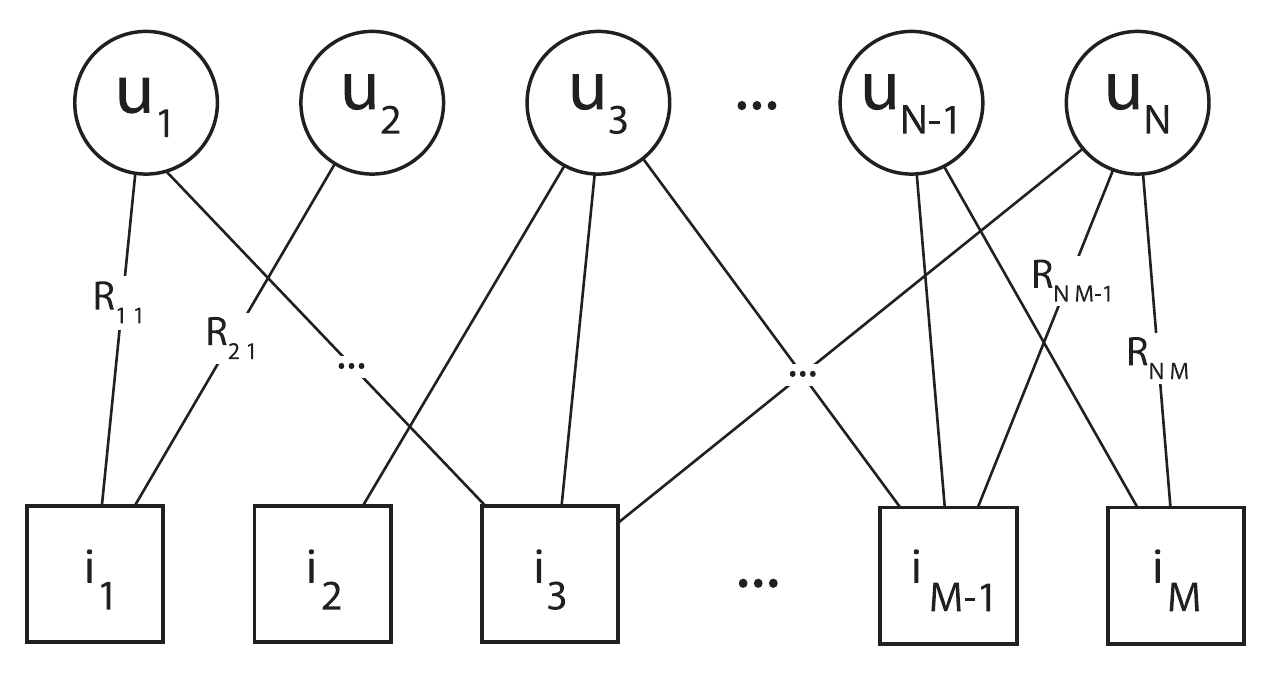}
 	 		 	\caption{Bipartite graph representing $N$ users, $M$ items, and the ratings given by user $u$ to item $i$, $R_{ui}$.}
	 	\label{fig:bipartite}
	\end{figure}

    These methods are more robust to spamming and attacks than the AA.
    Further, the methods above have a bipartite graph structure because there are two types of nodes, users, and items, with weighted edges (ratings) linking the two; see \figurename~\ref{fig:bipartite}.
    
    Subsequently, in~\cite{GrandiT16}, Grandi and Turrini study the resistance to bribery of two ranking systems, the first consists in computing ranking of items as the AA of users ratings, and the second considers the network of influence of each user, using the AA to compute rankings of items for each network.
    In their work, the authors showed that the first ranking system is bribable and that it is profitable to bribe users who did not rate the item.
    However, when using the network of influence, the effect of bribery is dimmed.
    The three main drawbacks of their work are that the set of users is assumed to be fixed, the users only have one item to rate, and, as we pointed out before, the AA does not capture possible ratings multimodal behavior. 
    In the group reputation (GR) method, the reputation of users is computed by the corresponding group sizes, resulting from grouping users according to their rating similarities. A user gets a high reputation if s/he belongs to a large rating group. 
    
    %\note{I think that it's a bit risky to mention~\cite{GrandiT16} (the work we just presented). It was published 1 year before the ICDM paper and it has the same goal, so a comparison with it seems necessary, but we want to avoid that. A possible solution would be to justify in the Experimental setup why it is more reasonable to compare with Li et al. and not with this.}

    In~\cite{GaoZ17}, Gao and Zhou propose an iterative group-based ranking (IGR) method by introducing an iterative reputation-allocation process into the original group-based ranking method. Specifically, ratings from higher reputation users are assigned with more substantial weights in calculating the corresponding group sizes. 
Both the user's reputation and the group sizes are computed as an iterative scheme.
    This work contrasts with ours since we propose to first group users based on their similarities and afterward compute the user reputations with information only from the group where the user belongs. Further, we propose a method that computes a ranking for an item for each group of users, which results in a ranking that better reflects the group preferences, instead of possibly presenting a single ranking weighted by group sizes and user reputations.

    In~\cite{AllahbakhshIMB15}, a set of novel algorithms for robust computation of product rating scores and reviewer trust ranks is introduced. The authors provide a framework consisting of three main components: product rating computation, reviewers' behavior analysis, and reviewers' trust computation.
    The proposed iterative algorithm takes into account the concurrence of votes on the quality of a product, as well as the helpfulness of the cast votes. 
In their method, the weight assigned to the votes of raters is computed from the concurrence of opinions, without any averaging. Our work aims at computing a reputation-based ranking without using any external source of information (like trust and product quality), to introduce a ranking that is based on the behavior of the users, thus making it more adaptive and efficient to compute. 

	%In prior work~\cite{symeonidis2011product}, its authors extended the bipartite graph approaches.
   % The authors considered implicit social networks, known as online Social Rating Networks (SRN) (that emerge from different users commenting similarly on a given set of products), and explicit social networks (built by users, through friendship or working relationship). 
%They used the SRN to recommend products by a weighted combination of users' similarities although not clustering users.

%    *** \cite{rezvani2014secure} ***
%A trust and reputation method ``In this paper we demonstrate that a number of existing iterative filtering algorithms, while significantly more robust against collusion attacks than the simple averaging methods, are nevertheless susceptible to a novel sophisticated collusion attack we introduce. To address this security issue, we propose an improvement for iterative filtering techniques by providing an initial approximation for such algorithms which makes them not only collusion robust, but also more accurate and faster converging. We believe that so modified iterative filtering algorithms have a great potential for deployment in the future wireless sensor networks.''

Rezvani et al.~\cite{rezvani2014secure} aim at improving robustness against collusion attacks by providing an approximation of the existing iterative filtering techniques. Experimental results show that, besides improving collusion robustness, the approach is more accurate and converges faster than state-of-the-art approaches. Our work is applied to a different domain (online ranking systems, instead of wireless sensors networks), so our reputation is associated with human behavior. Moreover, we deal with different types of attacks (love/hate, random spamming, and reputation), while also studying bribery.

%*** \cite{SU201755} ***
The approaches proposed in \cite{SU201755,xu2019meurep} aim at providing robustness when considering QoS (quality of service) data. The approach proposed in \cite{SU201755} introduces the use of unsupervised K-means clustering and Beta distribution-based methods to calculate the reputation of users. QoS is predicted by combining information about similar trustworthy users and similar services. The approach proposed in \cite{xu2019meurep}, named MeURep, is based on a combination of two algorithms that consider that if the QoS data provided by a user is very different from the median, then this user is probably not reliable.
%``We employ unsupervised K-means clustering and Beta distribution based methods to calculate the reputation of users and ensure the credibility of the QoS (quality of service) data. Sections 3.3 and 4 demonstrate that our approach is very efficient and robust without any extra parameter settings. (2) We propose a highly accurate and reliable QoS prediction approach by systematically combing the information of trustworthy similar users and similar services. (3) We conduct several extensive real-world experiments to study the prediction accuracy and robustness of our approach compared with other state-of-the-art approaches.'' 
The difference between our work and these two approaches is at multiple levels. At the domain level, we deal with the ranking of items, while these approaches compute reputation to measure QoS. At the algorithmic level, we distance from~\cite{SU201755}. We do not use the clustering to compute reputation, which is computed after clusters are formed; concerning the work by~\cite{xu2019meurep}, we do not consider the median, but a weighted average of the individual ratings of each item to measure user reliability.

The approach proposed by Tibermacine et al. in \cite{tibermacine2019reputation} is a HITS-based reputation evaluation process that allows to detect malicious users based on the majority voting model, and to assess service reputation after the exclusion of malicious users' feedback ratings. The approach is evaluated on a set of real-world web services to evaluate the proposed process against a selection of similar methods. Our approach uses a ranking algorithm that is not based on HITS, we compute user scores through a weighted average instead of using majority voting, and we compute the reputation of users and not of services.

\section{Notation}\label{notation}
	We now introduce some notation and definitions that we use in subsequent sections.
	Let $U$ be a set of users, $I$ a set of items, $R_\bot,R_\top\in\mathbb Z^+$ the minimum and maximum ratings, respectively. 
    We denote as $\mathcal R=[R_\bot,R_\top]\cap \mathbb Z^+$ the set of strictly positive integers, the allowed ratings as $\Delta_\mathcal R=R_\top-R_\bot$, and as $R\subseteq U\times I\times \mathcal R$ the set of ratings given by users to items.
	For instance, if user $u$ rates item $i$ with rating $R_{ui}$, then we write it as $(u,i,R_{ui})\in U\times I\times\mathcal R$, or simply $R_{ui}\in\mathcal R$.
	We denote the \emph{set of items rated by user} $u$ as $I_u=\{i\,|\,\exists {R_{ui}}\in\mathcal R$~s.t.~$(u,i,R_{ui})\in R\}$, and the \emph{set of users who rated item} $i$ as $U_i = \{u\,|\,\exists {R_{ui}}\in\mathcal R$~s.t.~$(u,i,R_{ui})\in R\}$.
    
     A reputation-based ranking system assigns a reputation, $c_u\in \mathbb R^+$, to each user, $u\in U$, and then utilizes it to weigh their ratings on products to compute the products' rankings.
    
	From now on, we consider normalized ratings (dividing by $R_\top$) and reputations, $R_{ui},c_u\in ]0,1]$ (i.e., the rankings and reputations take values in $]0,1]$). 

\section{Class of Bipartite Reputation-based Ranking Algorithms} % (fold) 
\label{sec:bipartite}
	%[GUILHERME, PLEASE REFINE]
     This section introduces a class of bipartite reputation-based ranking algorithms. 
    We present sufficient conditions for an algorithm in that class to converge and to be efficient.
   We also highlight the limitations of the state-of-the-art approach, presented in the literature.
    
	\subsection{Bipartite graph algorithms} % (fold)
	\label{sub:bi_partite_graph_algorithms}
    Let $\Bb = (U, I, R)$ be a bipartite graph, like the one presented in \figurename~\ref{fig:bipartite}, with two sets of vertices, $U$ and $I$, representing the users and the items. If a user rated an item, then there is an edge with the weight of the rating connecting the two in $\Bb$.   
	We generalize the iterative reputation-based ranking methods as:
	\begin{equation} \label{eq:rbra}
		\begin{cases}
			 r^{k+1} = g_R(c^k) \\%\qquad % \frac{1}{|O_j|} \sum_{u_i\in O_j}R_{ij}c_i^k\\
			 c^{k+1} = h_R(r^{k+1})%, \quad c_i^0 \in ]0,1], %1-\frac{f(\lambda,O_i)}{|O_i|} \sum_{o_j\in O_i}|R_{ij}-r_{j}^{k+1}|^p.
		\end{cases},
	\end{equation}
	where $k$ denotes the iteration index and $c^{0}$ the vector of initial reputation of users, %Coupled with the initial vector of users' reputation $c^0$, 
	with $c_u^0\in ]0,1]$.
	Here, $r=(r_1,\hdots,r_{|I|})$, where $r_i$ denotes the ranking of item $i$, computed by $g_R:[0,1]^{|U|}\to [0,1]^{|I|}$, with the set of ratings, $R$, as a parameter.
	The users' reputations, $c=(c_1,\hdots,c_{|U|})$, where $c_u$ denotes the reputation of user $u$, are determined by $h_R:[0,1]^{|I|}\to [0,1]^{|U|}$.	
	
	In~\cite{li2012robust}, the authors prove that their reputation-based ranking system converges with a certain convergence rate. In this section, we prove that a class of reputation-based ranking systems (more abstract) converges and with what convergence rate. Hence, we present more general results that subsume all the proofs of convergence and efficiency in~\cite{li2012robust}. 
	Hence, we can design a more extensive range of convergent and efficient reputation-based ranking systems.
    
	Consider a Banach space, $\mathcal X$, with an induced distance $d:\mathcal X^2 \to [0,1]$. 
	Using the Lipschitz condition~\cite{kreyszig1989introductory}, we prove what follows.
	\begin{lemma}\label{lemma:1}
		Consider the iterative scheme~\eqref{eq:rbra}. 
		Let $g_R$ and $h_R$ be $\eta_g$ and $\eta_h$-Lipschitz maps, respectively. 
		Then $g_R \circ h_R$ is an $\eta$-Lipschitz map, with $\eta = \eta_g \eta_h$.
		If $\eta<1$, then~\eqref{eq:rbra} is a contraction.~\hfill$\circ$
	\end{lemma}
	Because we are working in a Banach space, if the algorithm~\eqref{eq:rbra} converges, then it converges to a unique value.
	Using the previous lemma, we prove the following results:
	\begin{theorem}\label{th:1}
		The class of iterative reputation-based ranking algorithms~\eqref{eq:rbra} converges.\hfill$\circ$
	\end{theorem}

	\begin{theorem}\label{th:2}
		Let $d:X\to [0,1]$ be a normalized distance.
		Then the algorithm~\eqref{eq:rbra} has exponential rate of convergence. \hfill$\circ$
	\end{theorem}	
	To attain, at most, an error of $\varepsilon>0$, we need $\kappa=\log_{\eta}\varepsilon$ iterations, with $\eta$ the Lipschitz constant of~$g_R\circ h_R$.

	\subsection[]{Unintuitive Properties of Ranking System in~\cite{li2012robust}} % (fold)
	\label{sub:undesirable}    
	In~\cite{li2012robust}, \cite{li2012robust} propose an iterative reputation-based ranking system.
    Their iterative scheme to compute the ranking of item $i\in I$ for iteration $k$ with parameter $\lambda\in]0,1[$ for L1-AVG (for the others the computation of the reputation changes, but the properties still hold) is: \begin{equation}\label{eq:wtf}r_i^{k+1} = \frac{1}{|U_i|}\sum_{u\in U_i}c_u^{k}R_{ui}\,\text{, and }\,c_u^{k+1}=1-\frac{\lambda}{|I_u|}|R_{ui}-r_i^{k+1}|,\end{equation}
    where both rankings and reputations are in $]0,1]$, and $c_i^0$ is any initial value in that domain.
    This leads to the following two unintuitive properties:
    \begin{itemize}
    	\item if all $u\in U_i$ rate $i\in I$ with the same rating, $R_{ui}=R$ for all $u\in U_i$, then the ranking of $i$ is almost never $R$, $r_i\neq R$, unless all users $u\in U_i$ have the same reputation;
    	\item if all $u\in U_i$ rate $i\in I$ with $R_\bot$, $R_{ui}=R_\bot$ for all $u\in U_i$, then the ranking of $i$ is almost always smaller than $R_\bot$, $r_i< R_\bot$, unless all users $u\in U_i$ have the same reputation.     
    \end{itemize}
    We exemplify the previous assertions with the following.
    Let $U=\{u_1,u_2,u_3\}$, $I=\{i_1,i_2,i_3\}$, $R_\bot=1$, and $R_\top=5$, and the following ratings are given by users (rows) to items:
    $$
    R={\small
    \begin{pmatrix}
    1 & 4 & 1\\
    1 & 2 & 5\\
    1 & 1 & 5
    \end{pmatrix}}.
    $$
    In Table~\ref{tab:wtf}, we summarize the values of $r_{i_1}$ that result from applying the ranking system in~\eqref{eq:wtf}, using different values of $\lambda$.
    We first divide the ratings by $R_\top$ so that they are normalized, and, at the end, we multiply again by $R_\top$ to rescale to the original rating domain.
    Note that it would make sense for the value of $r_{i_1}$ to be at least $R_\bot=1$ and would be exactly $1$ because all users rated item $i_1$ with $1$.
    \begin{table}[h!]
    {\small
    \begin{tabular}{ | c | c c c c c | }
    \hline
    $\lambda$ & 0.1 & 0.3 & 0.5 & 0.7 & 0.9 \\ \hline
    $r_{i_1}$ & 0.981 & 0.941 & 0.900 & 0.856 & 0.806\\
    \hline
    \end{tabular}
    }
    \caption{$r_{i_1}$ for different values of the parameter $\lambda$.}
    \label{tab:wtf}
    \end{table}
    
    It is worth noting that these undesirable properties come from the fact that the ranking should be computed as the weighted average by the reputations, instead of~\eqref{eq:wtf}. Moreover, the use of the expression~\eqref{eq:wtf} makes the proofs of convergence in~\cite{li2012robust} more trivial than if we actually use the weighted average, as we detail in Section~\ref{convergence}, because we no longer have a constant factor $|U_i|$ that pops out of the norm, but instead a summation of reputations.
    
\section{A Multipartite Reputation-Based Ranking System}\label{sec:multipartite}
    To enable the multimodal rating behavior of the users in the ranking, this section presents our group ranking system, which employs a multipartite graph instead of a bipartite one.
    Our approach works in four steps:
    \begin{enumerate}
      \item {\em User similarity extraction.} Considering the ratings of two users for the items rated by both, we introduce three measures to compute their similarity. 
      \item {\em Extraction of the multipartite graph.} Given the similarities between the users, we build a new graph that connects two users if their similarity is above a threshold. We merge this graph with the bipartite graph that models the rating of the users, to form a multipartite graph.
      \item {\em Group detection.} The multipartite graph is split into subgraphs, considering the connected components in it.
      \item {\em Reputation-based ranking computation.} Given the users in a subgraph, we propose an algorithm that iteratively updates both the reputation of the users, according to the ratings they gave and the ranking estimations, and the ranking of the item, according to the ratings users gave and their reputations.
    \end{enumerate}
    All these tasks are further described below.
    \subsection{User similarity extraction}
    To group the users, we need to quantify how similar they are.
		For each pair of users that rated, at least, one item in common, we compute a similarity, based on the rating information.
		We specify three similarities: one linear and two non-linear.
		In the following, let $I_{u,v}=I_u \cap I_v$ denote the set of items that both users $u$ and $v$ rated.

		\textbf{Linear similarity.}
			\emph{We define the \emph{linear similarity} as: $\mathit{LS}(u,v)=0$ if $I_{u,v} = \varnothing$, and otherwise
			\begin{equation*}
				\mathit{LS}(u,v) = \ell(|I_{u,v}|)\left[1- \frac{1}{|I_{u,v}|} \sum_{i\in I_{u,v}}\frac{|R_{ui} - R_{vi}|}{\Delta_\mathcal R}\right], 
			\end{equation*}
			where the function $\ell:\mathbb Z^+\to [0,1]$ penalizes on how confident we are in the users' similarity.}
		LS is a linear function of the absolute rating difference, encoding the similarity between users, based on ratings of common rated items.
		If two users used the same rating for an item, the rating difference is zero, hence the similarity is 1, on the other hand, if the rating difference is $\Delta_R$ then the similarity is 0.	
			
		Next, we propose two compression-similarities based on Kolmogorov complexity~\cite{cover2012elements}.
		Given the description of a string, $x$, its \emph{Kolmogorov complexity}, $K(x)$, is the length of the smallest computer program that outputs $x$.
		In other words, $K(x)$ is the length of the smallest compressor for $x$. 
		Although the Kolmogorov complexity is non-computable, there are efficient and computable approximations by compressors.
		Let $C$ be a compressor, and $C(x)$ denote the length of the output string resulting from the compression of $x$ using $C$.
		 % such that $|K(x)- C(x)|\leq |x|+\mathcal O(\log|x|)$.
		%Let $u$ and $v$ denote a description of items and the respective ratings given by two different users. 		
			
		\textbf{Compression similarity.}
		 	\emph{Based on the \emph{normalized compression distance}~\cite{li2004similarity}, we define the \emph{compression similarity} as 1 minus the distance, i.e., $\mathit{CS}(u, v)=0$ if $I_{u,v} = \varnothing$, and otherwise
		 	\begin{equation*}
		 		\mathit{CS}( u, v) = 1 - \frac{ C(\tilde u\tilde v) - \min\{C(\tilde u),C(\tilde v)\}}{ \max\{C(\tilde u),C(\tilde v)\} },
		 	\end{equation*}
		 	for the string $\tilde u\tilde v$, the concatenation of $\tilde u$ and $\tilde v$. For each user, $u$, we denote by $\tilde u$ the string composed by the concatenation of the pairs $(item, rating)$ of her/his rated items.}
Intuitively, we measure the information (rating pattern) that users $u$ and $v$ have in common and normalize it, by subtracting the minimum and dividing by the maximum. If $u$ and $v$ have the same rating pattern, then $C(\tilde u\tilde v)\approx C(\tilde u) = C(\tilde v)$ and $\mathit{CS}( u, v) \approx 1$, while if the rating patterns are completely different (they have nothing in common) we have that $C(\tilde u\tilde v)\approx C(\tilde u) + C(\tilde v)$ and $\mathit{CS}( u, v) \approx 0$.
	Trivially, when the distance is maximum, 1, the similarity is minimum, 0, and vice-versa.

		The main drawback is that CS needs to compute the compression of each possible pair of users with common rated items.	
		To overcome this, we propose a nonlinear function of the absolute disparity of users descriptions' compressions, with lower time complexity.
			
		\textbf{Kolmogorov similarity.}
			\emph{We define the \emph{Kolmogorov similarity} as: $\mathit{KS}( u, v)=0$ if $I_{u,v} = \varnothing$, and otherwise
			% otherwise $\mathit{KS}( u, v)=\left(1+|C(\tilde u)-C(\tilde v)|\right)^{-1}$.}
			\begin{equation*}
				\mathit{KS}( u, v)=\frac{1}{1+|C(\tilde u)-C(\tilde v)|}.
			\end{equation*}
			}
            When the size of the compression of user $u$ and user $v$ rating patterns is the same, $KS( u, v)=1$, and $KS( u, v)$ goes to $0$ when the absolute value of the compression sizes difference goes to infinity.
			% If we think of $\tilde u$ as a training set and the compressor $C$ as a taste/preferences profiler of the users, then we can aggregate users $u$ and $v$ by computing the difference of their preferences $|C(\tilde u)-C(\tilde v)|$.		
		%For users that share, at least one, rated item note that the $\mathit{LD}$ and $\mathit{CD}$ distances only evaluate the information contained in the common rated items. % that both users rated, 
		%Whilst $\mathit{KD}$ accounts for the information of all rated items that each user rated.
		%This marks a clear distinction between those distances that we explore in the experimental results.
		% paragraph compression_similarity_distance (end)
	% subsection similarity_measures (end)
    \subsection{Extraction of the multipartite graph}
    Given a similarity measure $\mathit{SM}$ and a specified affinity level threshold, $\alpha$, we build a graph $\Gg = (U, E)$, with the set of users as vertices and where two users are connected if $\mathit{SM}(u,v)>\alpha$. More specifically, let $S$ be the (possible sparse) adjacency matrix and $S_{u,v}=1$ if $\mathit{SM}(u,v)>\alpha$ and $0$ otherwise. Then $S$ characterizes the undirected graph $\Gg \equiv \Gg(S)$. A large $\alpha$ means that users need to be more strongly related in order to be connected, translating to a larger number of clusters (automatically computed).
    
    Given the bipartite graph $\Bb$ that models the ratings of the users and the graph $\Gg$ that associates similar users, we generate a multipartite graph $\Mm$\footnote{A multipartite graph is a graph such that two vertices that are connected by an edge have different colors~\cite{bollobas2013modern}. Here, we need one color for the items and at least two more whenever there is a cluster with more than one user.}. An example of multigraph is depicted in \figurename~\ref{fig:multipartite}. 
    
    \subsection{Group detection}
    This step groups together users with similar preferences, extracting them from $\Mm$. To do so, we compute the subnetworks of $\Mm$, $\Mm_j$ for $j\in\mathcal J$, which are the $|\mathcal J|$ connected components of $\Mm$. %Each time a new user enters the system or an existing user changes his/her rating behavior, all have to be recomputed.
	%However, the update may be performed only from time to time, not assigning new users to clusters until either some time has passed or there is a sizable amount of new information. 
    
    \subsection{Reputation-based ranking computation}\label{sub:ranking}
     For each subnetwork $\Mm_j$, we apply a reputation-based ranking algorithm to compute the reputation of users and the ranking of items. 
    Here, we show how to compute the ranking of an item and the reputation of a user. %, by considering the algorithms defined in equations~\eqref{eq:g} and~\eqref{eq:h} below.
	We compute the ranking of the item, $r_i$, as a weighted average.
	That is, the rating of user $u$ to item $i$ is weighted by the user reputation, $c_u$, and therefore $g_R$ in~\eqref{eq:rbra} becomes:
	\begin{equation}\label{eq:g}
		r_i^{k+1}= \sum_{u\in U_i}R_{ui}c_u^{k+1}\bigg/\sum_{u\in U_i}c_u^{k+1}.
	\end{equation}
    It is worth highlighting that our formulation of ranking differs from the one presented in~\cite{li2012robust} that, instead of normalizing by the sum of the users' reputations, divides by the number of users that rated the item $i$, $|U_i|$. 
    Our definition allows us to have a ranking that is based on the reputation of the users (thus more robust), but it makes more challenging to prove the convergence of the method, which we present in Section~\ref{convergence} (indeed, having a sum, instead of a constant value, means that we cannot simply get the constant $|U_i|$ out of the norm).
    
	For $h_R$ in~\eqref{eq:rbra}, we tested three functions, parametrized by $f_{\lambda,s}$:
	 \begin{equation}\label{eq:h}
	 	c_u^{k+1}=1-f_{\lambda,s}(I_u) e_{R,u}(r),
	 \end{equation}
	 where
	 \begin{equation*}
	 	e_{R,u} = \begin{cases}
	 		\frac{1}{|I_u|} \displaystyle\sum_{i\in I_u}|R_{ui}-r_{i}^{k}|^p \\
	 		\underset{i\in I_u}{\max}|R_{ui}-r_{i}^{k}|^p \\
	 		\underset{i\in I_u}{\min}|R_{ui}-r_{i}^{k}|^p
	 	\end{cases}.
	 \end{equation*}
     
%	\begin{equation}\label{eq:h}
%		c_i^{k+1}=1-f_{\lambda,s}(O_i) \cdot
%		\left\{
%			\frac{1}{|O_i|} \displaystyle\sum_{o_j\in O_i}|R_{ij}-r_{j}^{k}|^p,\quad
%			\underset{o_j\in O_i}{\max}|R_{ij}-r_{j}^{k}|^p,\quad
%			\underset{o_j\in O_i}{\min}|R_{ij}-r_{j}^{k}|^p
%		\right\}.
%	\end{equation}
	The users' reputation is chosen as a function of the average, maximum, or minimum disagreement of individual user's ratings, $R_{ui}$, and the rankings of the rated items, $r_i$.
	In order to control the penalization a user incurs on, for not rating according to the ranking, we define a \emph{decay function} $f_{\lambda,s}$. 
	We consider four decay functions:\\
	
	\begin{tabular}{l l}
	    $ i)\,$ $f_{\lambda,s}^1(x) = \lambda$,
	    & $\qquad ii)\,$ $f_{\lambda,s}^2(x) = \lambda \left(1 - e^{-\frac{x}{2}}\right)$,  \\
	    $iii)\,$ $f_{\lambda,s}^3(x) = \lambda \left[ 1-\frac{1-\upsilon}{1+e^{s-x}} \right]$,
	    & $\qquad iv)\,$ $f_{\lambda,s}^4(x) = \begin{cases}
			1 & \text{ if }x\geq 10\\
			1/2  & \text{ otherwise}
		\end{cases}$,
	\end{tabular}
	\\
	where $\lambda \in[0,1[$ determines the penalization a user occurs in for rating differently than the ranking, $\upsilon\in ]0,1[$ is the lowest penalization an user can incur, and $s \in \Nn$ is a parameter based on the number of rated items such that the penalization is decreased by a half.	
	The role of the decay function is to control the penalization a user $u$ suffers if it does not rate the item, $R_{ui}$, close to its ranking $r_i$.
	The first, constant, function $f_{\lambda,s}^1$ above is proposed in \cite{li2012robust}, the second is an exponential decrease function, the third is a logistic function, while the fourth is a threshold function.
	In the second and third cases the penalization increases and decreases, respectively, with the number of rated products.
	In the remaining of the paper we fix for $h_R$ the average and for $f_{\lambda,s}$ the constant function, $f_{\lambda,s}^1$, denoting by \emph{bipartite weighted average} (BWA) the resulting iterative scheme in equations~\eqref{eq:g} and~\eqref{eq:h}. The choice to fix these two functions is because they are easy to compute and, considering the different $f_{\lambda,s}$ and the datasets used to evaluate our proposal, there is not much difference between the functions (more details are provided in Section~\ref{sec:experimental_results}). Hence, they represent a good trade-off between efficiency and effectiveness.
	\paragraph{Convergence}\label{convergence} % (fold)
	\label{ssub:convergence}
		Here, we prove the convergence of the proposed method.
		In what follows, for a given vector $x\in\mathbb R^n$ and $p\in\mathbb Z^+$, the $p$-norm of $x$ is $\|x\|_p=(\sum_{j=1}^n|x_j|^p)^{\frac{1}{p}}$, and the $\infty$-norm is $\|x\|_{\infty} = \max_{j\in\{1,\ldots,n\}}|x_j|$.
		\begin{lemma}\label{prop:convergence}
			For all $\lambda\in[0,(1+\Delta_R)^{-1}[$, the iterative method in~\eqref{eq:rbra} with functions $g_R$ and $h_R$ defined as in~\eqref{eq:g} and~\eqref{eq:h} converges.\hfill$\circ$
		\end{lemma}
		In this work, we consider $\Delta_\mathcal R = 1-0.2$. 
		Therefore, if $\lambda\leq \frac{5}{9}$ then the algorithm converges.
		However, we may ensure convergence for any $\lambda\in[0,1[$ changing the denominator of~\eqref{eq:h} to $\max\{\|c^{k+1}\|_1,1\}$.
    
	% subsection bi_partite_graph_algorithms (end)
		
	\subsection{Algorithmic summarization and computational complexity analysis} % (fold)
	\label{sub:summarization}
%		Consider a similarity measure, $\mathit{SM}$. 
		
%		Then, we select the most relevant subnetworks $\{\Mm_j\}_{j\in \mathcal J}$, with $\mathcal J\subseteq \mathcal I$. 
        
	% subsection implicit_user_based_subnetworks (end)	
	%Let $\mathit{rep\_rank}$ denote a reputation-based ranking algorithm~\eqref{eq:rbra}.

		Algorithm~\ref{alg:Rankingsubnet} summarizes our approach. Its time complexity is given by the sum of the complexities of each step.
		%Let $\Gg=(V,E)$ denote a graph, where $V$ is a set of vertices and $E$ a set of edges.
		Step 3 builds $\Mm$, where $V=U$, this is done computing its sparse adjacency matrix, $\mathcal M$, where each rating is used once. 
		Hence, the time complexity is $\mathcal O(|C||R|)$, where $|C|=O(1)$ for similarities LS and KS, and where, for the CS, $|C|$ is the worst case complexity of compressing the concatenation of pairs of users. 
		Step 4 can be performed using Tarjan's Algorithm~\cite{hopcroft1973algorithm}, with time complexity in the worst case of $\mathcal O(|V|+|E|)$.
		Step 5 has, in the worst case, the same time complexity of~\cite{li2012robust}, i.e., $\mathcal O(\kappa|R|)$.
		Hence, Algorithm~\ref{alg:Rankingsubnet} has worst case time complexity of $\mathcal O\left((\kappa+|C|)|R|+|V|+|E|\right)$.
		In theory $|E|$ can be, in the worst case $|U|^2$, leading to a time complexity of $\mathcal O\left((\kappa+|C|)|R|+|U|^2\right)$.
		In practice, since users are often sparsely connected in $\Gg$, $|E|=\mathcal O(|U|)$, so the time complexity is $\mathcal O\left((\kappa+|C|)|R|+|U|\right)$.
		In all cases, the space complexity of Algorithm~\ref{alg:Rankingsubnet} is $\mathcal O(|R|)$.
        \begin{algorithm}
		\caption{Clustering reputation-based ranking algorithm.}
		\label{alg:Rankingsubnet}
		\begin{algorithmic}[1]
			\STATE{\textbf{input}: $\alpha$, \emph{dataset}}
			\STATE{\textbf{build} $S$ from dataset and apply threshold $\alpha$}
			\STATE{\textbf{build} $\mathcal M$, computing its adjacency matrix $\Mm\equiv\Mm(S)$}
			%\STATE{\textbf{apply} threshold \alpha}
			\STATE{\textbf{find} the connected components of $\Mm$, $\{\Mm_j\}_{j=1}^m$}
			% \FOR{$i$ from $1$ to $m$}
			% 	\STATE{\textbf{compute} $\mathit{rep\_rank}(\Mm_i)$}
			% \ENDFOR
			\STATE{\textbf{output}: weighted average of $\{\mathit{rep\_rank}(\mathcal M_j)\}_{j=1}^m$}
		\end{algorithmic} 
		\end{algorithm} 
	% subsubsection complexity_analysis (end)	
% section implementation (end)

Notice that Algorithm~\ref{alg:Rankingsubnet} can be generalized to {\em any} ranking system. This means that, in step~5, we may replace the $rep\_rank$ method by any ranking system.

\section{Bribing in Ranking Systems} % (fold)
\label{sec:bribing_in_ranking_systems}
	Here, we analyze the effect of bribing for both bipartite and multipartite reputation-based ranking systems.
	This is an important and common scenario, because consumers rely on rankings to make decisions, see~\cite{sparks2011impact}.
	Henceforth, since the sellers are aware of how  ratings impact sales~\cite{chevalier2006effect,dellarocas2007exploring}, they need robust ranking systems.
	
	To simplify the analysis, we assume that in both scenarios, when a user rates an item or changes its previous rating on the item, the reputation of each user is not recomputed.
	Otherwise, it would render the computations much more difficult.
	In the experimental results, we compare this situation with the one where the reputations are recomputed.
	Suppose that the seller of item $i$ has, initially, a \emph{wealth} (and popularity) proportional to the ranking of the item and the number of users that rated the item.
	A company may invest its wealth to persuade users directly or indirectly to buy an item, for instances, by giving free samples of the item, by offering discount vouchers, by paying directly to users to rate/review the item. 
	In turn, the users start to like more the product, not necessarily loving it.
	\subsection{The setup} % (fold)
	\label{sec:the_setup}
	
    	 In the bipartite reputation-based ranking system (BRS), we denote the \emph{wealth} of seller $i$ by $J_i$, while in the multipartite reputation-based ranking systems (MRS) we denote it by $\bar J_i$. 
    	 The two quantities are calculated as follows:
        
		\begin{equation}\label{eq:weath_in}
				J_i = |U_i|r_i, \text{ and }\bar J_i=\sum_{n\in \mathcal N_i}J_i^{\mathcal M_n},
		\end{equation}
		where $\mathcal N_i=\{k:\exists_{u\in U_i}u\in \mathcal M_k\text{ and }k=1,\hdots,m\}$ and $J_i^{\mathcal M_n}=|U_i^{\mathcal M_n}|r_i^{\mathcal M_n}$.
        
        % \todo{In Section 4.1, $\mathcal X$ denotes the Banach space. Shall we denote $\mathcal X_i$ differently?}
			
		The strategy of seller of item $i$ consists in targeting a group of users and investing its wealth on those users so that they either rate or increase their ratings for item $i$. 
		We represent such a strategy as a vector $\sigma^i\in \mathcal S_i$, where $ \mathcal S_i\subseteq [0,1]^{|U|}\setminus\{\mathbf{0}\}$ and $\mathbf{0}$ is the null strategy where no user is bribed.
		The elementary strategy, $\sigma^i_u$, consists in investing some part of the wealth, denoted as $\rho_u$, in a single user, $u$.
		Further, $\rho_u+R_{ui}\leq 1$, whenever user $u$ rated item $i$, i.e., user $u$ can only change the given rating to the maximum allowed rating.
		We denote the set of strategies of item $i$'s seller that consist in bribing users that already rated item $i$ by $\Xi_i=\{\sigma^i\in \mathcal S_i:\sigma^i(u)=0\text{ for all }u\notin U_i\}$.
		To easy notation, instead of $\sigma^i(u)$ we write $\sigma^i_u$ to denote the effect of strategy $\sigma^i$ on user $u$.
		Analogously, we denote the set of strategies, of the seller of item $i$, that consists in bribing users that did not rate the item $i$ by $\bar \Xi_i=\mathcal S_i\setminus\Xi_i$. 
		
		The wealth that the seller of item $i$  spends by playing strategy $\sigma^i$ is $\|\sigma^i\|_1=\sum_{u\in U}\sigma^i_u$. 
		A strategy $\sigma^i$ is \emph{elementary} whenever there is $u\in U$ s.t. $\sigma^i_u\geq 0$ and for all $v\in U$ with $v\neq u$ $\sigma^i_v=0$.
		A strategy is \emph{compound} whenever it is not elementary.
		
		Let $U_{\sigma^i}$ denote the set of users that rate item $i$ and $r_{\sigma^i}$ the ranking of item $i$ after strategy $\sigma^i$ being played, respectively. 
		%\note{In the previous sentence is $r_{\sigma^i}$ the rating, as we state, or the ranking? If it is the rating, we used a different notation, with a capital R.}
		Similarly, for $n\in\mathcal N_i$, let $U_{\sigma^i}^{\mathcal M_n}$ and $r_{\sigma^i}^{\mathcal M_n}$ denote the set of users in $\mathcal M_n$ that rate item $i$ and the rating of item $i$ in $\mathcal M_n$, respectively.  
		After playing strategy $\sigma^i$, the wealth of item $i$ seller, in the bipartite and multipartite cases, becomes:
		\begin{equation}\label{eq:wealth_fin}
			J_{\sigma^i}=|U_{\sigma^i}|r_{\sigma^i} - \|\sigma^i\|_1 \quad \text{ and } \quad \bar J_{\sigma^i}=\sum_{n\in \mathcal N_i}|U_{\sigma^i}^{\mathcal M_n}|r_{\sigma^i}^{\mathcal M_n} - \|\sigma^i\|_1,
		\end{equation}
		respectively.
		Therefore, the \emph{profit}, or \emph{return}, of playing strategy $\sigma^i$, in the bipartite and multipartite cases, is 
		\begin{equation}\label{eq:profit}
			\pi_{\sigma^i}=J_{\sigma^i}-J_i \quad \text{ and } \quad \bar \pi_{\sigma^i}=\bar J_{\sigma^i}-\bar J_i,
		\end{equation}
		respectively. 
		
		Hence, when designing an optimal bribing strategy for item $i$ seller, we should maximize the profit, among allowed strategies of $\Upsilon_i$, by addressing the following optimization problem: 
		\begin{equation} \label{eq:gen_opt_prob}
			\begin{aligned}
				\text{maximize: } & \pi_{\sigma^i} 
				\\ 
				\text{subject to: } & \|\sigma^i\|_1 \leq J_i, \quad \sigma^i\in \Upsilon_i,
				\end{aligned}
			\end{equation}
			where $\Upsilon_i$. For the multipartite scenario, we replace $\pi_{\sigma^i}$ by $\bar{\pi}_{\sigma^i}$ and $J_i$ by $\bar J_i$ in~\eqref{eq:gen_opt_prob}.
	% subsubsection the_setup (end)
    \subsection{Optimal bribing in bipartite ranking systems} % (fold)
	\label{sec:bipartite_ranking_systems}
	    In this section, we present the optimal strategies that a seller should use in order to maximize its profit, when subject to a bipartite ranking system.
	   
	    We analyze three distinct cases; in the first one the seller of item $i$ bribes users that already rated the item, so that the buyer changes its rating. 
	    In the second case the seller bribes users that did not buy a product $i$ previously. 
	    Finally, as last case, we study the mixed one, where the seller bribes both types of buyers.
	    
	    % \subsubsection{Changing opinion of raters, $u\in U_i:$\\}
	    % \*** In this section we study the cases ***
	    
	    We start by exploring what are the conditions that the seller of item $i\in I$ needs to verify such that  bribing users that already rated item $i$ to increase the rating yield a profitable bribing strategy.
	    \begin{prop}[Bribing users that rated the item]\label{prop:1}
	        Let $U_i$ denote the set of users that rated the item $i$.
	        Consider that the seller of item $i$ bribes $n$ buyers to increase their ratings on item $i$, where $n \leq |U_i|$.
	        For a single user $u$, this strategy is profitable when its reputation is larger than the average reputation of users that already rated item $i$:
	        \[c_u > \bar c_{U_i}.\] 
	        Furthermore, when several users are bribed $U_b\subseteq U_i$, the profit is given as the sum of the profits of elementary strategies:
	        %of bribing each user:
	        \[
	            \sum_{u\in U_b} \pi_{\sigma_u^i}, \quad \text{ where } \quad \pi_{\sigma_u^i} = \left(\frac{c_u}{\bar{c}_{U_i}} -1 \right) \rho_u.
		    \]\hfill$\circ$
		    %is an elementary strategy.
	    \end{prop}
	    
\begin{remark}
	\textbf{Observation from Proposition~\ref{prop:1}}. \textit{In order to be profitable, a bribing strategy that bribes users which rated item $i$ should only select users $u\in U_i$ such that $c_u>\bar c_{U_i}$.}
\end{remark}

	    %Observe that Proposition~\ref{prop:1} states that, in order to be profitable, a bribing strategy that bribes users which rated item $i$ should only select users $u\in U_i$ such that $c_u>\bar c_{U_i}$. 
	    
        Subsequently, we check what are the conditions that the seller of item $i\in I$ needs to verify such that bribing users that did not rate item $i$ yet have to fulfill to obtain a profitable bribing strategy. 
        \begin{prop}[Bribing users that did not rate the item]\label{prop:2}
            Let $U_i$ denote the set of users that rated the item $i$.
	        Consider that the seller of item $i$ bribes $m$ user that did not rate item $i$ yet to rate it, where $m \leq |U\setminus U_i|$. 
	        For one user $v\in U\setminus U_i$, this strategy is profitable when 
	        \[
	            \lb c_v < \bar c_{U_i} \bigwedge r_i > \rho_v \rb \quad \bigvee \quad \lb c_v > \bar c_{U_i} \bigwedge r_i < \rho_v \rb.
	        \]
	        When a set of $M$ users that did not rate item $i$, denoted as $V_b$, are bribed to rate it, the profit is given as 
	   %     \[
	   %     \sum_{v \in V_b} \pi^i_v + \frac{1}{\tilde \alpha} \sum_{v\in V_b} \lb c_v \lsb        (M-1) \rho_v - \sum_{w \neq v} \rho_w \rsb \rb,
		  %  \]
		    \[
		        \pi_{\sigma_i} = \frac{1}{\tilde \alpha} \sum_{v \in V_b} \lb \alpha + c_v \rb \pi^i_v + \frac{1}{\tilde \alpha} \sum_{v\in V_b} \lb c_v \lsb        (M-1) \rho_v - \sum_{w \neq v} \rho_w \rsb \rb,
		    \]
		    where $\tilde \alpha = \displaystyle\sum_{u\in U_i} c_u + \sum_{v\in V_b}c_v$. \hfill$\circ$
	   \end{prop}

       \begin{remark}
       	\textbf{Observation from Proposition~\ref{prop:2}}. \textit{If the bribed user $v$ has bigger influence than the average rating of the raters, then her/his opinion must be better then the actual average, otherwise it would drag the rating down by himself. If the influence of user $v$ is smaller than the average, then we might not spend too much effort ($\rho_v < r_i$) to persuade her/him, since his contribution to the rating will not be big enough.}
       \end{remark}

        %According to this proposition, if the bribed user $v$ has bigger influence than the average rating of the raters, then her/his opinion must be better then the actual average, otherwise it would drag the rating down by himself. If the influence of user $v$ is smaller than the average, then we might not spend too much effort ($\rho_v < r_i$) to persuade her/him, since his contribution to the rating will not be big enough.
        
        Finally, we analyze under which conditions the seller of item $i\in I$ obtains profit if it bribes both users that rated and did not rate item $i$. 
        
        % {\color{red} To easy notation and without loss of generality, from now on, we identify $N$ users that rated the item $i$ in the summations by the variable $u$ ranging from $1$ to $M$ and, similarly, $M$ users that did not rate the item $i$ in the summations by the variable $v$ that ranges from $1$ to $M$. This would correspond to apply a permutation of the users labels, and it allow us to avoid a cumbersome notation in the following results.} 
        \begin{prop}[Bribing both users that rated and did not rate the item]\label{prop:3}
            Let $U_i$ denote the set of users that rated  item $i$.
	        Consider that the seller of item $i$ bribes $n+m$ users, $n\leq |U_i|$ users that rated the item and $m\leq |U\setminus U_i|$ that did not rate the item. 
	        If $n=m=1$, $u\in U_i$ and $v\notin U_i$ then the profit of the strategy is
	        \[
    	        \pi_{\sigma_i}=\pi_{\sigma^i_u} + \pi_{\sigma^i_v} + \lb 1- \frac{c_v}{\bar c_{U_i}} \rb \frac{c_u \rho_u}{\alpha+c_v}. 
	        \]
	        In the general case, $n=N$ and $m=M$,  yielding the profit given by  
	        \[
	        \resizebox{.999\hsize}{!}{$
	                \pi_{\sigma_i} = \displaystyle\sum_{u \in U_b} \pi_{\sigma_u^i} + \frac{1}{\tilde \alpha} \displaystyle\sum_{v \in V_b} \lb \alpha +c_v \rb \pi_{v}^i 
	                                    + \frac{1}{\tilde \alpha} \displaystyle\sum_{v\in V_b} \lb c_v \lsb (M-1) \rho_v - \displaystyle\sum_{w \neq v} \rho_w \rsb \rb 
	                                    + \frac{1}{\tilde \alpha} \lsb \sum_{v \in V_b} \lb 1 - \frac{c_v}{\bar c_{U_i}} \rb\rsb \displaystyle\sum_{u\in U_b}c_u \rho_u,
	            $}
	        \]
	       % \[
	       % \begin{split}
	       %     \pi_{\sigma_i}=\displaystyle\sum_{u\in U_b} \pi_{\sigma_u^i} + \sum_{u\in U_b} c_u \rho_u \displaystyle\lb \frac{\alpha M - |U_i| \sum_{v\in V_b} c_v}{\alpha (\alpha+\beta)} \rb +\lb M - \frac{|U_i|}{\alpha} \sum_{v\in V_b} c_v \rb \lb \frac{\alpha r_i + \sum_{v\in V_b} c_v \rho_v}{\alpha + \beta} \rb\\ + \frac{|U_i|}{\alpha} \sum_{v\in V_b} c_v \rho_v - \sum_{v\in V_b} \rho_v,
	       % \end{split}
	       % \]
	        where $U_b\subseteq U_i$ is the set of $N$ bribed users that already rated the item, and $V_b\not\subseteq U_i$ is the set of $M$ bribed users that did not rate the item. \hfill$\circ$ %, and   $\alpha = \displaystyle\sum_{u\in U_b} c_u$ and $\beta = \displaystyle\sum_{v\in V_b} c_v$. \hfill$\circ$
    \end{prop}
	\begin{remark}
		\textbf{Observation from Proposition~\ref{prop:3}}. \textit{In this case, we obtain an expression that decouples the profit in the sum of profits of elementary strategies that correspond to bribe raters (first term from Proposition \ref{prop:1}, second and third terms from Proposition \ref{prop:2}) plus an additional term that weights the bribed users on $V_b$ by the reputation-weighted cost of bribing each user $u \in U_b$. The profitability conditions are just the join profitability conditions in \ref{prop:1} and \ref{prop:2}, since we obtained a decomposition and the signal of last term in the above expression only depends on the summation of the terms $1-c_v / \bar c_{U_i}$.}
	\end{remark}

	\subsection{Optimal bribing in multipartite ranking systems} % (fold)
	\label{sec:multipartite_ranking_systems}
	%	{\color{red} 
		Now, we explore the profit of bribing on the MRS case.
		To simplify the analysis, we assume that, when a user is bribed and changes her/his rating for an item, her/his reputation remains unchanged.
		This assumption is not unrealistic, since not only whenever the user has rated several items her/his reputation change is small if only one of her/his ratings changes, but also because in real systems the re-computation of the reputations is often performed only from time to time.
		We assume that the users' ratings and reputations are publicly available, but the network of users, i.e., the clusters' partition is private. 
		
		The first scenario that we explore is the one of bribing a user that rated the item.  
		\begin{prop}[Bribing a user in a cluster that already rated the item]\label{prop:AllVoteCluster}
			Suppose that $u\in U_i^{\Mm_s}$, for some cluster $s\in\{1,\ldots,N\}$. 
			If $ c_u> \bar c_{U_i^{\Mm_s}}$, then any $\sigma_u \in \Xi_u$ is profitable.\hfill$\circ$
		\end{prop}	

  	  	This result is a corollary of  Proposition~\ref{prop:1} applied to $\mathcal M_s$. 
  	  	Next, we explore the case where a user that did not rate an item is bribed, but the user is in a cluster with users that rated the item. 
  	 
		\begin{prop}[Bribing a user in a cluster to rate a non-rated item in the cluster]\label{prop:7}
			Suppose that $v\in \Mm_s$, for a cluster $s\in\{1,\ldots,N\}$, and consider an item, $i$, that was not rated by any member of the cluster, that is $i\notin I^{\Mm_s}$. 
			In this case, any $\sigma_v \in \Xi_v$ is non-profitable.\hfill$\circ$
		\end{prop}
	
	Last, we explore the scenario where a user that did not rate an item is bribed, and the user is in a cluster without users that rated the item. 
		
		\begin{prop}[Bribing a user in a cluster to rate an item that he did not rate before, but $i\in I^{\Mm_s}$]\label{prop:8}
			Suppose that we want to bribe a user that did not rate item $i$ and the user belongs to a cluster where some user already rated item $i$, in other words, $v\in \Mm_s$, $v\notin U_i^{\Mm_s}$ and $i\in I^{\Mm_s}$.
			The strategy $\sigma_v^i$ is profitable whenever one of the following holds:\\
			$1)$ $c_v < \bar c_{U_i^{\Mm_s}}$ and  $\rho_v < r_i^{\Mm_s}$, $2)$ $c_v > \bar c_{U_i^{\Mm_s}}$ and $\rho_v > r_i^{\Mm_s}.$\hfill$\circ$
		\end{prop}

        \begin{remark}
        	\textbf{Observation from Proposition~\ref{prop:8}}. \textit{In this case, we rediscover the result of Proposition~\ref{prop:2} for one user, but where the average reputation and the ranking of item $i$ are relative to the cluster where the user belongs, $\Mm_s$.}
        \end{remark}
		
	The more general setups in which $N$ users that rated the item are bribed and/or $M$ users that did not rate the item are also bribed are left as future work. Indeed, since in the multipartite case we would need to account for the membership of clusters that may or may not have rated the item, it becomes challenging to obtain closed expressions for the profit of such strategies.
	%	}
	% subsection multipartite_networks (end)
	\subsection{Optimal bribing strategies in multipartite ranking systems} % (fold)
	\label{sub:optimal_strategies_mult}
	   % {\color{red}
		Next, we study the optimal bribing strategies for the MRS, as we did in Section~\ref{sec:bipartite_ranking_systems} for the bipartite ranking systems.
		Again, we consider three scenarios: 
		($i$) bribing users that rated the item; 
		($ii$) bribing users that did not rate the item; 
		($iii$) bribing users from the set of all users.
		We compute the close form of the optimal strategies for some cases, for the others Linear Programming can be used.
	
		To model these problems, we assume that the seller of item $i$ disposes of an initial wealth given by $\bar J_i$, and we consider two reference customers, $u$ and $v$, with reputations s.t. $c_u>c_v$. 
		Notice that we are using here $u,v\in U$ to easy notation, but both users may or may not have rated the item. 
		
		First, we look at the scenario where the users rated the item. 
		\begin{prop}[Bribing users that rated item $i$] % (fold)
		\label{par:bribing_users_that_rated_item_i}
			Consider that the seller wants to bribe users that already rated item $i$, \emph{i.e.} $u\in U_i$, \emph{i.e.,}, recalling~\eqref{eq:gen_opt_prob}, 
			\begin{equation*}
			\begin{aligned}
				\text{maximize: } & \bar\pi_{\sigma^i} 
				\\ 
				\text{subject to: } & \|\sigma^i\|_1 \leq \bar J_i, \quad \sigma^i\in \Upsilon_i,
				\end{aligned}
			\end{equation*}
			where $\Upsilon_i = \Xi_i$. Then, the following hold: 
			\begin{enumerate}
			    \item[$(i)$] if $u,v\in\Mm_s$ are two users that already rated item $i$, then 
			the optimal strategy is: to bribe users by decreasing reputation, investing all the wealth until either the lack of available profitable users ($c_u > \bar c_{U_i^{\Mm_s}}$) or the exhaustion of funds to bribe profitable users.
			
			\item[$(ii)$] if each reference user belongs to distinct clusters, $u\in \Mm_s$, $v\in \Mm_t$ and $s\neq t$, if $|U_{i}^{\Mm_s}|< |U_i^{\Mm_t}|$ then the profit per unit of invested wealth is larger for user $u$ if $|U_i^{\Mm_s}|>(c_u-c_v)^{-1}$ and $|U_i^{\Mm_t}|<(| U_i^{\Mm_s}|c_u-1 )/c_v$, and larger for user $v$, otherwise.\hfill$\circ$
			\end{enumerate}
			\end{prop}
		% paragraph bribing_users_that_rated_item_i (end)
		
		Second, we explore the case where users did not rate the item. 
		\begin{prop}[Bribing users that did not rate the item $i$] % (fold)
		\label{par:bribing_users_that_did_not_rate_the_item_i}
			Under the same conditions for item $i$ seller, suppose that s/he wants to bribe users that did not rate $i$, \emph{i.e.,} $u\notin U_i$.
			We formulate this as \eqref{eq:gen_opt_prob} with $\Upsilon_i = \bar\Xi_i$.
		In this case, the following hold:
		\begin{enumerate}
		    \item[$(i)$] If the users belong to clusters without users that rated item $i$, then the profit is zero. 
		    \item[$(ii)$] If users are in the same cluster, then the optimal strategy is to bribe users by decreasing order reputation, investing all the available wealth until either the exhaustion of profitable users ($c_u > \bar c_{U_i^{\Mm_s}}$) or funds. \hfill$\circ$
		\end{enumerate}
		\end{prop} 
		Notice that in the case that there are  users in distinct clusters and the users  did not rate item $i$, we cannot derive simple conditions and we need to solve a linear program for each instance. 
		
		Next, we explore the general case. 
		% paragraph bribing_users_that_did_not_rate_the_item_i (end)
		\begin{prop}{[General case]} % (fold)
		\label{par:general_case}
			Under the same conditions for item $i$ seller, we consider that all users, $u\in U$, can be bribed.
			The problem of finding the best bribing strategy is written as \eqref{eq:gen_opt_prob} with $\Upsilon_i = \mathcal S_i=\Xi_i\cup\bar\Xi_i$.
			For users in the same cluster $\Mm_s$, 
			the optimal strategy is to order bribable users by decreasing reputation for each of the sets $U_i^{\Mm_s}$ and $U\setminus U_i^{\Mm_s}$, and start allocating wealth to $U_i^{\Mm_s}$ and, afterward, to $U\setminus U_i^{\Mm_s}$. \hfill$\circ$
			\end{prop}
			
		Notice that if the users are in different clusters, we cannot draw simple conditions, and again we need to solve the linear program for each instance.
		%}
	% subsubsection multipartite_ranking_systems (end)	
	\subsection{Bipartite versus multipartite networks} % (fold)
		\label{sub:bipartite_vs_multipartite_networks}

% 		\subsection{Bipartite RS vs. Multipartite RS} % (fold)
% 	\label{sub:bipartite_vs_multipartite_networks}
%    {\color{red}
		Here, we compare the profits obtained in the MRS case and BRS, for same conditions.
		In the case where the user rated the item, we have the following result:
		\begin{theorem}\label{prop:deutica}
			Suppose that the seller of item $i$ wants to bribe a user $v$ that already rated the item, \emph{i.e.,} $v\in U_i$.
			Let the user $v$ be in cluster $\mathcal M_s$, then the profit is larger in the BRS, $\bar \pi_{\sigma^i}<\pi_{\sigma_i}$, if and only if $\bar c_{(U_i\setminus U_i^{\mathcal M_s})}<\bar c_{U_i^{\mathcal M_s}}$, the average of the reputations in $(U_i\setminus U_i^{\mathcal M_s})$ and $U_i^{\mathcal M_s}$, respectively.\hfill$\circ$
		\end{theorem}

        \begin{remark}
        	\textbf{Observation from Theorem~\ref{prop:deutica}}. \textit{There are cases where bribing a user in MRS is more profitable than in BRS. 
        		Since the clusters' partition is assumed to be unknown for the sellers, they cannot determine the users that verify the previous condition. 
        		Unlike users' reputations that are often public.
        		Now, we compare the profit of bribing a user that did not rate the item $i$ in the case the bribed user $v$ belongs to a network where no users rated the item, $v\in \Mm_s$ and $i\notin I^{\Mm_s}$.
        		In this case, bribing user $v$ in MRS yields zero profit, but in BRS the strategy can be profitable, as we showed in Proposition~\ref{prop:2}.
        		In the case that the bribed user did not rate the item, but he belongs to a cluster where some user rated the item, we cannot draw simple conditions as in the previous cases.
        		We need to check for each concrete case which one is the most profitable.}
        \end{remark}

    It should be noted that it is very challenging to compare these two classes of systems theoretically because when the ratings change in the multipartite networks, the user might change its cluster. 
    For this reason, we make this comparison in the experimental results, see Section~\ref{sec:experimental_results}.
		
		%Hence, there are cases where bribing a user in MRS is more profitable than in BRS. 
		%Since the clusters' partition is assumed to be unknown for the sellers, they cannot determine the users that verify the previous condition. 
		%Unlike users' reputations that are often public.
		%Now, we compare the profit of bribing a user that did not rate the item $i$ in the case the bribed user $v$ belongs to a network where no users rated the item, $v\in \Mm_s$ and $i\notin I^{\Mm_s}$.
		%In this case, bribing user $v$ in MRS yields zero profit, but in BRS the strategy can be profitable, as we showed in Proposition~\ref{prop:2}.
		%In the case that the bribed user did not rate the item, but he belongs to a cluster where some user rated the item, we cannot draw simple conditions as in the previous cases.
		%We need to check for each concrete case which one is the most profitable. 
	
%	}
		
		% subsubsection other_ranking_systems (end)
% section bribing_in_ranking_systems (end)
\section{Experimental strategy and setup} % (fold)
\label{sec:experimental_setup}
Next, we detail the metrics we use to evaluate the ranking systems we propose.
Further, we detail the type of attacks and spam, and the bribing strategies that we consider and explore in two datasets. 
In the following experiments, we compare the proposed  reputation-based bipartite ranking system with the closest approach to ours~\cite{li2012robust}. 
Further, we illustrate how the proposed general multipartite ranking system performs when using the bipartite ranking system that we design, but this could be extended to {\em any} ranking system, because the clustering part of the proposed system is independent of the the ranking algorithm that we select. 
Moreover, the study of bribing in reputation-based ranking systems is also more general than the proposed ranking systems that we used to illustrate the theoretical results, and can be applied to any ranking system that is computed as a weighted average of the ratings. 

		\subsection{Evaluation metrics} % (fold)
		This section introduces our strategy to evaluate the robustness against spamming and attacks, and against briebery.
	\label{sub:evaluation_metrics}
		To assess the quality of the ranking systems, we compute the Kendall rank correlation coefficient, a.k.a. \emph{Kendall's tau}\footnote{Given two sets $X$ and $Y$, let $C$ and $D$ denote the sets of concordant and discordant pairs of elements in $X\times Y$, respectively. The Kendall's tau is defined as $\tau=(|C|-|D|)/(|C|+|D|)$.}, $\tau$~\cite{kendall1938new}.
		This statistic measures the ordinal association between two quantities.
		Intuitively, the Kendall correlation between two variables is higher when observations are identical and lower otherwise.
			
		The \textbf{effectiveness} is given by the Kendall tau of the rankings' vector, $r$, versus a ground truth, $\hat r$, that is $\tau(r,\hat r)$.
%		The selection of the ground truth is an intricate matter, for practical reasons we choose the AA.
		Usually the used ground truth is the AA, due to its simplicity and its popularity among ranking systems, \cite{de2010iterative,jurczyk2007discovering}.
		% Because it is commonly used, for its simplicity, and it is a popular method for ranking systems, \cite{jurczyk2007discovering,de2010iterative}.
		However, evaluating the discrepancy between the ranking vector, $r$, and the AA might not be very informative, since it does not capture the possible multimodal behavior of ratings, and therefore might not be very useful to evaluate the quality of a ranking system.
		% When using $f_s(x)=\lambda=0.1$, as in~\cite{li2012robust}, we see that it yield high effectiveness values ($\tau \approx 1$).
		% The reason is that in this case $c\in[0.9,1]$, and we obtain rankings very close to AA, and in the case $c=1$ it yields the AA.
		% This in turn plays against what we want to achieve, a departure from the simplicity of AA, in order to gain insight through an intelligent weighting of ratings.
		%Therefore the effectiveness might not be that important to evaluate the quality of a ranking system.
		%In such a heterogeneous environment the AA do not carry useful information. 
		For this reason, we opt for the robustness metric.
		Notice that, in the multipartite case, the effectiveness is helpful to check for homogeneity within the clusters. % classes/clusters/groups/subnetworks.
		We generalize the Kendall tau as 
		\begin{equation*}
			\begin{split}
				\bar \tau = \frac{1}{|\Mm|} \sum_{j=1}^N |\Mm_j| \tau_{\Mm_j}, \quad
				\Mm = \bigcup_{j=1}^N \Mm_j, \quad \Mm_j \bigcap \Mm_q = \varnothing, 
			\end{split}
		\end{equation*}
		where $\Mm_j$ is a subgraph of $\Mm$. % denotes the number of vertices of the graph.
		We denote the effectiveness of a cluster, $\Mm_j$, by $\tau(r_{\Mm_j},r_{AA|\Mm_j})$. %, measuring its homogeneity.
		
		The \textbf{robustness} evaluates the ability of the system to cope with noise or spamming attacks.
		A noisy user gives random ratings to a random set of products~\cite{aggarwal2016recommender}.
		A spamming attacker targets a set of items with the intent of increasing (\emph{Push Attack}) or decreasing (\emph{Nuke Attack}) their rankings.
		%In our simulations, we run our algorithm in the original dataset and in spammed ones.
		%After, we compute the Kendall tau between the rankings of the original dataset, that we assume as the ground truth, against the rankings obtained with the spammed dataset.
			
		For the multipartite case, the robustness Kendall tau is $\bar \tau = \tau(\bar r,\bar r_{\text{spam}})$, where $\bar r$ is a vector of $\bar r_i$'s given by
		\begin{equation*}
			\bar r_i = \frac{1}{|\hat \Mm|} {\sum_{m} |\hat \Mm_m| r_{i,\mathcal M_m}}, \,\,\text{where }\,\, \hat \Mm = \bigcup_m \hat \Mm_m
		\end{equation*}
		is the union of subnetworks where users rated item $i$, and $r_{i,\mathcal M_m}$ denotes the ranking of item $i$ for the subnetwork $\mathcal M_m$ (if any user in the subnetwork rated the item, otherwise it is undefined).
		%follows:
		% \begin{equation*}
			% \begin{split}
			% 	\Mm = \bigcup_{i} \Mm_i, \quad \Mm_i \bigcap \Mm_j = \varnothing, \quad
			% 	\bar \tau = \frac{1}{|\Mm|} \sum_{i=1}^N |\Mm_i| \tau_{\Mm_i},
			% \end{split}
		% \end{equation*} 
		% where $\Mm_i$ are the subgraphs of $\Mm$, and $|\cdot|$ is the number of vertices of the graph.
		% The effectiveness within a subgroup, $\Mm_i$, is given by $\tau(r_{\Mm_i},r_{AA})$, and it measures the homogeneity of the it.
		This measure is useful to assess the quality of the partition of the original network, and it can be used to tune the affinity level, $\alpha$, between users so that they are in the same cluster.
		For items not ranked in a subnetwork, or for new users, we average the rankings among subnetworks, $\bar r$, using weights proportional to the size of the subnetworks.
		Because the weighted average is not a sufficient statistic, this protects the system against attacks.
        
        We do not present the analysis of the personalization perspective in this paper. That is the analysis of how much closer to the real user preferences our cluster-based ranking system is, compared to the ranking systems that use just the AA or a weighted average of the ratings.
This is because it is trivial to notice that a ranking produced by considering the preferences of highly similar users is more personalized than a global one. Hence, in this work, we focus on the robustness perspective and leave the personalization aspect as future work.

    \subsection{Experimental Strategy}
    In this section, we present our experimental strategy, to evaluate robustness against spamming and attacks, and against bribery.
    
	\paragraph{Robustness against spamming and attacks} % (fold)
	\label{sub:spamming_and_attacks}
		In the bipartite graph scenario, the information available to a new user is every products' rankings.
		This information can be used by malicious users to tamper with the ranking of an item in a malicious way (push or nuke it.)
		For instance, in a reputation-based system, an attacker can give ratings matching the ranking of items to increase its reputation, before attacking an item.
		
		When allowing for subnetworks, either the user is already classified into a cluster and s/he accesses the item's ranking within that cluster, or s/he is a new user. 
		In this case, the displayed ranking, $\bar r_j$, of the item, $j$, is the weighted average of its ranking within each subnetwork. 		
		Both of these scenarios mitigate the spamming effect. 
		Since the information made available is not a sufficient statistic, a user cannot fully recover all the information to efficiently attack the underlying ranking system.
		
		In this section, we discuss the robustness of the algorithm to different kinds of spamming/attacks:
\begin{itemize}

		\item \emph{Random spamming}: A set of spammers gives random ratings, uniformly distributed on $\mathcal R$, to a random number of items, following a Poisson distribution, starting at $1$ with parameter $\lambda_{P} = 5$.
			The rated items are randomly sampled from the initial dataset distribution of  ratings' number per item. 					
		\item \emph{Love/hate attack}: A set of spammers targets one item to push/nuke and selects another set of items to nuke/push.
			In our simulations, each attacker nukes the most voted item and pushes another random set of nine filler items. 
		\item \emph{Reputation attack}: In this case, a set of spammers targets one item to push/nuke its ranking. 
			They randomly select another fixed number of items, from the initial dataset, typically the most popular ones, and give them the closest ratings to their rankings. 
			% We discuss the nuke version, because the push attack is similar.
			% \item Add spamming users that give random ratings to a random set of items.
			% \item Use a bot to nuke one item's rating 
\end{itemize}

		In all experiments, we set $\lambda=0.3$, $\alpha=0.8$, and for $\text{LS}$  the \emph{confidence level function} $\ell(|I_{u,v}|)=\theta^{-1}$ if $|I_{u,v}|\leq \theta$ and $1$ otherwise.
		The parameter $\theta$ sets the number of common rated items of users $u$ and $v$ from which we are confident that they can be similar. We choose $\theta=3$. 
		To evaluate the effect of the attacks/spamming, we compute the robustness Kendall tau, $\tau(\bar r,\bar r_{spam})$.		
% subsection spamming_and_attacks (end)		
% This protects the system against attacks since the weighted average is not a sufficient statistic.
		 % in the same community. 
         
         \paragraph{Robustness against bribery}
         %\todo{I think here we are missing a subsection that states how we are going to evaluate the robustness against bribery, like we just did when talking about the robustness against spamming and attacks. I don't know if this is possible, let's think about this.}
         The robustness against bribery will be evaluated in both synthetic and real data. We compute the attainable profit of bribing for a set of bribing strategies.

         \subsection{Experimental Setup}
         
We run all experiments on MATLAB~2016, using macOS 10.11 (2.8 GHz Intel Core 2 Duo and 4 GB RAM).%\footnote{The code will be released as soon as the anonymity restrictions are no longer required.}
\\	
	\textbf{Datasets.} % (fold)
    In this work, we use two real world datasets obtained from the Stanford Large Network Dataset Collection, \cite{snapnets}.
		% We used the ``Amazon rating movies an TV'' dataset, with $XXX$ users, $YYY$ items and $4,607,047$ ratings.
		% We also used the  ``Automotive'' dataset, with $XXX$ users, $YYY$ items and $1,373,768$ ratings.
		We use, as the first dataset, the $5$-core version of ``Amazon Instant Video'' dataset (\emph{Dataset A}) that consists of users that rated at least $5$ items, as in~\cite{mcauley2015inferring}.
		It has $5,130$ users, $1,685$ items and $37,126$ ratings, with $R_\bot=1$ and $R_\top=5$, see Table~\ref{tab1}.
		We use, as the second dataset, the $5$-core version of ``Tools and Home Improvement'' (\emph{Dataset B}), also in~\cite{mcauley2015inferring}, see Table~\ref{tab1}.
		This dataset has $16,638$ users, $10,217$ items and $134,476$ ratings, also with $R_\bot=1$ and $R_\top=5$.
		Both datasets consist in tuples of the form (user, item, rating, timestamp), and for both we normalize the ratings by dividing them by $R_\top$.
        
        The choice to employ the $5$-core version of the datasets is intrinsically related to the scenario we consider in the evaluation, i.e., a ranking system where users are clustered based on their similarity. Therefore, having information about the user preferences is key to measure effective similarities (indeed, if two users did not rate any common items, their similarity would be 0). 
		\begin{table}
	 	\centering
        {\small
	 	\begin{tabular}{ | c | c | c | }
	 		\hline
	 		 & \textsc{Dataset A} & \textsc{Dataset B} \\ \hline
	 		 \textsc{Users}   & $5,130$  & $16,638$ \\ 
	 		 \textsc{Items}   & $1,685$  & $10,217$ \\ 
			 \textsc{Ratings} & $37,126$ & $134,476$ \\ \hline
	 	\end{tabular}
        }
	 	\caption{Details of the datasets.}
	 	\label{tab1}
	 	\end{table}%
	
%		These datasets include only tuples of the form (user, item, rating, timestamp) and no metadata or reviews.
%		We normalized all ratings, dividing them by $R_{\top}$.	\\
		% $426,922$ users, $23,965$ items and $583,933$ ratings.
		% The majority of the users ($\approx 70\%$) in this dataset only rated a single item.
		% This may not allow us to fully explore the advantages of the decay functions and the clustering.
		% Thereupon, we used the $5$-core version that consist of users that rated at least $5$ items.
		% This dataset has $xxx$ users, $yyy$ items and $37,126$ ratings.
	% subsection datasets (end)		
	\noindent\textbf{Benchmarks.} % (fold)
		We compare our results with the reputation-based ranking system in~\cite{li2012robust}.
		The authors already compared their algorithm with the state-of-the-art algorithms. % several other algorithms.
		Namely, the HITS~\cite{kleinberg1999authoritative}, the Mizz~\cite{mizzaro2003quality}, the YZLM~\cite{yu2006decoding} and the dKVD~\cite{de2010iterative} algorithms, showing that their algorithm outperforms them, in the standard metrics.
	% subsection benchmarks (end)
	% subsection evaluation_metrics (end)
% section experimental_setup (end)
\section{Experimental results} % (fold)
\label{sec:experimental_results}
	In this section, we test the robustness of the ranking systems against spamming (noise) and attacks, and evaluate their resistance to bribery using the two real datasets.
    First, we analyze the behavior of the ranking system in the presence of noise for the two datasets, in Section~\ref{sub:robustness_against_noise}.
    Next, we evaluate the robustness of the algorithms against Love/Hate and Reputation attacks, in Section~\ref{sub:robustness_against_attacks}.
    We discuss how the robustness of the proposed ranking systems responds to changes in the parameters of the system, namely the parameter $\alpha$, in Section~\ref{sub:sensivity_to_parameters}. 
    We also test this response for the different decay functions $f_{\lambda,s}$ and different parameters $\lambda$, but since the gains are small, we omit the tests.
    Finally, in Section~\ref{sub:robustness_against_bribery}, we study the robustness of the ranking systems against bribery.
    	\begin{figure*}
			\centering			
		    \begin{subfigure}[b]{.38\textwidth}
		        \includegraphics[width=\textwidth]{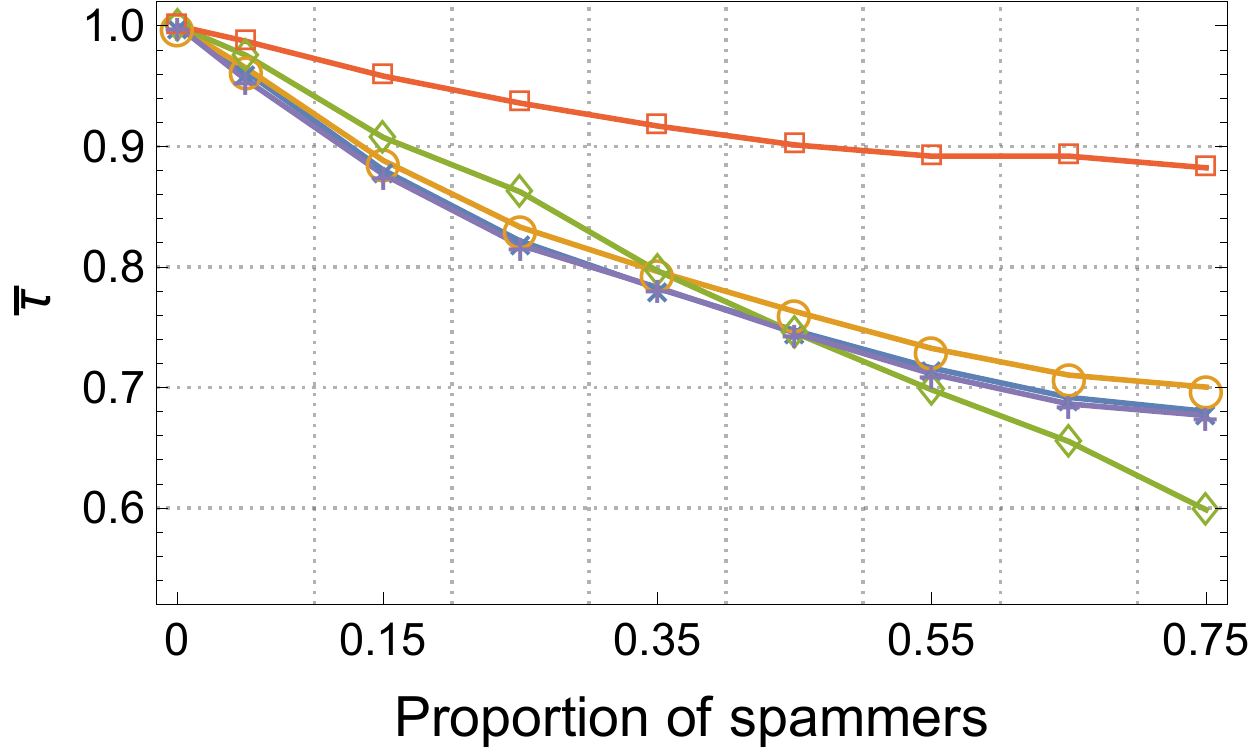}
		        %\caption{Tau evolution for random spamming.}
		        \caption{Dataset ``Amazon Instant Video''.}
				\label{fig:rand_tau}
		    \end{subfigure}		    
		    \begin{subfigure}[b]{0.38\textwidth}
		        \includegraphics[width=\textwidth]{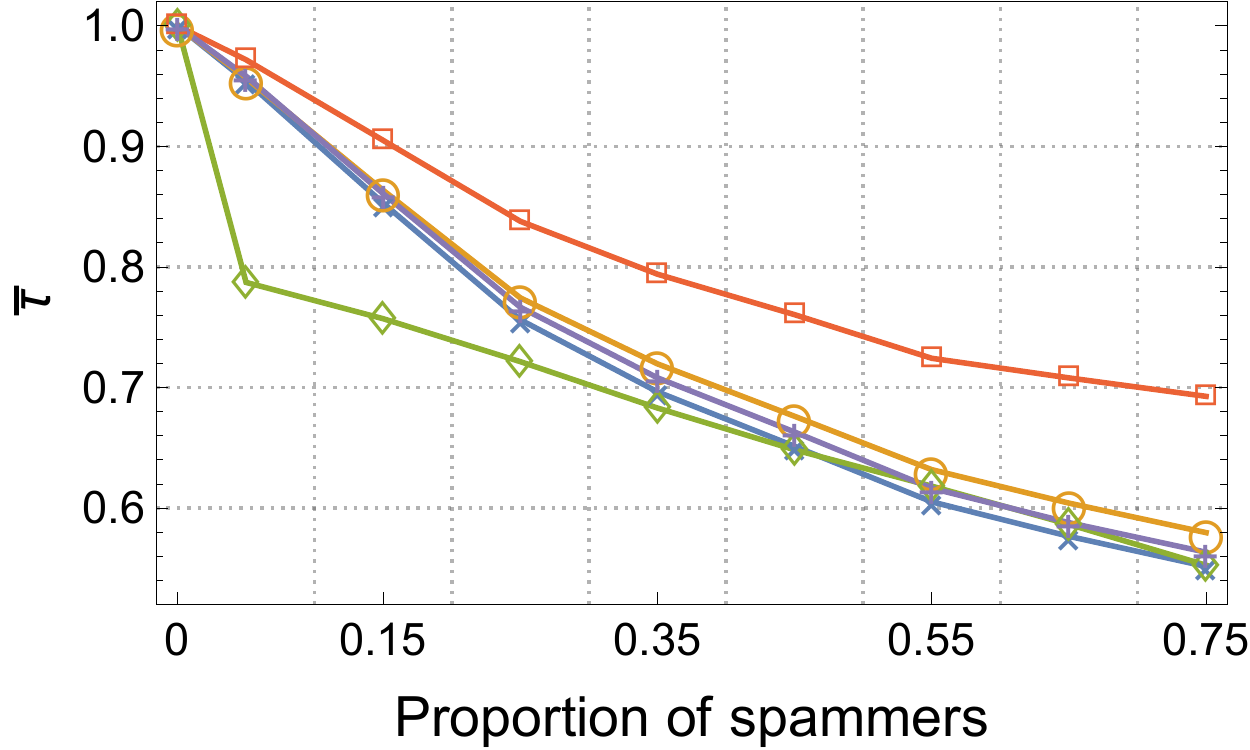}
		        %\caption{Tau evolution for random spamming.}
		        \caption{Dataset ``Tools and Home Improvement''.}
				\label{fig:rand_tau2}
		    \end{subfigure}
		    \begin{subfigure}[b]{0.37\textwidth}
		        \includegraphics[width=\textwidth]{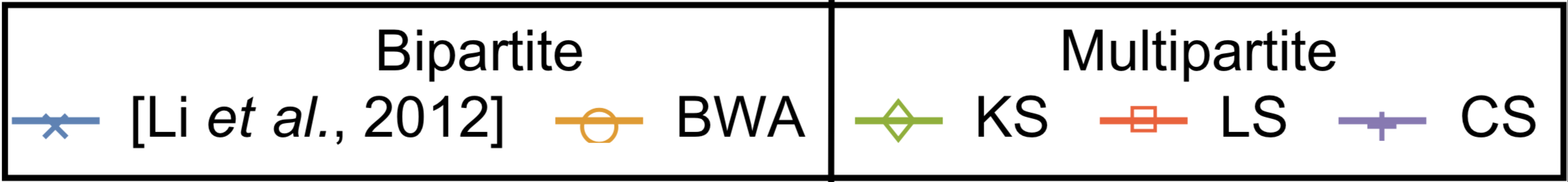}
		    \end{subfigure}
			\caption{Evolution of the $\bar \tau$ for random spamming with the proportion of spammers.}
			\label{fig:random_tau}
		\end{figure*}
		\begin{figure*}
			\centering
			\begin{subfigure}[b]{0.40\textwidth}
				\includegraphics[width=\textwidth]{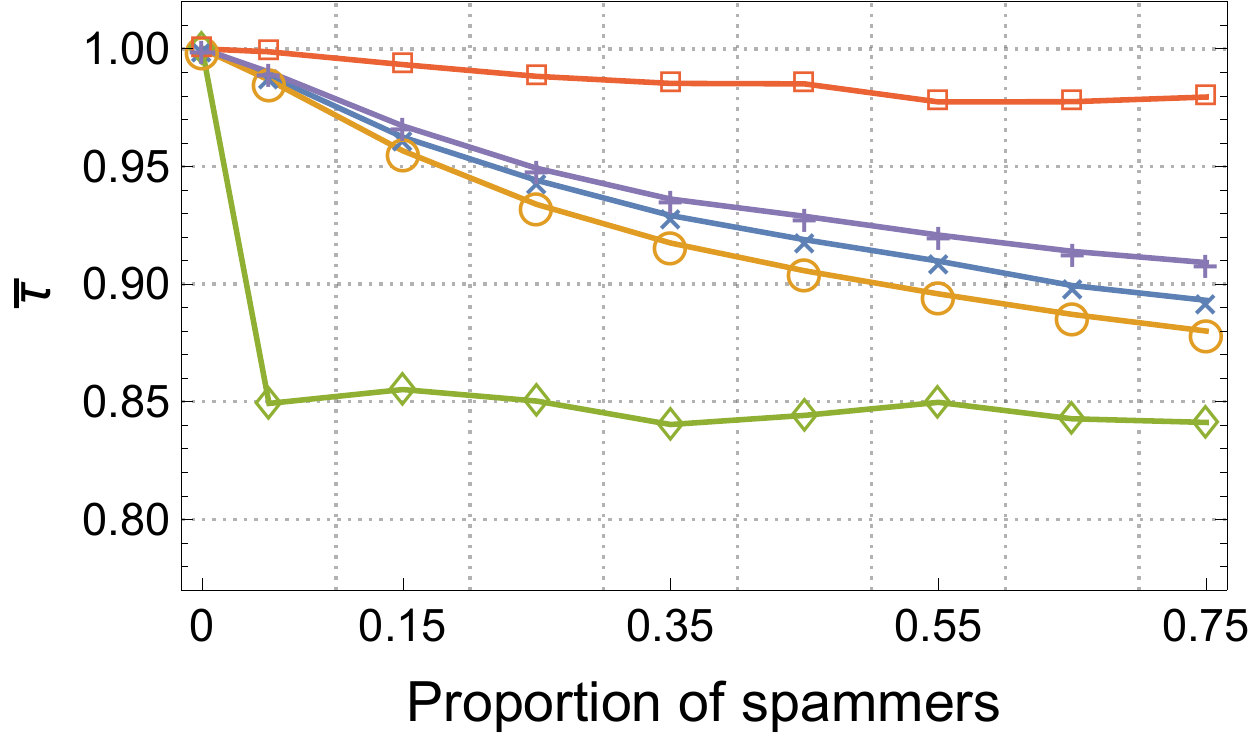}
			    %\caption{Tau evolution for love/hate attack.}
			    \caption{Dataset ``Amazon Instant Video''.}
				\label{subfig:lha:tau:dataset1}
			\end{subfigure}
			\quad %add desired spacing between images, e. g. ~, \quad, \qquad, \hfill etc. 
			\begin{subfigure}[b]{0.40\textwidth}
				\includegraphics[width=\textwidth]{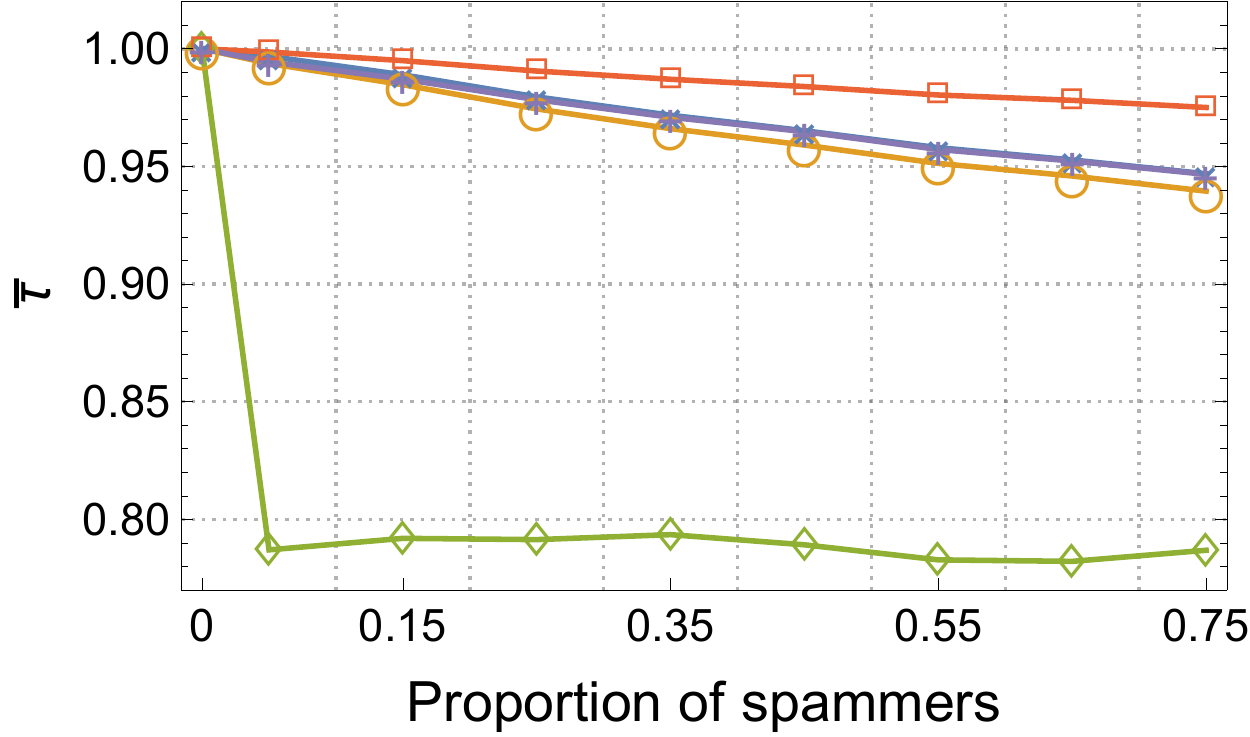}
				  %\caption{Tau evolution for love/hate attack.}
				\caption{Dataset ``Tools and Home Improvement''.}
				\label{subfig:lha:tau:dataset2}
			\end{subfigure}
  		    \begin{subfigure}[b]{0.37\textwidth}
  		        \includegraphics[width=\textwidth]{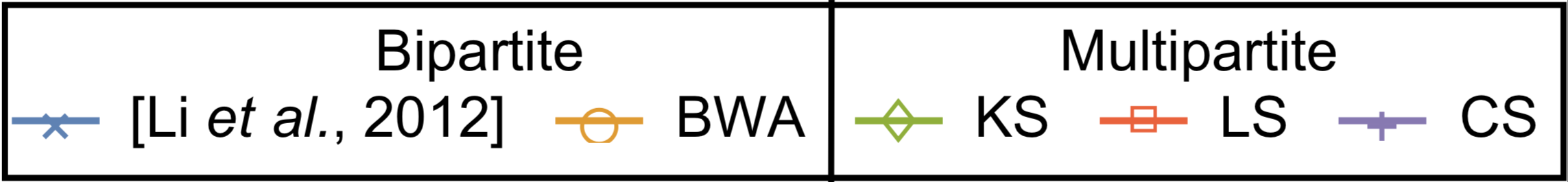}
  		    \end{subfigure}
			\caption{Evolution of the $\bar \tau$ for the love/hate attack with proportion of spammers.}
			\label{fig:lh_attack:tau}
		\end{figure*}
		\begin{figure*}
			\centering			
			\begin{subfigure}[b]{0.40\textwidth}
				\includegraphics[width=\textwidth]{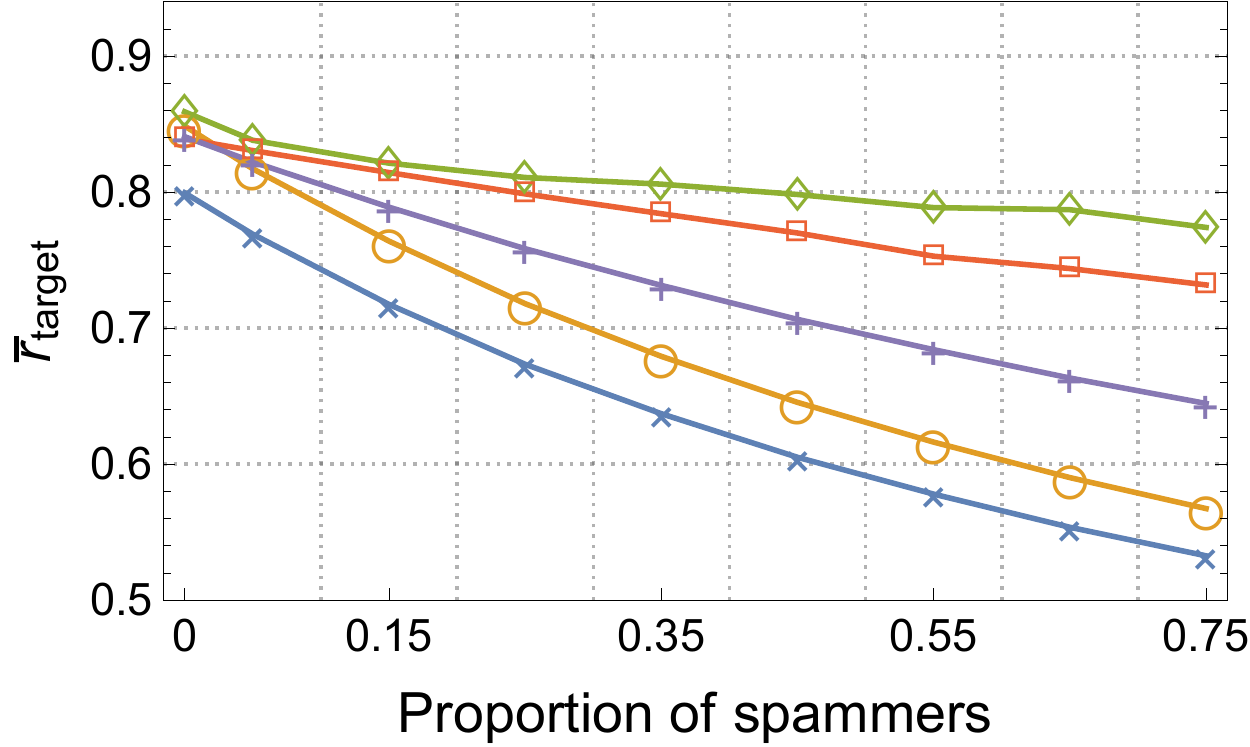}
			    %\caption{Rank evolution of the targeted item, for love/hate attack.}
			    \caption{Dataset ``Amazon Instant Video''.}
				\label{subfig:lha:tau_r:dataset1}
			\end{subfigure}
			\quad  % add desired spacing between images, e. g. ~, \quad, \qquad, \hfill etc.
			\begin{subfigure}[b]{0.40\textwidth}
				\includegraphics[width=\textwidth]{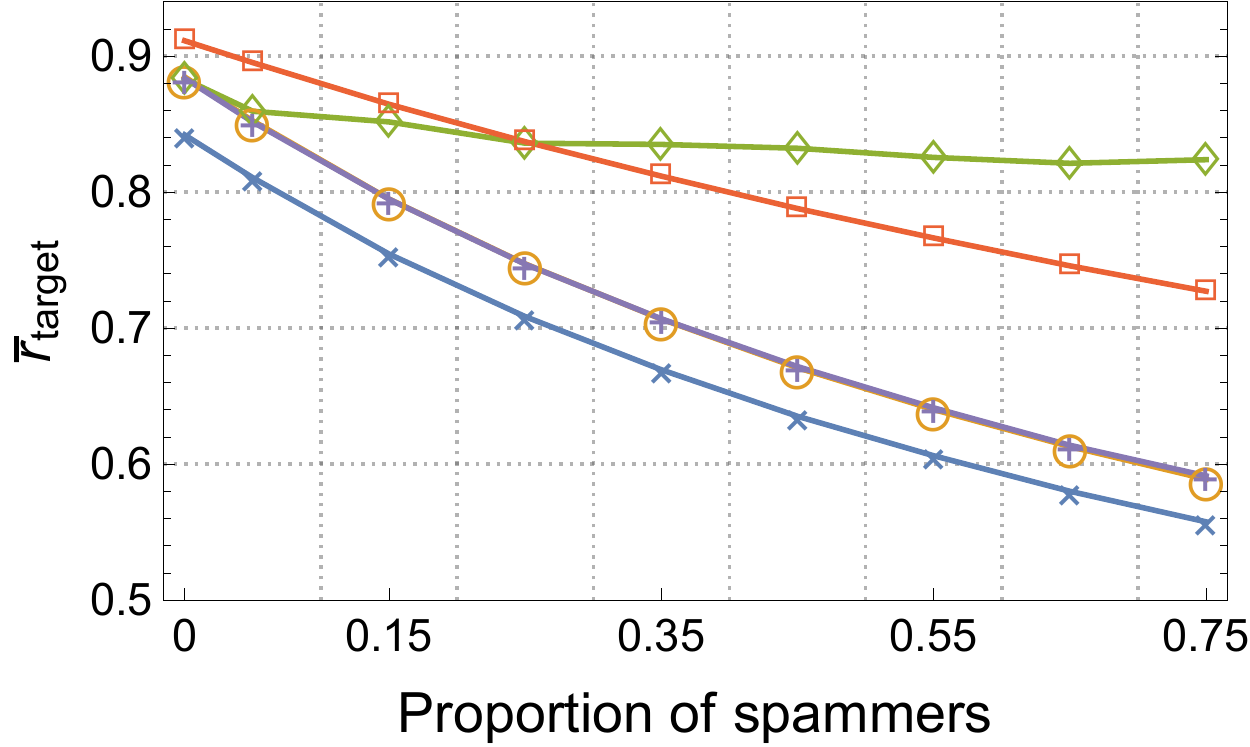}
				%\caption{Rank evolution of the targeted item, for love/hate attack.}
				\caption{Dataset ``Tools and Home Improvement''.}
				\label{subfig:lha:tau_r:dataset2}
			\end{subfigure}
		    \begin{subfigure}[b]{0.37\textwidth}
		        \includegraphics[width=\textwidth]{Figures/label}
		    \end{subfigure}
			\caption{Evolution of the ranking of the targeted item, $\bar r_{target}$, for love/hate attack  with proportion of spammers.}
			\label{fig:lha:tau_r}
		\end{figure*}
		\begin{figure*}
			\centering			
		    \begin{subfigure}[b]{0.40\textwidth}
		        \includegraphics[width=\textwidth]{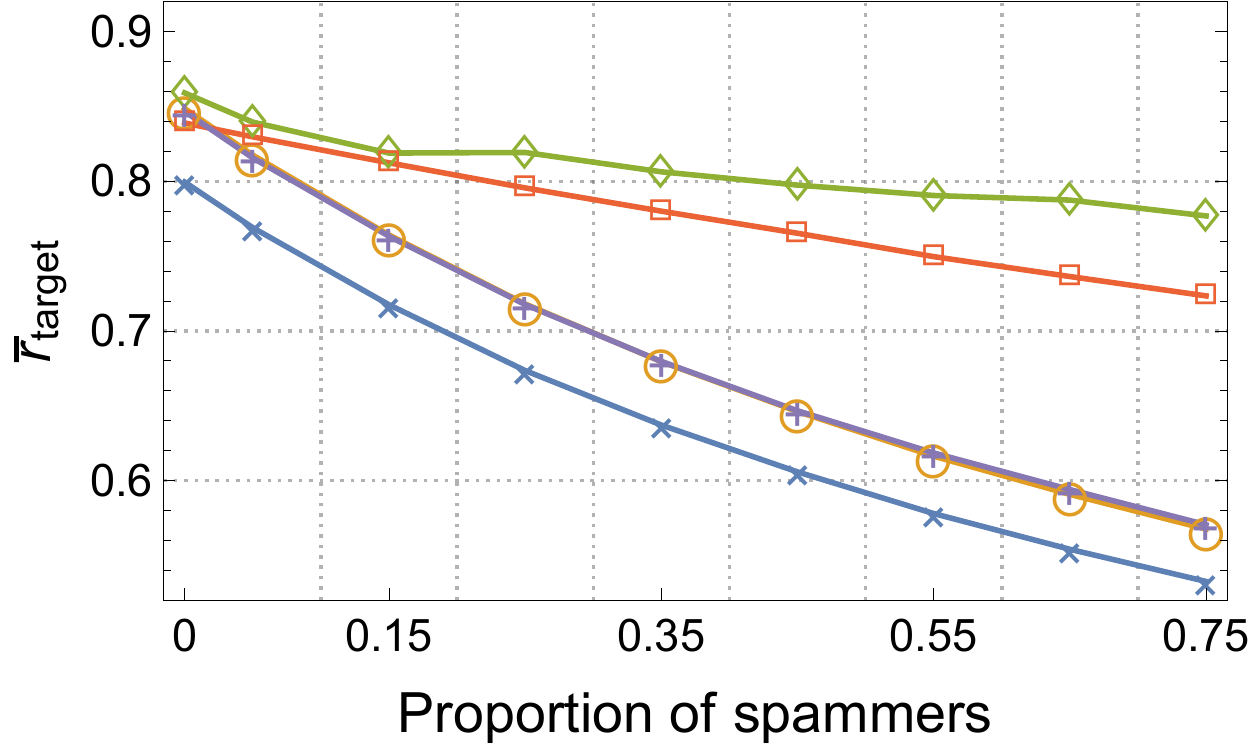}
		        \caption{Dataset ``Amazon Instant Video''.}
				%\caption{Ranking evo2ution of the targeted item.}
		        \label{fig:rep_r1}
		    \end{subfigure}
		    \begin{subfigure}[b]{0.40\textwidth}
		        \includegraphics[width=\textwidth]{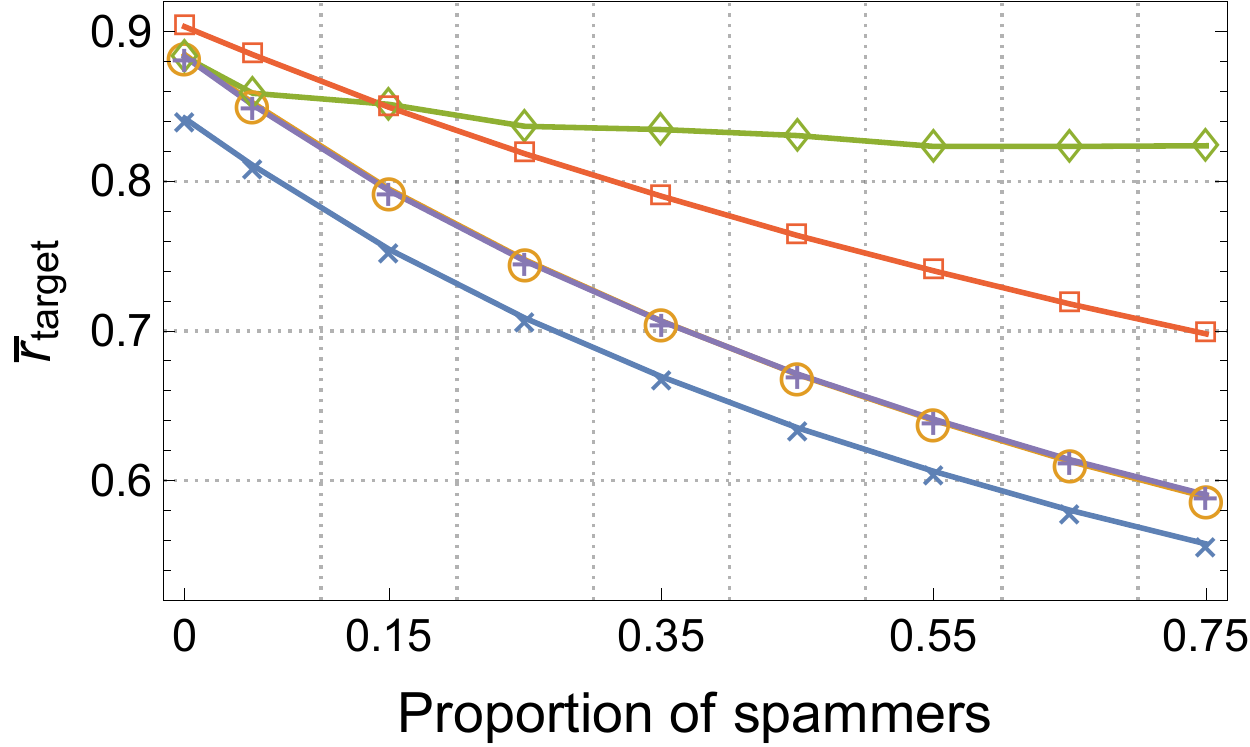}
		        \caption{Dataset ``Tools and Home Improvement''.}
				%\caption{Ranking evo2ution of the targeted item.}
		        \label{fig:rep_r2}
		    \end{subfigure}
		    \begin{subfigure}[b]{0.37\textwidth}
		        \includegraphics[width=\textwidth]{Figures/label}
		    \end{subfigure}
			\caption{Evolution of the ranking of the targeted item, $\bar r_{target}$, for reputation attack with proportion of spammers.}
			\label{fig:rep_att:r}
		\end{figure*}
		\begin{figure*}	
			\centering
		    \begin{subfigure}[b]{0.40\textwidth}
		        \includegraphics[width=\textwidth]{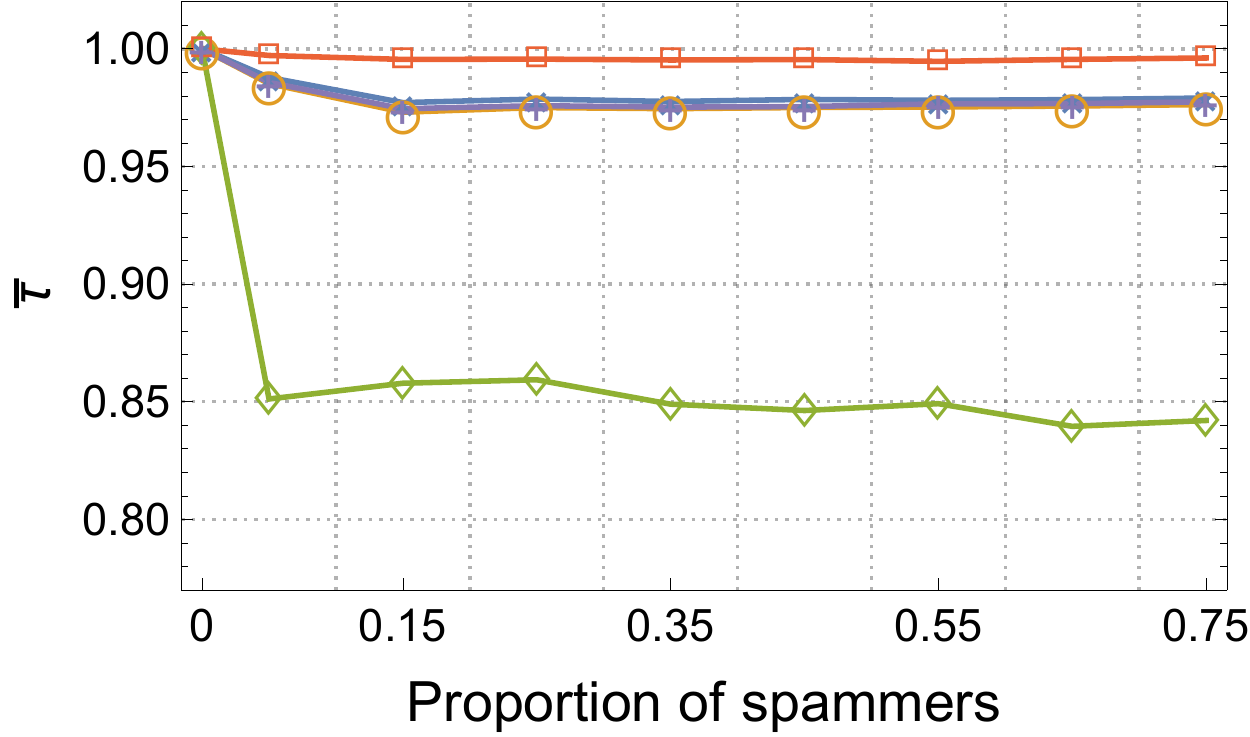}
		        \caption{Dataset ``Amazon Instant Video''.}
				%\caption{Tau evolution for love/hate attack.}
		        \label{fig:rep_t1}
		    \end{subfigure}
		    \begin{subfigure}[b]{0.40\textwidth}
		        \includegraphics[width=\textwidth]{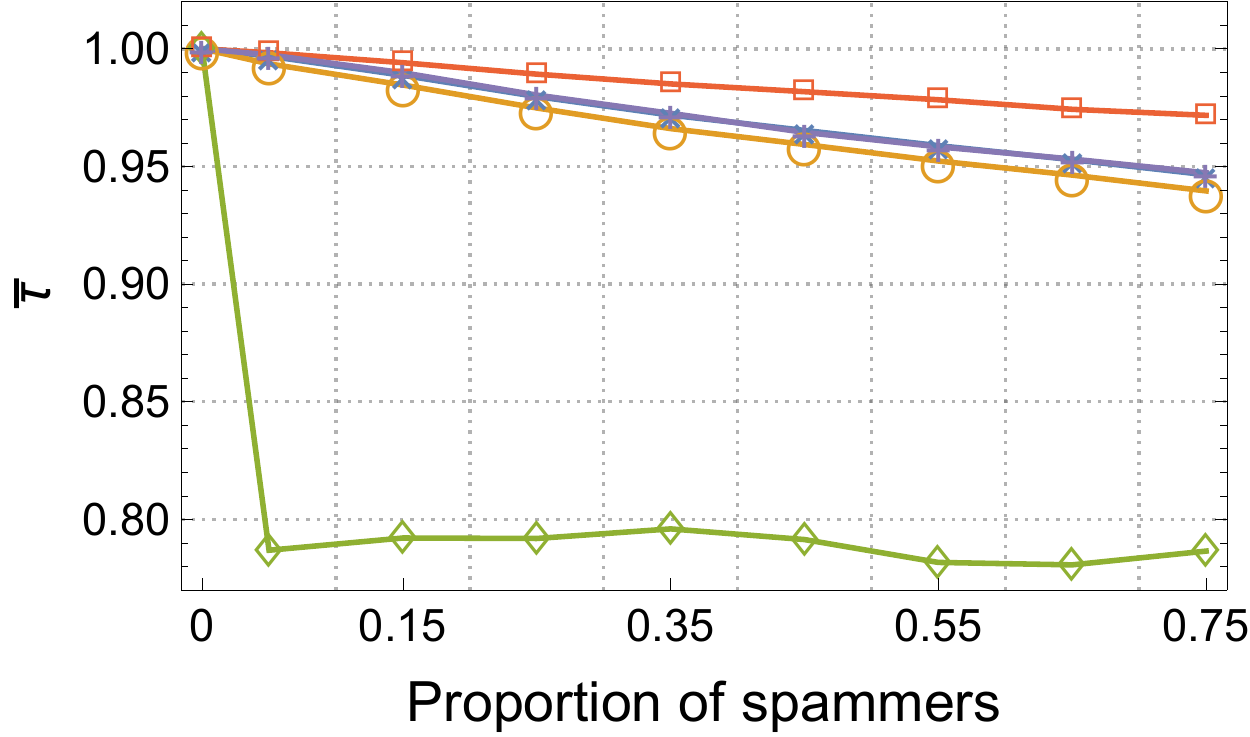}
		        \caption{Dataset ``Tools and Home Improvement''.}
				%\caption{Tau evolution for love/hate attack.}
		        \label{fig:rep_t2}
		    \end{subfigure}
		    \begin{subfigure}[b]{0.37\textwidth}
		        \includegraphics[width=\textwidth]{Figures/label}
		    \end{subfigure}
			\caption{Evolution of the $\bar \tau$ for the reputation attack with proportion of spammers.}
			\label{fig:rep_att:tau}
		\end{figure*}

	\subsection{Robustness against random spamming} % (fold)
	\label{sub:robustness_against_noise}
		We test the random spamming (noise) by simulating a proportion of spammers ranging from $0$ to $0.75$ of the total number of ratings.
        The results are reported in \figurename~\ref{fig:random_tau}.
		Using the multipartite ranking systems, we notice an increase of robustness for the LS, whereas for the KS and CS we obtain similar robustness to the bipartite methods, because these similarities accommodate new users by rearranging the clusters, and this degrades the $\bar \tau$.
	% subsection robustness_against_noise (end)
	\subsection{Robustness against attacks} % (fold)
	\label{sub:robustness_against_attacks}
		Now, we simulate two different attacks to the most voted item, $\bar r_{\text{target}}$, ranging the proportion of attackers from $0$ to $0.75$ of the total number of voters, in this case, of the target item.
		\paragraph{Love/Hate attack} % (fold)
		\label{par:love_hate_attack}
		In Figures~\ref{fig:lh_attack:tau} and~\ref{fig:lha:tau_r}, we can see that, using the multipartite ranking systems, the attack is less effective on both datasets.
		In the case of the variation of $\bar \tau$, in \figurename~\ref{fig:lh_attack:tau}, the results are significantly better when we perform the clustering with the LS.
		In both datasets, the effect of the attack on the ranking of the target item, $r_{\text{target}}$, in \figurename~\ref{fig:lha:tau_r}, is more dimmed in the multipartite scenario, thus less effective.
		The best similarity measure to avoid the effect of the attack on the target item's ranking is the KS. 
		
 		While the KS is effective to deter the attack on the item's ranking, it has the most nefarious effect on $\bar \tau$.  
    	This is a consequence of the reorganization of the subnetworks, to minimize the effect of the attack on $r_{\text{target}}$, and our generalization of the Kendall tau does not account for this repercussion.
		% Moreover, in the larger clusters with users who rated the targeted item the ranking of the item is kept unchanged for both LS and CS, yielding the same effect as for the reputation attack, see Figure~\ref{fig:rep_r1}.
		This indicates that the attackers are not grouped with normal users and thus do not affect the rankings of items in the cluster. 
		%For KS and CS the ranking oscillates (specially for CS) due to the reorganization of clusters, but 
		The ranking is not nuked in the multipartite cases as when using the ranking system in~\cite{li2012robust} and BWA.	
		% paragraph love_hate_attack (end)
		\paragraph{Reputation attack} % (fold)
		\label{par:reputation_attack}
		In both datasets, using subnetworks, the effect of the reputation attack on the ranking of the targeted item is dimmed, see \figurename~\ref{fig:rep_att:r}. Since the intelligent attacker chooses the closest rating to the ranking of the filler items, it should not affect drastically the rankings of the filler items, while the attackers increase their reputation. 
		In fact, in the multipartite ranking systems, the ranking of the nuked item drops less than in the bipartite ranking systems, and the best case is when using LS.
		The robustness $\bar \tau$ (\figurename~\ref{fig:rep_att:tau}) has a similar behavior as in the love/hate attack, and the best result is achieved in the multipartite scenario when using LS.

\begin{figure*}	
			\centering
		    \begin{subfigure}[b]{0.40\textwidth}
		        \includegraphics[width=\textwidth]{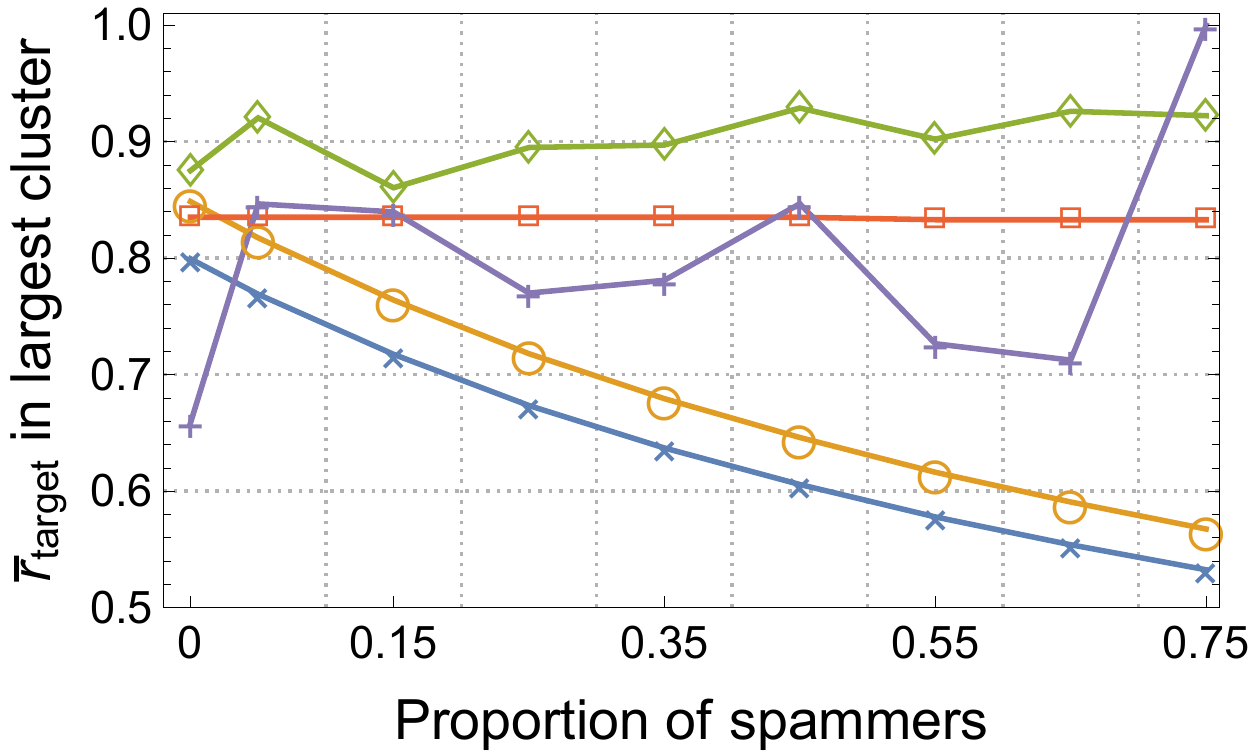}
		        \caption{Dataset ``Amazon Instant Video''.}
				%\caption{Tau evolution for love/hate attack.}
		        \label{fig:rep_r11}
		    \end{subfigure}
		    \begin{subfigure}[b]{0.40\textwidth}
		        \includegraphics[width=\textwidth]{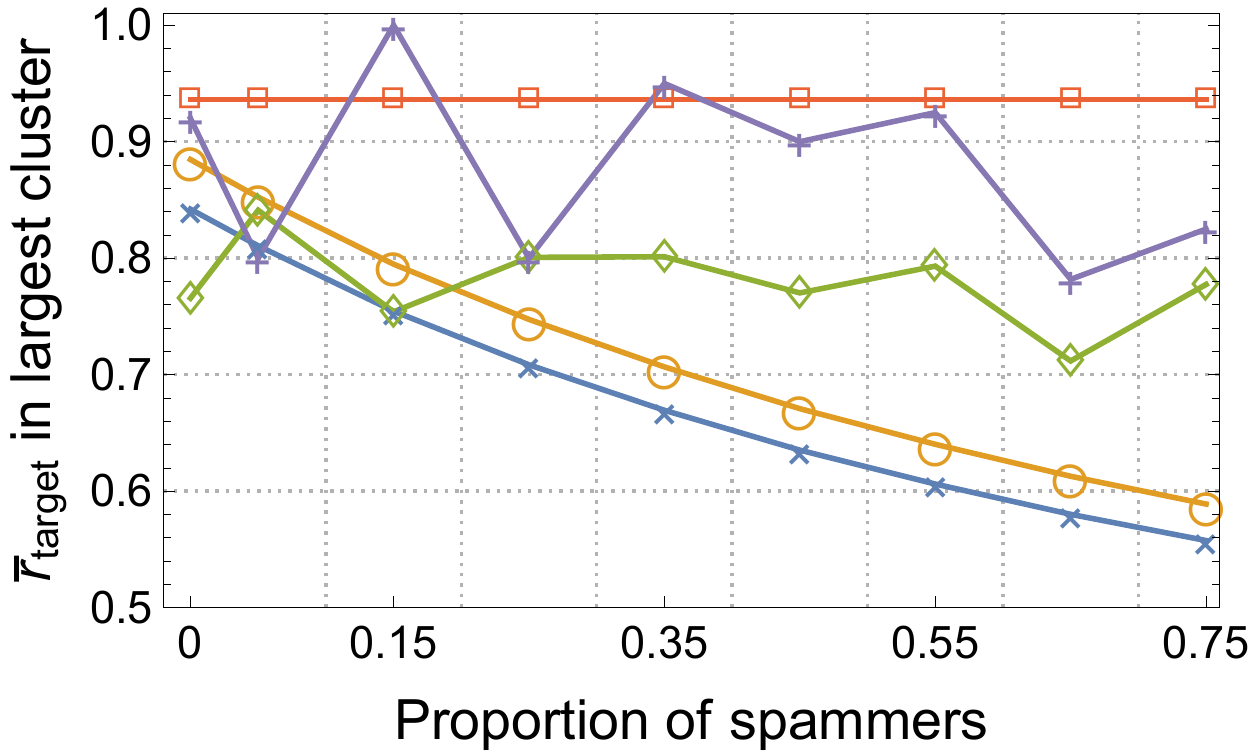}
		        \caption{Dataset ``Tools and Home Improvement''.}
				%\caption{Tau evolution for love/hate attack.}
		        \label{fig:rep_r12}
		    \end{subfigure}
		    \begin{subfigure}[b]{0.38\textwidth}
		        \includegraphics[width=\textwidth]{Figures/label}
		    \end{subfigure}
			\caption{Evolution of $\bar r$  with proportion of attackers, for reputation attack, in the largest cluster.}
			\label{fig:rep_att_com1}
		\end{figure*}

		The organization of subnetworks changes with the increasing number of spammers, this is a collateral effect of the system, that helps to cope with the attack.
		Thus, this effect produces a bigger change in $\bar\tau$, because it reduces drastically the effect of the targeted attack.
		Since the ranking of the filler items does not change drastically (the attackers rate those items with their weighted average ranking), this is not an important side effect. 
		Moreover, in the larger cluster, containing users who rated the targeted item, the ranking of the item is almost kept unchanged, when using LS.
		For the KS, it has a small variation and has a large variation for the CS.
		Both variations reflect the opposite effect on the ranking of the targeted item as what is intended by the attacker, see \figurename~\ref{fig:rep_att_com1}.
		The clustering produced by the KS and the CS aggregate attackers with legit users (that gave smaller ratings to the target item) on a separated cluster, leaving raters who gave high ratings on the biggest cluster.
        Recall that for new users, the displayed rankings are a weighted average of the ratings by user's reputations, whereas in each cluster they are the weighted average within the users of the cluster.

    		\begin{figure*}
			\centering
			\begin{subfigure}[b]{0.31\textwidth}
				\includegraphics[width=\textwidth]{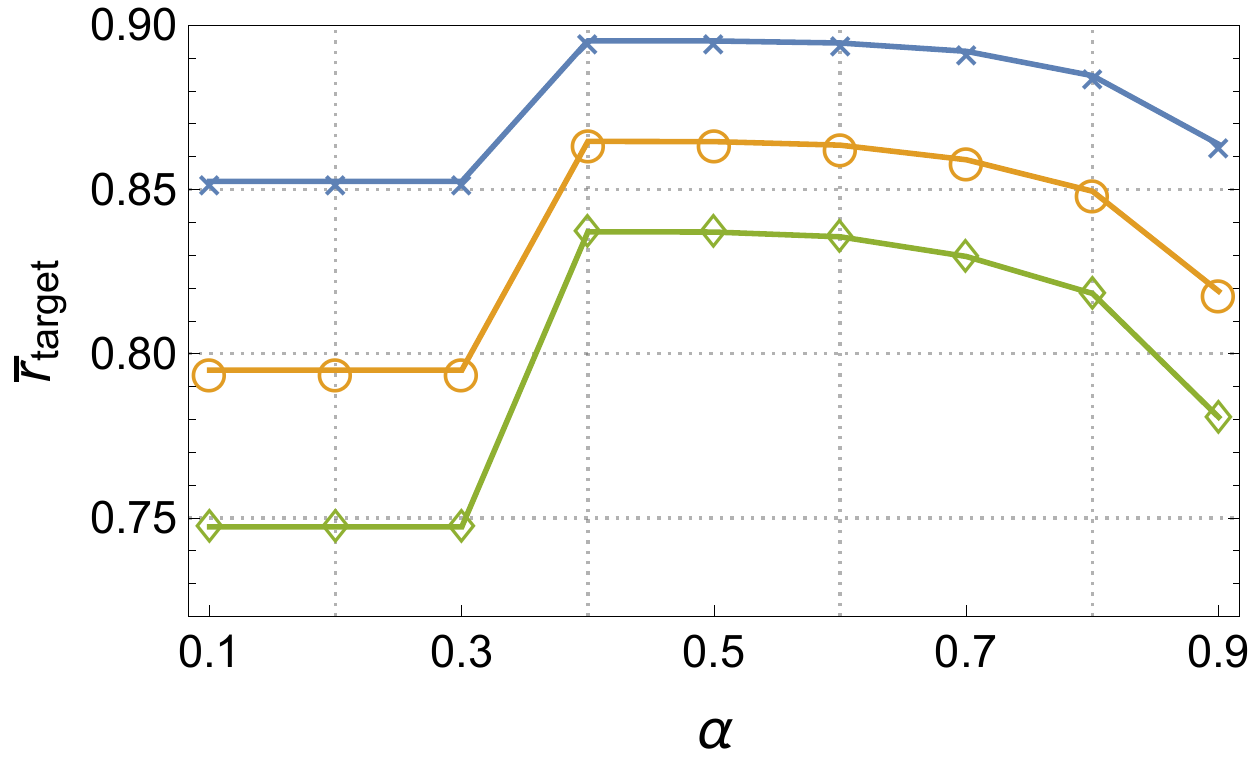}
			    %\caption{Tau evolution for love/hate attack.}
			    \caption{$\bar r_{\text{target}}$ versus $\alpha$, using LS.}
				\label{fig:lh_t1}
			\end{subfigure}
			% ~ %add desired spacing between images, e. g. ~, \quad, \qquad, \hfill etc. 
			\begin{subfigure}[b]{0.31\textwidth}
				\includegraphics[width=\textwidth]{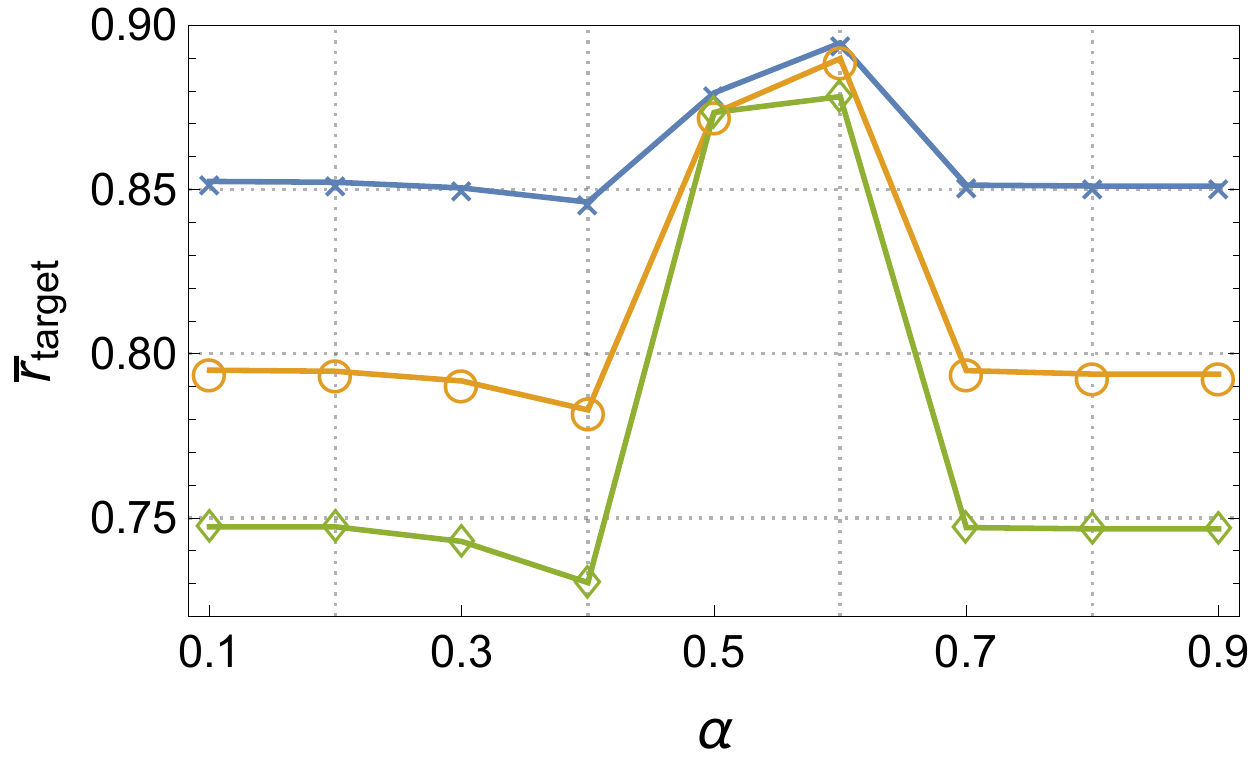}
			    %\caption{Tau evolution for love/hate attack.}
			    \caption{$\bar r_{\text{target}}$ versus $\alpha$, using CS.}
				\label{fig:lh_t2}
			\end{subfigure}
			% ~ %add desired spacing between images, e. g. ~, \quad, \qquad, \hfill etc. 
			\begin{subfigure}[b]{0.31\textwidth}
				\includegraphics[width=\textwidth]{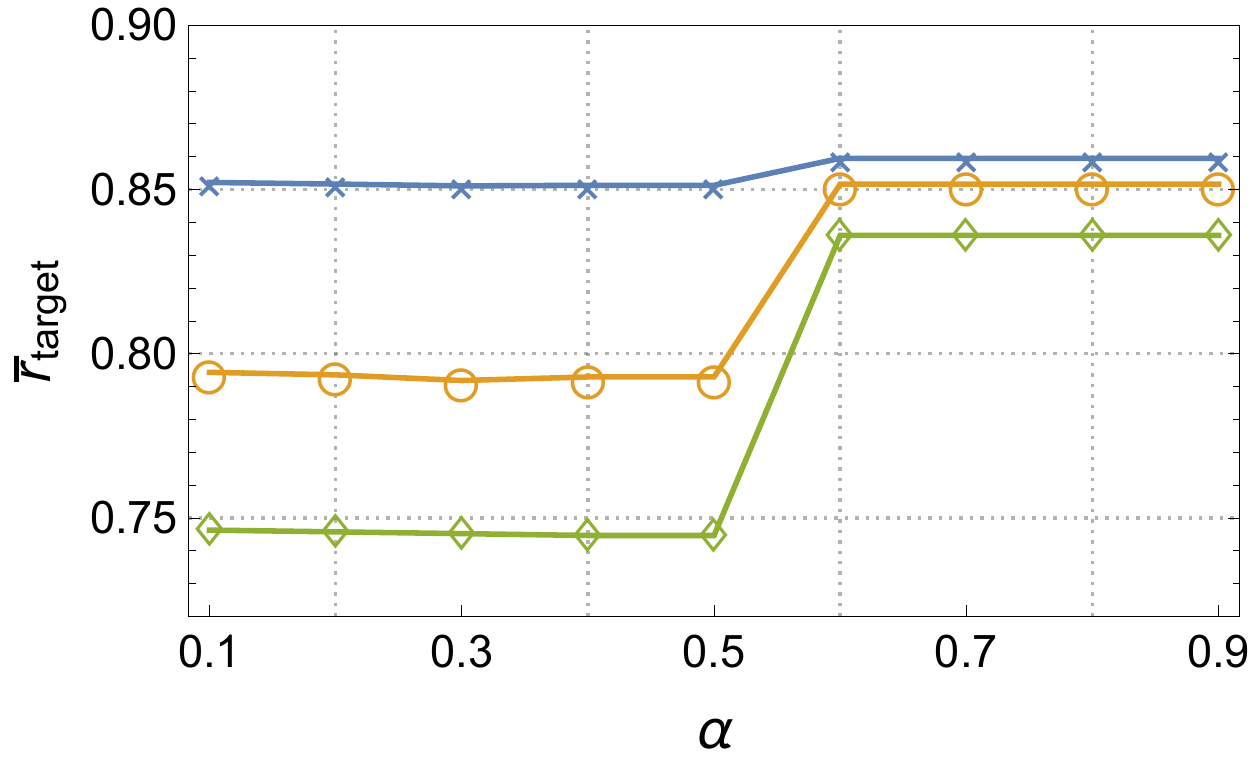}
			    %\caption{Tau evolution for love/hate attack.}
			    \caption{$\bar r_{\text{target}}$ versus $\alpha$, using KS.}
				\label{fig:lh_t3}
			\end{subfigure}
		    \begin{subfigure}[b]{0.21\textwidth}
		        \includegraphics[width=\textwidth]{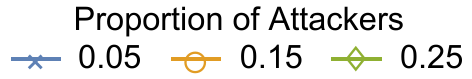}
		    \end{subfigure}
			\caption{Variation of $\bar r_{\text{target}}$ with the affinitity parameter, $\alpha$, for different proportions of attackers.}
			\label{fig:var_par:a_rtarg}
		\end{figure*}
        \begin{figure*}
			\begin{subfigure}[b]{0.31\textwidth}
				\includegraphics[width=\textwidth]{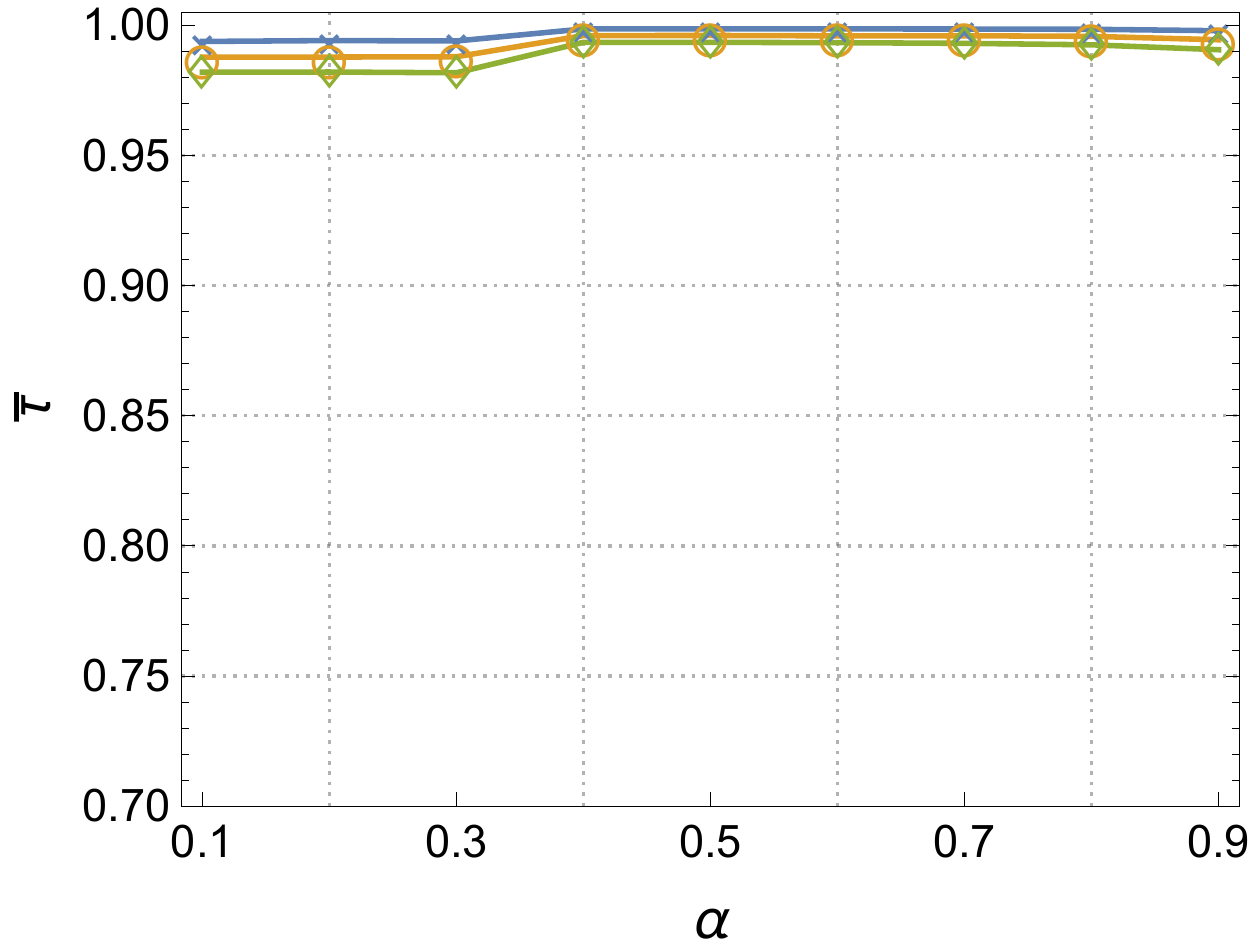}
			    %\caption{Tau evolution for love/hate attack.}
			    \caption{$\bar \tau$ versus $\alpha$, using LS.}
				\label{fig:var_par:lh_tau1}
			\end{subfigure}
			% ~ %add desired spacing between images, e. g. ~, \quad, \qquad, \hfill etc. 
			\begin{subfigure}[b]{0.31\textwidth}
				\includegraphics[width=\textwidth]{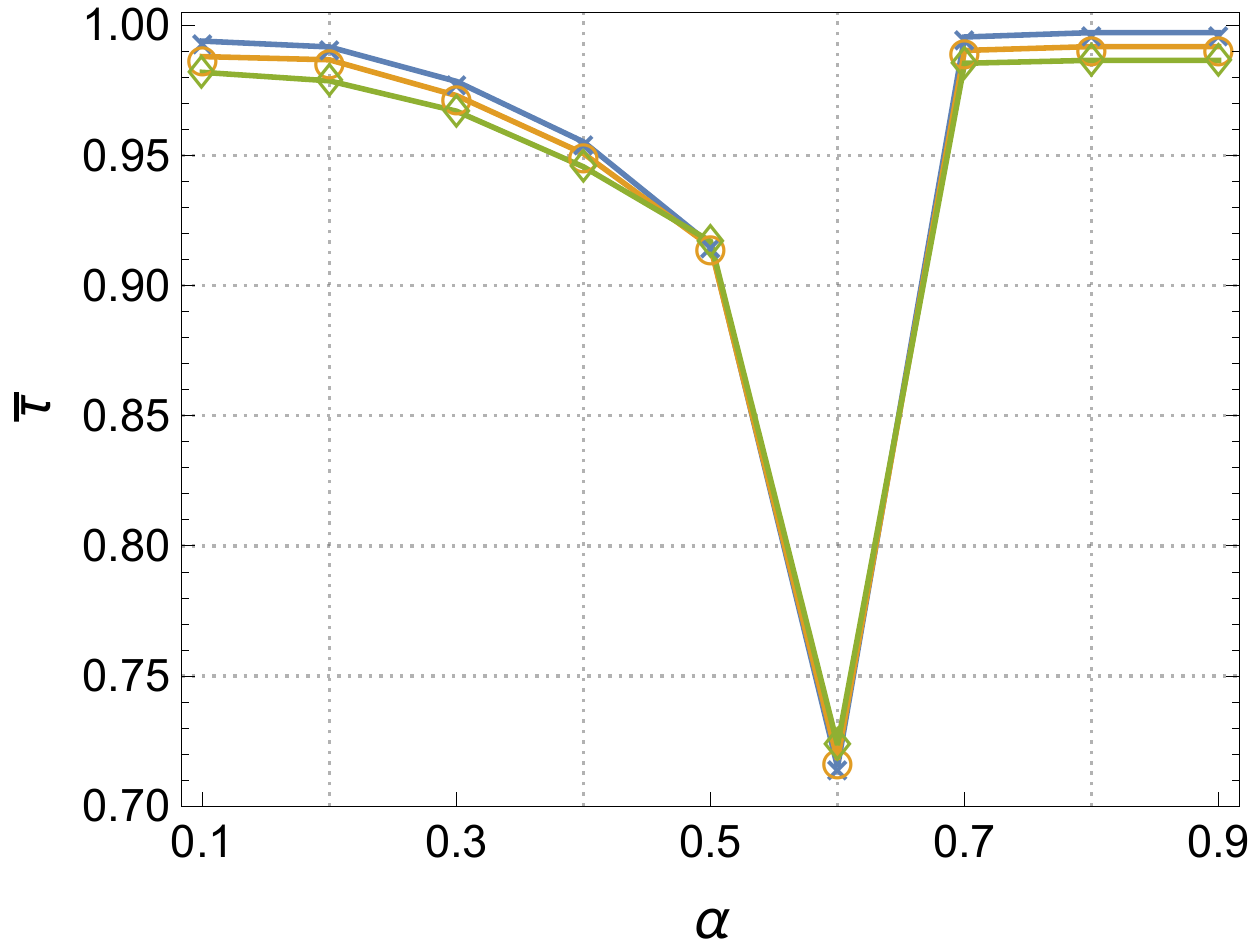}
			    %\caption{Tau evolution for love/hate attack.}
			    \caption{$\bar \tau$ versus $\alpha$, using CS.}
				\label{fig:var_par:lh_tau2}
			\end{subfigure}
			% ~ %add desired spacing between images, e. g. ~, \quad, \qquad, \hfill etc. 
			\begin{subfigure}[b]{0.31\textwidth}
				\includegraphics[width=\textwidth]{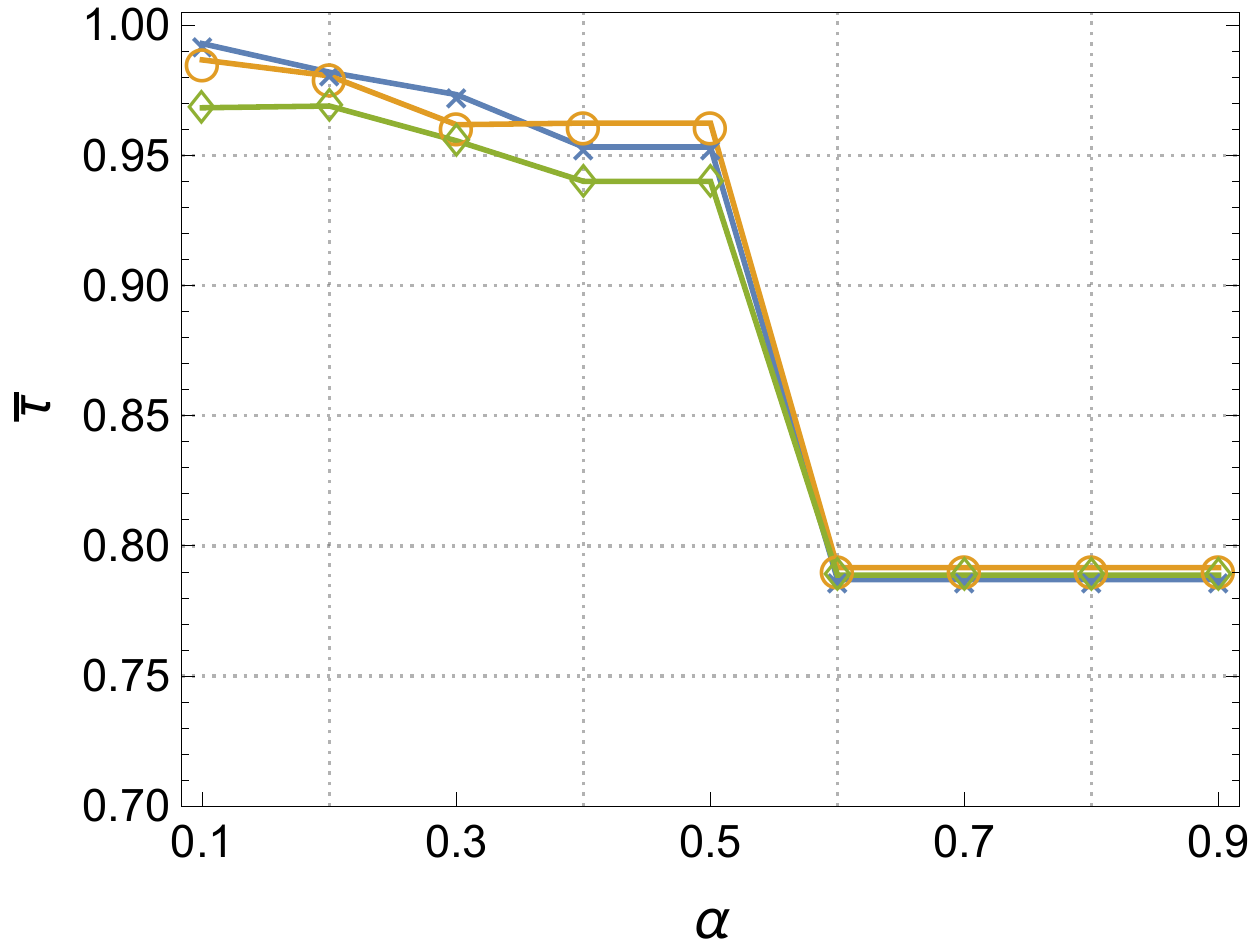}
			    %\caption{Tau evolution for love/hate attack.}
			    \caption{$\bar \tau$ versus $\alpha$, using KS.}
				\label{fig:var_par:lh_tau3}
			\end{subfigure}
		    \begin{subfigure}[b]{0.21\textwidth}
		        \includegraphics[width=\textwidth]{Figures/Parameters/Label.pdf}
		    \end{subfigure}
			\caption{Variation of $\bar \tau$ with the affinitity parameter, $\alpha$, for different proportions of attackers.}
		\end{figure*}

		An important aspect is that the experiments, in both datasets, are coherent in the sense that we obtain similar results for the different attacks and spam. This indicates that when we scale the size of the dataset, we expect to get similar robustness to the attacks for the evaluated metrics, $\bar \tau$ and $\bar R_{\text{target}}$.
		As we pointed out, the multipartite scenario allows us to present rankings of items to users that allow a multimodal behavior, and this can also be explored for the item recommendation scenario, where we expect to get recommendations more tailored to the users.

			% 	\caption{The plots depict the variation of the Kendall $\bar \tau$ and of the ranking of the targeted item, $\bar r_{\text{target}}$ (the most voted), with respect to the fraction of spammers from $0$ to $0.75$.
		% 	(a)-(c) with Dataset ``Amazon Instant Video''
		% 	(d)-(f) with dataset ``Tools and Home Improvement''.
		% 	For the random spam, (a)-(c), we consider the fraction of the total number of users.
		% 	For the love/hate attack, (b)-(c), and for the reputation attack, (d)-(e), we consider the fraction of the raters of $\bar r_{\text{target}}$.
		% 	Figure (f) shows the variation of $\bar \tau_{\text{target}}$ in the biggest cluster along no clustering systems.}
		 	
		% \end{figure*}
		% paragraph reputation_attack (end)
	% subsection robustness_against_attacks (end)
        
	\subsection{Sensivity to parameters} % (fold)
	\label{sub:sensivity_to_parameters}
		Here, we discuss the response of our system to the variation of its parameters, using Dataset B (the results for Dataset A are almost the same, so they have not been reported to facilitate the readability of the paper).

		In \figurename~\ref{fig:var_par:a_rtarg}, we look for the threshold $\alpha$ that leads to a smaller variation of $\bar r_{\text{target}}$ when the number of attackers increases, for the love/hate attack.
		For LS (\figurename~\ref{fig:lh_t1}), we have the best results for $\alpha \in [0.4,0.6]$.
		For CS (\figurename~\ref{fig:lh_t2}), we have the best results for $\alpha \in [0.5,0.6]$.
		Finally, for KS (\figurename~\ref{fig:lh_t3}), the best range is $\alpha\in[0.6,0.9]$.		% \begin{figure}[ht]
		% 	\centering
		%     \begin{subfigure}[b]{0.31\textwidth}
		%         \includegraphics[width=\textwidth]{Figures/Parameters/tauVSlambda.pdf}
		%         %\caption{Tau evolution for random spamming.}
		%         \caption{$\bar \tau$ vs $\lambda$, using BWA.}
		% 		\label{fig:rand_tau}
		%     \end{subfigure}
		% 	\begin{subfigure}[b]{0.31\textwidth}
		% 	        \includegraphics[width=\textwidth]{Figures/Parameters/RTargetVsLambda.pdf}
		% 	        %\caption{Rank evolution of the targeted item, for love/hate attack.}
		% 	        \caption{$\bar \tau$ vs $\lambda$, using LD.}
		% 			\label{fig:lh_r}
		% 	\end{subfigure}
		%     \begin{subfigure}[b]{0.25\textwidth}
		%         \includegraphics[width=\textwidth]{Figures/Parameters/Label.pdf}
		%     \end{subfigure}
		% \end{figure}
		To chose a good threshold, we need to analyze the effect of $\alpha$, not only on the ranking of the attacked item, $\bar r_{\text{target}}$, but also on the robustness, $\bar \tau$.
		In this case, we look for values of $\bar \tau$ close to 1.
		
		When using LS for clustering and comparing Figures~\ref{fig:lh_t1} and \ref{fig:var_par:lh_tau1}, we see that the best affinity level lays in the interval $\alpha\in[0.4,0.6]$.
		Choosing some $\alpha$ in this interval allows the systems to protect the ranking of an item, $\bar r_{\text{target}}$, maintaining the robustness of the system, $\bar \tau$, close to 1.
		In both the CS and KS cases, the affinity level that protects better the ranking of the attacked item produces worst robustness to attacks, not only within the clustering method, but also when compared to LS.
		These results are in line with those presented in Section~\ref{sub:robustness_against_attacks}.
		The effect of parameter $\alpha$ on the $\bar \tau$ metric might be due to the fact that CS and KS produce more clusters (without a bigger one) and the users tend to be regrouped as the proportion of attackers change, and this effect 
		is not captured by $\bar\tau$.
		
			% subsection sensivity_to_parameters (end)
            
			\subsection{Robustness against Bribery} % (fold)
	\label{sub:robustness_against_bribery}
		In this section, we explore the robustness against bribery.
		First with synthetic data and second with real data.
		\subsubsection{Synthetic data} % (fold)
		\label{sub:synthetic_data}
			Here, we explore the main results of this paper using synthetic generated data.
		%	We start by exploring {\color{red}Proposition~\ref{eq:1rated1notrated}}.
			\begin{example} \label{exp1}
				Consider a scenario where $U_i=\{v\}$, $c_v=1$, $c_w=0.8$ and $R_{vi}=R_{wi}=0.5$.
				Consider strategy $\sigma^i$ s.t. $\sigma_v^i = \sigma_w^i=0.5$.
				We start by computing the profit of each elementary strategy.
				We have that $\pi_{\sigma_w^i} = (c_v-|U_i|c_w)\frac{r_i-\rho_w}{c_v+c_w} = 0$, and (after this strategy is applied) we have that $\pi_{\sigma_v^i} = (\frac{c_v}{c_v}|U_i|-1)\rho_v = 0$.
				This yields a sum of the elementary strategies profit of $0$.
				Whilst in the case of strategy $\sigma^i $, we have that $\pi_{\sigma^i}=\pi_{\sigma_v^i+\sigma_w^i} = \frac{c_v}{c_v+c_w}\pi_{\sigma_v^i} + \pi_{\sigma_w^i}+\frac{1}{c_v+c_w}\rho_v(c_v-c_w) = \frac{1}{18}$.
				The final reward is, hence, different for the two sequence of strategies.
			\end{example}
			
			For the next examples, we consider 5 users, 2 items and 2 clusters of users.
			The ratings given by users to the items are presented in the first two rows of the users's columns in the first table of each example.
			\begin{example}\label{exp2}
				Consider $I = \{i,j\}$, $U = \{u_1,\hdots,u_5\}$ and two subnetworks $\mathcal M_1=\{u_1,u_2,u_3\}$ and $\mathcal M_2=\{u_4,u_5\}$. 
				The users' reputations, the ratings given by users to items and the ranking of items for both BWA case and MRS are summarized in Table~\ref{table1}.
				\begin{table}
	 				\centering
	 				\resizebox{0.48\textwidth}{!}{\begin{tabular}{ | c |cc|ccc| c | c c c | }
	 				\hline
	 				& \multicolumn{5}{ c|}{\textsc{Users}} & \textsc{Bipartite} &\multicolumn{3}{|c | }{\textsc{Multipartite}} \\ \hline
	 				\textsc{Items} & $u_1$ & $u_2$ & $u_3$ & $u_4$ & $u_5$ & $r_x$ & $r_{x,\mathcal M_1}$ & $r_{x,\mathcal M_2}$ & $\bar r_x$      \\[0.1cm]  
					$x=i$ & 0.4 & 0.6 & 0.5 & 0.7 & 0.5 & 0.539 & 0.506 & 0.590 & 0.540 \\
					$x=j$ & 0.7 & 0.3 & 0.5 & 0.5 & 0.6 & 0.514 & 0.488 & 0.555 & 0.515 \\	 
					$c_u$ & 0.4 & 0.5 & 0.8 & 0.5 & 0.6 & & & & \\	 	\hline
	 				\end{tabular}}
	 				\caption{Ratings given by users to items, users' reputations and items' rankings for both BWA and MRS.}
	 				\label{table1}
				\end{table}

				Suppose the sellers of item $i$ want to bribe a user in order to increase its ranking.
	  			In the first strategy, the sellers bribe user $u_1$, with $\sigma_{u_1}^i = 0.6$. In the second strategy, the sellers bribe user $u_3$, with $\sigma_{u_3}^i = 0.5$.
		 	   	The profits of each strategy, for both BWA and MRS cases, are represented in Table~\ref{table2}. 
				\begin{table}
	 				\centering
	 				\resizebox{0.48\textwidth}{!}{
					\begin{tabular}{ | c | c c | c | c c| }
	 					\hline
	 					& \multicolumn{2}{ c}{\textsc{Profit $\pi$}} & \multicolumn{3}{| c|}{\textsc{Ranking of} $o_1$} \\ \hline
	 					\textsc{Strategy} & \textsc{BWA} & \textsc{MRS} & $r_i$ & $r_{i,\mathcal M_1}$ & $\bar r_i$\\ 
						$\sigma_{u_1}^i$ & -0.171 & -0.176 & 0.625 & 0.647 & 0.624  \\
						$\sigma_{u_3}^i$ & 0.214 & 0.206 & 0.682 & 0.741 & 0.706 \\ \hline	 
	 				\end{tabular}}
	 				\caption{Profits of bribing strategies $\sigma_{u_1}^i$ and $\sigma_{u_3}^i$ and new ranking, after applying the strategies, in both BWA and MRS.}
	 		 	   \label{table2}
				\end{table}% 
	 			Using either the bipartite or the multipartite schemes, the strategy $\sigma_{u_1}^i$ is not profitable and strategy $\sigma_{u_3}^i$ is profitable. 
	 			Further, the profit obtained in the BWA case is larger that in the MRS cases for both strategies. 
				Figure~\ref{fig:barchart} depicts the profit of item $i$ sellers when bribing each user, for both ranking systems.
		   	 	\begin{figure}[ht]
		      		\centering
		      		\includegraphics[width=0.47\textwidth]{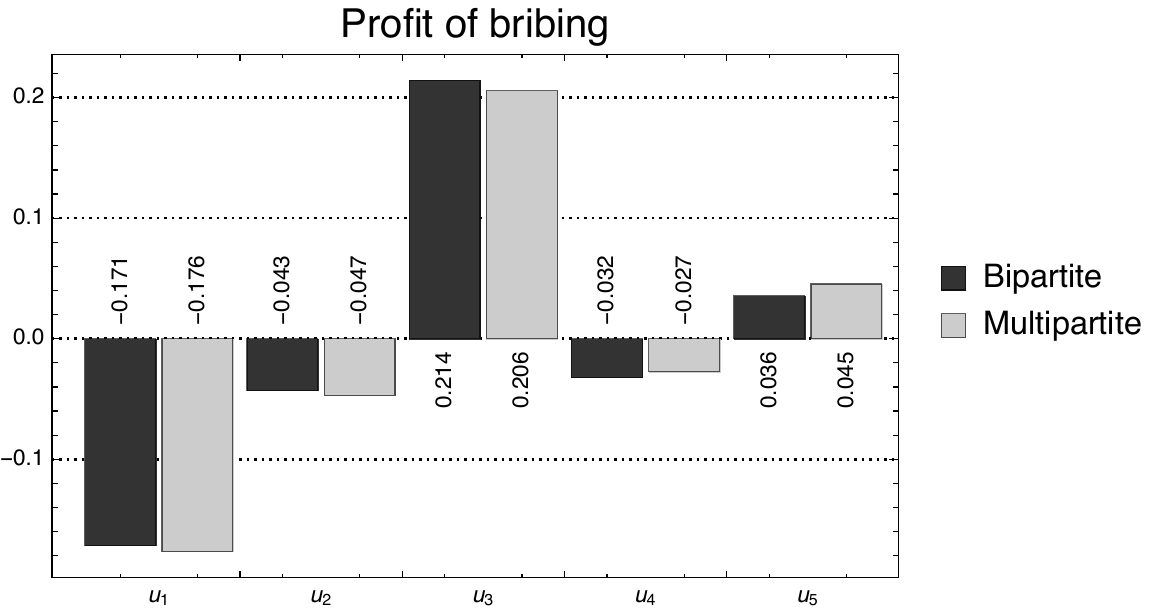}
		     	 		 	\caption{Profit of $i$ seller when bribing each user in the setup of Table~\ref{table1}. The black and gray bars correspond to use the BWA and MRS, respectively.}
		    	 	\label{fig:barchart}
		    	\end{figure}
			\end{example}
			\begin{example}\label{exp3}
				Consider the same users, items and subnetworks as in Example~\ref{exp1},  
				now with users' reputations, ratings given by users to items and ranking of items, for both the BWA case and the MRS case, summarized in Table~\ref{table3}.
				\begin{table}
			 	\centering
			 	\resizebox{0.48\textwidth}{!}{\begin{tabular}{ | c | c c | c c c | c | c c c | }
			 		\hline
			 		  & \multicolumn{5}{ c|}{\textsc{Users}} & \textsc{BWA} &\multicolumn{3}{|c | }{\textsc{MRS}} \\ \hline
			 		 \textsc{Items} & $u_1$ & $u_2$ & $u_3$ & $u_4$ & $u_5$ & $r_x$ & $r_{x,\mathcal M_1}$ & $r_{x,\mathcal M_2}$ & $\bar r_x$      \\[0.1cm]  
					 $x=i$ & 0.4 & 0.6 & -- & -- & -- & 0.511 & 0.511  & -- &	0.511	 \\
					 $x=j$ & -- & 0.3 & 0.5 & 0.8 & 0.6 & 0.546 & 0.423 & 0.691 & 0.530 \\	 
					 $c_{u}$ & 0.4 & 0.5 & 0.8 & 0.5 & 0.6 & & & & \\	 	\hline
			 	\end{tabular}}
			 	\caption{Ratings given by users to items, users' reputations and items' rankings for both BWA and MRS.}
			 	\label{table3}
			 	\end{table}
				Suppose the sellers of item $i$ want to bribe a user in order to get a larger ranking.
				In the first strategy, the sellers bribe user $u_4$, with $\sigma_{u_4}^i = 1$. In the second strategy, the sellers bribe user $u_5$, with $\sigma_{u_5}^i= 1$.
				The profit of each strategy for both BWA and MRS is represented in Table~\ref{table2}. 
				\begin{table}
			 	\centering
			 	\resizebox{0.48\textwidth}{!}{\begin{tabular}{ | c | c c | c | c c| }
			 		\hline
			 		  & \multicolumn{2}{ c}{\textsc{Profit $\pi$}} & \multicolumn{3}{| c|}{\textsc{Ranking of} $i$} \\ \hline
			 		 \textsc{Strategy} & \textsc{Bipartite} & \textsc{Multipartite} & $r_i$ & $r_{i,\mathcal M_2}$ & $\bar r_i$\\ 
					 $\sigma_{u_4}^i$ & 0.035 & 0 & 0.686 & 1 & 0.707  \\
					 $\sigma_{u_5}^i$ & 0.098 & 0 & 0.707 & 1 & 0.707 \\ \hline	 
			 	\end{tabular}}
			 	\caption{Profits of bribing strategies $\sigma_{u_4}^i$ and $\sigma_{u_5}^i$ and new ranking, after applying the strategies, in both BWA and MRS.}
			 	\label{table4}
			 	\end{table}% 
			 	Both strategies $\sigma_{u_4}$ and $\sigma_{u_5}$ are profitable for the BWA.
				However, both are not profitable for the MRS. 
				The profit of bribing each user is depicted in Figure~\ref{fig:barchart2}.
			 	\begin{figure}[ht]
		   		\centering
		   		\includegraphics[width=0.48\textwidth]{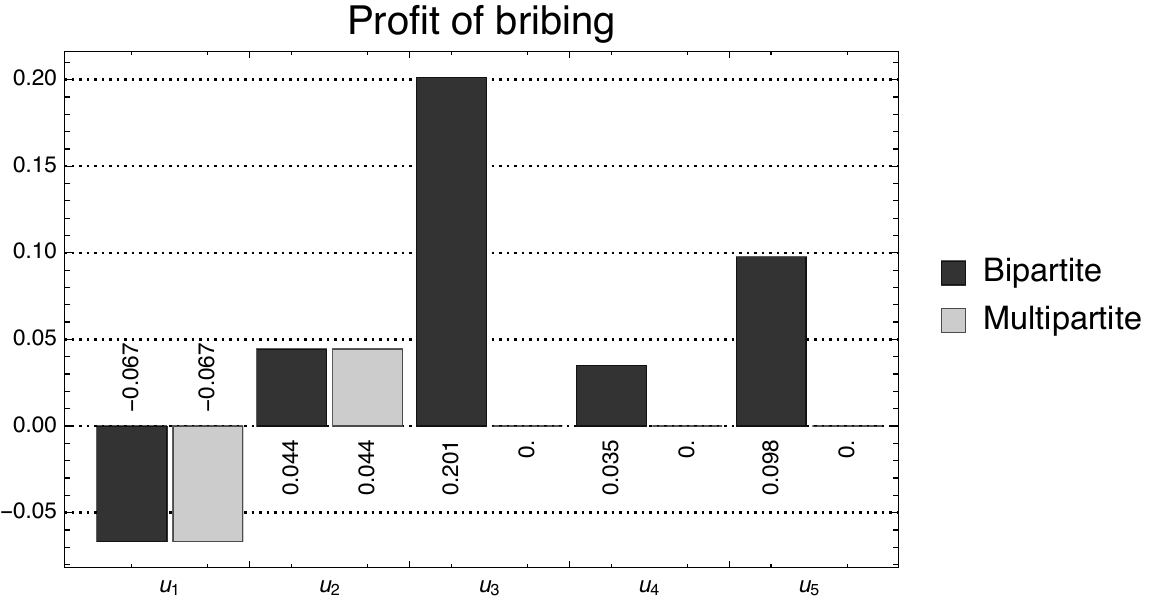}
		  	 		 	\caption{Profit of $i$ seller when bribing each user in the setup of Table~\ref{table3}. The black and gray bars correspond to use BWA and MRS, respectively.}
		 	 	\label{fig:barchart2}
		 		\end{figure}
			\end{example}
			Note that in both Examples~\ref{exp2} and~\ref{exp3}, the average and maximum of the profits for all elementary strategies are greater in BWA.
	 	    Therefore, in these examples, MRS is more robust to bribing, as expected.
		% subsection synthetic_data (end)
		\subsubsection{Real data} % (fold)
		\label{sub:real_data}
			We illustrate the results in Section~\ref{sec:bribing_in_ranking_systems} for the $5$-core version of ``Amazon Instant Video'' data set.
			As in Section~\ref{sub:robustness_against_noise}, we study bribing strategies for the seller of the most rated item, target, with $455$ ratings. 
		
			Under the described scenario, we study the effect of four strategies in BWA and two in MRS, which are:
			\begin{description}
				\item[$\sigma_1$] bribe users that rated the item, by a random order;
				\item[$\sigma_2$] bribe users that rated the item, by decreasing reputation;
				\item[$\sigma_3$] bribe users uniformly at random, from all users (only for BWA);
				\item[$\sigma_4$] bribe users in decreasing order of reputation (only for BWA).
			\end{description} 
			  
\begin{figure}
				\centering
				\begin{subfigure}[b]{0.27\textwidth}
			        \includegraphics[width=\textwidth]{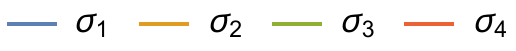}
			    \end{subfigure}	
			    	
			    \begin{subfigure}[b]{0.36\textwidth}
			        \includegraphics[width=\textwidth]{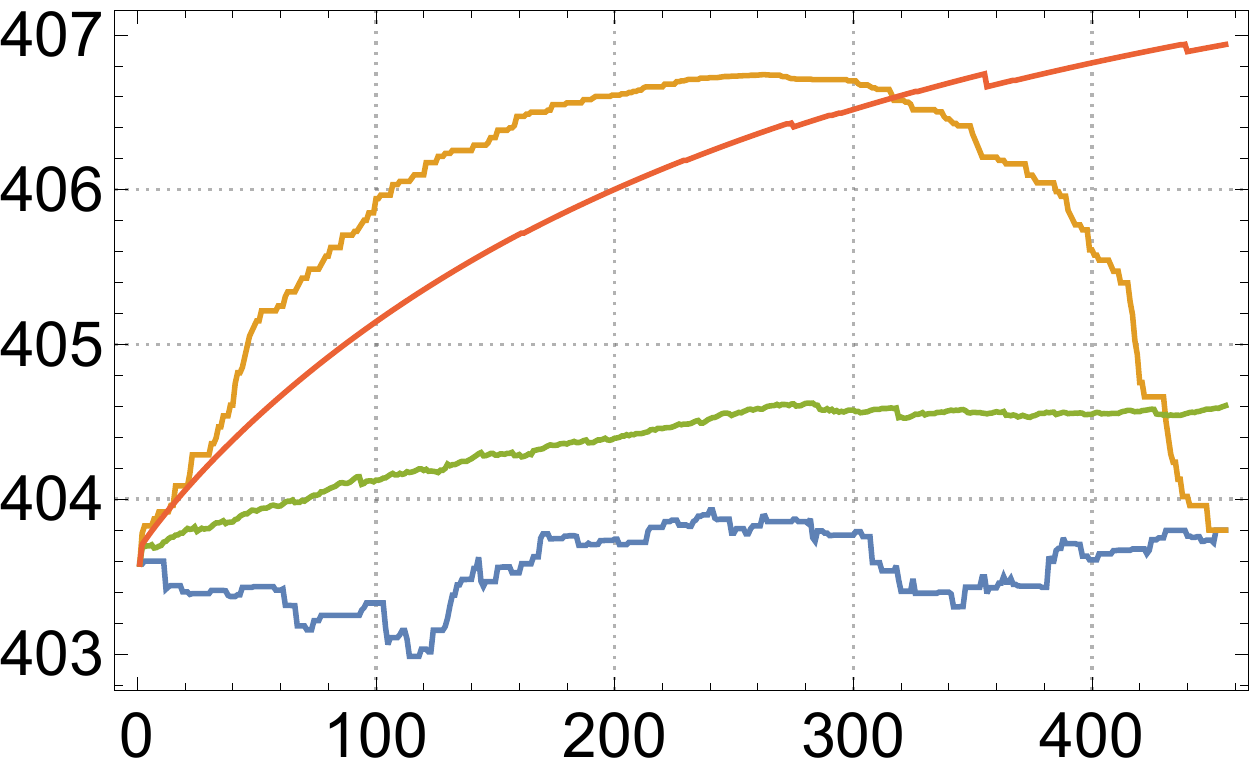}
			        \caption{}%{Bribing on bipartite system.}
					\label{fig:a}
			    \end{subfigure}
				\begin{subfigure}[b]{0.36\textwidth}
					\includegraphics[width=\textwidth]{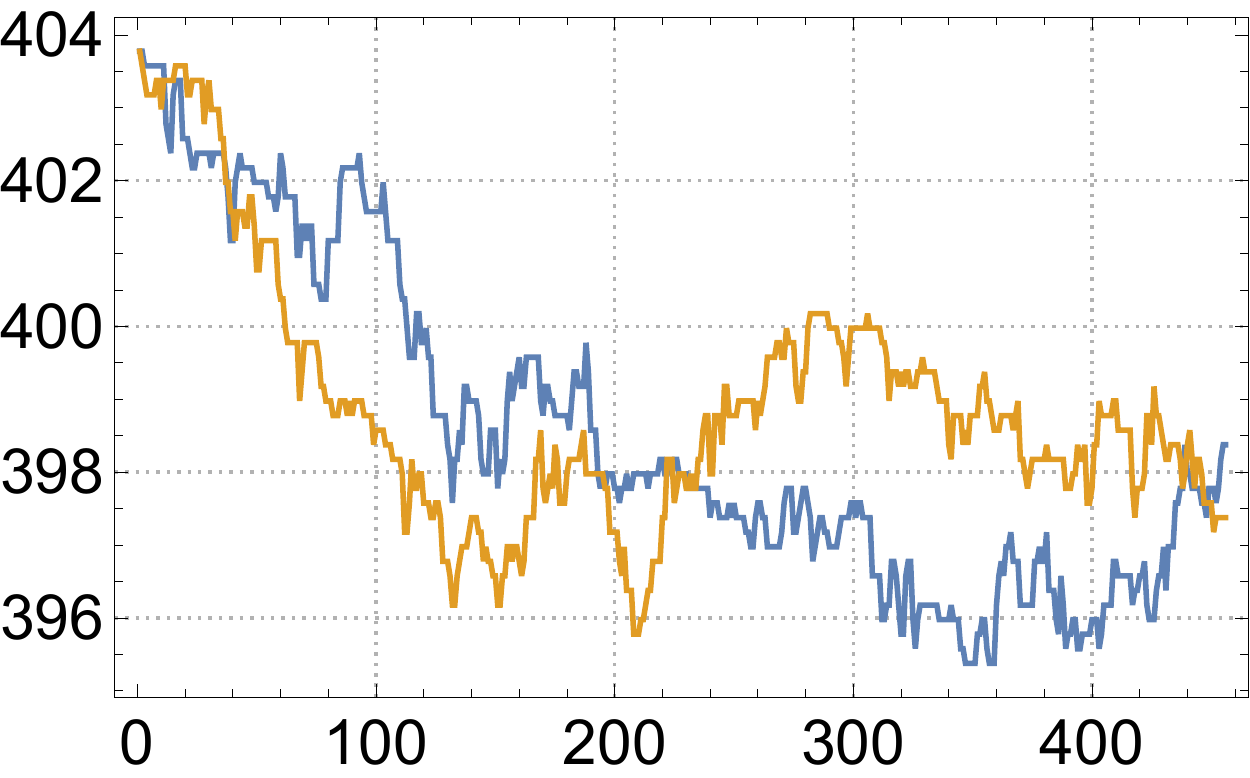}
				    \caption{}%{Bribing on multipartite system.}
					\label{fig:b}
				\end{subfigure}
				\caption{Profit of bribing strategies of the most rated item's sellers in (a) BWA ($\sigma_1$ -- $\sigma_4$), and (b) MRS ($\sigma_1$ and $\sigma_2$).}
				\label{fig:results}
			\end{figure}

			Figures~\ref{fig:results}~(a) and~(b) depict the wealth evolution of item $i$ seller for the different bribing strategies, respectively for BWA and MRS.
			In Figures~\ref{fig:results}~(a) and~(b), the steps where the wealth is constant, correspond to choosing users that rated the item with the maximum allowed rating, hence both the invested wealth and the profit are zero.
			As expected, for the BWA (Figure~\ref{fig:results} (a)), after bribing the same users in strategies $\sigma_1$ and $\sigma_2$, both yield the same wealth, as noticed in  Proposition~\ref{prop:1}.
			As conjectured, bribing users with larger reputation, among the ones who rated the item, yields a faster increase of reward, whereas random bribing among the item's raters has an expected profit close to zero, and does not increase wealth.
			In strategy $\sigma_3$ (Figure~\ref{fig:results} (a)) the seller bribes users from the set of all users, by decreasing reputation.
			This time, the wealth mostly increases for all bribed users. This occurs because a good amount of the chosen users did not rate the	item and yield a positive profit, comparing to the strategy $\sigma_1$. 
			The most profitable strategy is $\sigma_4$.
			However, the profit is larger then in $\sigma_2$ only after a large number of users.
			In summary, the profit is positive the four bribing strategies.
			
			In the MRS scenario, Figure~\ref{fig:results}~(b), for both $\sigma_1$ and $\sigma_2$, at the end of the bribing strategy, the wealth is strictly smaller than in the BWA case.
			Moreover, both strategies are not profitable. 
			This meets the discussion in Section~\ref{sub:bipartite_vs_multipartite_networks}, where we point that MRS is more robust to bribery than the BWA.
			
\begin{figure}
			\centering
		    	\includegraphics[width=.42\textwidth]{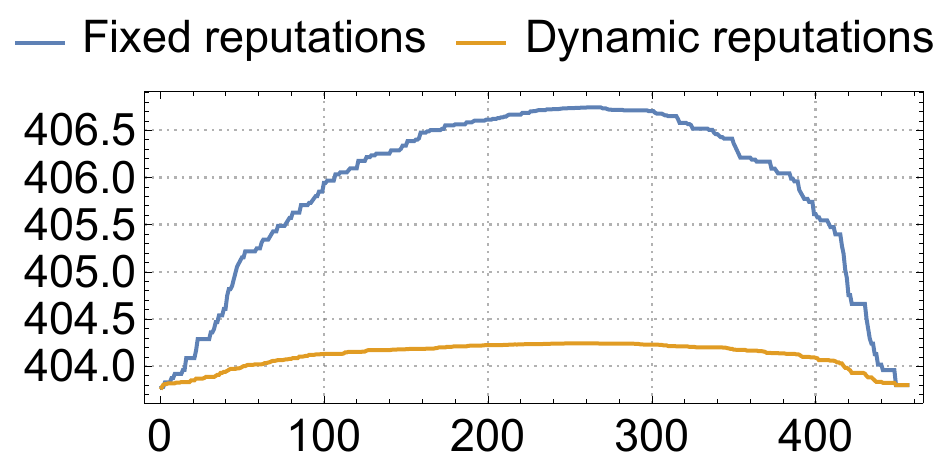}
		    	\caption{Profit of bribing strategy $\sigma_2$ in BWA, fixed users' reputations versus reputations recomputed after each user being bribed.} 
				\label{fig:dyn} 
			\end{figure}

			Lastly, to study the effect of the re-computation of the user's reputations, we compare applying strategy $\sigma_2$ to the BWA, assuming that the users' reputations are fixed, with the case where each time a user is bribed, both rankings and reputations updated as in Section~\ref{sub:ranking}. 
			This comparison is depicted in Figure~\ref{fig:dyn}.
			In fact, we see that the reputations of the bribed users decrease.
			This means that the impact of the bribing strategy is, actually, dimmed when the reputations are recomputed, i.e., the return of the strategy is smaller for dynamic reputations than when the reputations are fixed.
		% subsection real_data (end)
	% subsection robustness_against_bribery (end)
% section experimental_results (end)
\section{Conclusions} % (fold)
\label{sec:conclusions}
	%Compare the robustness of our algorithm to the bipartite one.
    In this paper, we advanced state of the art in ranking systems, both theoretically and algorithmically.
    We developed a new multipartite ranking system that allows the coexistence of multiple preferences by enabling different rankings for the same item for different users.
    This is achieved by automatically clustering similar users, based on their given ratings.
    For each cluster, we used a bipartite reputation-based ranking system, for which we proved convergence and efficiency in a more general setting than previous results. 
    Our method favors the creation of bubbles, \emph{i.e.}, segregates users into groups, which we show that makes the ranking system more robust to attacks and spamming.
    
	Further, we model bribing in the BWA and MRS.
	In both scenarios, we study which users are profitable to bribe.
	Also, we show that clustering users, the MRS case, decreases the profitable bribing strategies. 
	We illustrate our main results, the effect of attacks and spamming, and the effect of bribing, with real world datasets.
	
	One possible future direction is to explore the use of steadiness functions, based on a timestamp, so that established clusters do not change so easily, in order to reduce the rate of change in the clusters. 
	%Another possible extension of the proposed MRS is its use for recommendation systems.
	Moreover, we would like to study the interactions between big and
small players, as well as the scenario where sellers bribe users to degrade a competitor item's ranking through a game theory model, with the sellers as players.
	Finally, another aspect we want to explore and incorporate into the bribery analysis is the impact on the profit of strategies when the reputations are dynamic.
	Hence, exploring new conditions to design bribing strategies with positive return.
% section conclusions (end)

% \begin{acks}
% %\todo{To be filled}
% G. Ramos is with Department of Electrical and Computer Engineering, Faculty of Engineering, University of Porto, Portugal. This work was supported in part by FCT project POCI-01-0145-FEDER-031411-HARMONY. 
% Further, this work was developed under the scope of R\&D Unit 50008, financed by the applicable financial framework (FCT/MEC through national funds and when applicable co-funded by FEDER - PT2020 partnership agreement). 
% G. Ramos further acknowledges the support of the DP-PMI and Funda\c{c}\~ao para a Ci\^encia e a Tecnologia (Portugal), through scholarship SFRH/BD/52242/2013 and the support of Instituto de Telecomunica\c{c}\~oes through the research grant - BIM/N\textsuperscript{o}154 - 16/11/2017 - UID/EEA/50008/2017. 
% Research funded by FCT/MCTES through national funds and when applicable co-funded by EU under the project UIDB/EEA/50008/2020. 
% \end{acks}

% Bibliography
\bibliographystyle{unsrt}
\bibliography{ref}

\pagebreak

\appendix\label{sec:Appendix_rank_algo}

\section{Proofs} 
    In this appendix, we state the proofs regarding the properties of both bipartite and multipartite reputation-based ranking algorithms presented in Section \ref{sub:proof_ranking_algo}.
    In Section \ref{sub:proof_brib} we state the proofs regarding the robustness of the algorithms to attacks, namely bribing.
    \subsection{Proofs related to ranking algorithms} \label{sub:proof_ranking_algo}
    	\begin{proof}{[Lemma~\ref{lemma:1}].}
    		Since the domain of $g$ contains the codomain of $h$ and both are Lipschitz the composition, $g\circ h$, is also Lipschitz.
    		Let $d$ be a distance, we prove the induction's basis:
    		\begin{equation*}
    			\begin{split}
    				d(r^2,r^1) & = d\lb g_R(c^{1}),g_R(c^0) \rb \\
    						   & = d\lb (g_R\circ h_R)(r^{1}),(g_R\circ h_R) (r^0)\rb 
    						   \leq \eta d(r^{1},r^0),
    			\end{split}
    		\end{equation*}
    		where $\eta\in[0,1[$ is the Lipschitz constant for $g_R \circ h_R$.
    		The induction step then reads
    		\begin{equation*}
    			\begin{split}
    				d(r^n,r^{n-1}) &= d\lb g_R(c^{n-1}),g_R(c^{n-2}) \rb \\
    							   & = d\lb (g_R\circ h_R)(r^{n-1}),(g_R\circ h_R) (r^{n-2})\rb \\
    							   &\leq \eta d\lb r^{n-1}, r^{n-2} \rb 
    							    = \eta d\lb g_R(c^{n-2}),g_R(c^{n-3}) \rb \\
    							   & \leq \eta^{n-1}d(r^{1},r^0),
    			\end{split}
    		\end{equation*}
		    and the last inequality holds by the induction hypothesis.
	    \end{proof}
    	\begin{proof}{[Theorem~\ref{th:1}].}
    		Let $m,n \in \Nn$. 
    		For any $\varepsilon>0$, there exists an order, $N$, from which $\eta^N < (1-\eta)\varepsilon/d(r^1,r^0)$.
    		Using the triangle inequality we have
    		\begin{equation*}
    			\begin{split}
    				d(r^n,r^m) &\leq \sum_{k=m+1}^n d(r^k, r^{k-1}) \leq \sum_{k=m+1}^n \eta^{k-1} d(r^1,r^0) \\
    						   & \leq \eta^m d(r^1,r^0)\sum_{k=0}^{+\infty} \eta^k \leq \frac{\eta^N d(r^1,r^0)}{1-\eta} < \varepsilon,			
    			\end{split}
    		\end{equation*}
    		since $0<\eta<1$, therefore the algorithm~\eqref{eq:rbra} converges.
    	\end{proof}	
        \begin{proof}{[Theorem~\ref{th:2}].}
    		The basis of the induction reads:
    		\begin{equation*}
    			\begin{split}
    				d(r^*,r^1) &= d\lb g_R(c^{*}),g_R(c^0) \rb	\\
    						   &= d\lb (g_R\circ h_R)(r^{*}),(g_R\circ h_R) (r^{0})\rb \\
    						   &
    						   \leq \eta d\lb r^{*},r^0 \rb\leq\eta.
    			\end{split}
    		\end{equation*}
    		Assume that the induction hypothesis holds, for $k=n$, then it follows that 
    		\begin{equation*}
    			\begin{split}
    				d(r^*,r^{n+1}) &=  d\lb (g_R\circ h_R)(r^{*}),(g_R\circ h_R) (r^{n}) \rb\\	
    							   &\leq \eta d\lb r^{*}, r^{n} \rb \leq \eta^{n+1} d\lb r^{*}, r^{0} \rb \leq \eta^{n+1}.\qedhere
    			\end{split}
    		\end{equation*}
    	\end{proof}
        \begin{proof}{[Lemma~\ref{prop:convergence}].}
			Between iterations, $r^{k+1}$ and $r^k$, we get
			\begin{equation*}
				\|r_i^{k+1}-r_i^k\|_{\infty} = \left\| \frac{R_{i} \cdot c^{k+1}}{\|c^{k+1}\|_1} -  \frac{R_{i} \cdot c^{k}}{\|c^{k}\|_1} \right\|_{\infty}.
			\end{equation*}
			Here, $R_i \in [0,1]^{|U|}$ denotes a vector that contains the rating $R_{ui}$ that each user $u$ gave to item $i$ (the element corresponding to a user is 0 if s/he did not rate the item). 
            
            Without loss of generality, assume that $\|c^{k+1}\|_1 \geq \|c^{k}\|_1$, then the above difference is equal to
			\begin{equation*}
				\begin{split}
				%\|r_j^{k+1}-r_j^k\|_{\infty} &= 
					\left\| \frac{R_{i} \cdot c^{k+1}}{\|c^{k+1}\|_1}  - \frac{R_{i} \cdot c^{k}}{\|c^{k+1}\|_1} + \frac{R_{i} \cdot c^{k}}{\|c^{k+1}\|_1} - \frac{R_{i} \cdot c^{k}}{\|c^{k}\|_1} \right\|_{\infty}
					\\ \leq \left\| \frac{R_{i} \cdot c^{k+1}}{\|c^{k+1}\|_1} - \frac{R_{i} \cdot  c^{k}}{\|c^{k+1}\|_1} + \frac{R_{i} \cdot  c^{k}}{\|c^{k}\|_1} - \frac{R_{i} \cdot  c^{k}}{\|c^{k}\|_1} \right\|_{\infty} \\
					\\ \leq \frac{R_\top}{\|c^{k+1}\|_1} \left| c_{\gamma}^{k+1}-c_{\gamma}^k \right|,
				% &= \frac{R^T}{\|c^{k+1}\|_1} \left\| c^{k+1}-c^k \right\|_1
				\end{split}
			\end{equation*}
			where $\left| c_{\gamma}^{k+1}-c_{\gamma}^k \right| = \max_{u\in U_i} \left| c_{u}^{k+1}-c_{u}^k \right|$.
			The iteration step for the reputation, $c$, gives us
			\begin{equation*}
				\begin{split}
					|c_u^{k+1}-c_u^k| %=\left| 1- \frac{f(\lambda,O_i)}{|O_i|} \sum_{i} \left| R_{ji} - r_j^{k+1} \right|^p - 1 + \frac{f(\lambda,O_i)}{|O_i|} \sum_{i} \left| R_{ji} - r_j^{k} \right|^p\right| \\
						&\leq \frac{|f_{\lambda,s}(I_u)|}{|I_u|} \sum_{i} \left|  \left| R_{ui} - r_i^{k} \right|^p - \left| R_{ui} - r_i^{k-1} \right|^p\right| 
						\\
						&\leq \lambda |r_{\beta}^{k}-r_{\beta}^{k-1}|,
				\end{split}
			\end{equation*}
			where $\left| r_{\beta}^{k+1}-r_{\beta}^k \right| = \max_{i\in I_u} \left| r_{i}^{k+1}-r_{i}^k \right|$, and using the triangular inequality, the translation invariance of norms and the fact that $|f_{\lambda,s}(I_u)|\leq 1$.
			Combining the previous inequalities we get
			\begin{equation}\label{eq:pr:conv}
				|r_i^{k+1}-r_i^k| \leq \frac{\lambda}{\|c^{k+1}\|_1} |r_{\beta}^{k}-r_\beta^{k-1}|,
			\end{equation}
			which is a contraction
			for $\lambda<(1+\Delta_R)^{-1}$, since $1-\Delta_\mathcal R\lambda \leq \|c\|_1 \leq 1$. 
			Therefore~\eqref{eq:rbra} converges.
		\end{proof}

    \subsection{Proofs related to robustness of algorithms to bribing} \label{sub:proof_brib}
        \begin{proof}{[Proposition~\ref{prop:1}].} \label{proof:prop1}
            \underline{Case $n$=1}:
                When one user is bribed the ranking of the item $i$ changes according to 
		    	\[
			    	r_{\sigma_u^i} = r_i + \frac{c_u}{\alpha} \rho_u,
			    \]
			    where $\alpha = \sum_{u\in U_i} c_u$.
		    	The profit of the elementary strategy is given by
		    	\begin{equation} \label{eq:profit_elem_strat}
		    	    \pi_{\sigma_u^i} = |U_i| r_{\sigma_u^i} - \rho_u - |U_i| r_i 
		    		                 = \left(\frac{c_u}{\bar{c}_{U_i}} -1 \right) \rho_u,
		    	\end{equation}
		    	where $\bar c_{U_i} = \frac{\alpha}{|U_i|}$ is the average reputation of the users that rated item $i$.
			    This strategy is profitable when the reputation of user $u$ is bigger than the average reputation of users that already rated item $i$, \emph{i.e.}, $c_u > \bar c_{U_i}$.
			    
		    \underline{Case $n=N$}:
			    %Case where $n$-users are bribed and $\{u_1,\dots, u_N\} \in U_i$:
			    After bribing $N$ buyers the ranking of the item $i$ changes according to 
			    \[
				    r_{\sigma_i} = r_i +\frac{1}{\alpha} \sum_{u=1}^N c_u \rho_u.
			    \]
			    The profit can be written as a sum of the profit of elementary strategies, \eqref{eq:profit_elem_strat}, as
			    \begin{equation*}
			    	\pi_{\sigma^i} = \sum_{u\in U_b} \lb \frac{|U_i|}{\alpha} c_u -1 \rb \rho_u 
			    	               = \sum_{u\in U_b} \pi_{\sigma_u^i}.%,
			    \end{equation*}
		    	%where $\pi_{\sigma_u^i}$ is the profit of an elementary strategy.
        \end{proof}
        \begin{proof}{[Proposition~\ref{prop:2}.]}
			Here, we consider the case when the seller of item $i$ bribes users that did not rate its product to do so.
			Let $V_i = U\setminus U_i$ denote the set of users that did not rate item $i$, and $V_b \subseteq V_i$ be the set of users bribed by the seller of item $i$.
			
			\underline{Case $m=1$}:
			    Let $v\in V_b$, bribing user $v$ changes the rating of product $i$ as
			    \[
			        r_{\sigma^i_v} = \frac{c_v \rho_v + \sum_{u\in U_i} c_u R_{ui}}{c_v + \sum_{u\in U_i}c_u}
			                       = \frac{\alpha r_i +c_v \rho_v}{\alpha + c_v}.
			    \]
			    Using the above we compute the profit of this elementary strategy:
			    \[
			        \begin{split}
			            \pi_{\sigma^i_v} &= \lb |U_i| +1 \rb r_{\sigma^i_v} -\rho_v - |U_i| r_i \\
			                             &= \frac{|U_i|+1}{\alpha+c_v}(\alpha r_i + c_v \rho_v) - \rho_v - |U_i| r_i\\
			                             &= \frac{|U_i|c_v}{\alpha+c_v} (\rho_v - r_i) + \alpha \frac{r_i-\rho_v}{\alpha+c_v}\\
			                             &= \frac{\alpha - |U_i| c_v}{\alpha + c_v} (r_i - \rho_v)\\
			                             &= (\alpha-|U_i|c_v) \frac{r_i-\rho_v}{\alpha+c_v}.
			        \end{split}
		    	\]
		    	This strategy is profitable if
			    \[
			    	\lb c_v < \bar c_{U_i} \bigwedge r_i > \rho_v \rb \quad \bigvee \quad \lb c_v > \bar c_{U_i} \bigwedge r_i < \rho_v \rb,
		    	\]
		    	where $\bar c_{U_i} = \frac{\alpha}{|U_i}$ is the average reputation of users that had previously rated item $i$.
		    	
			\underline{Case $m=M$}:
			    Here, we look to the case where the seller of item $i$ bribes $M$ users $\{v_1,\ldots,v_M\} = V_b$ that have not rated the item $i$ previously.
		    	We have
		    	\begin{align*}
			    	\pi_{\sigma_i} &= \lb |U_i|+|V_b| \rb r_{\sigma^i} - \sum_{v\in V_b} \rho_v - |U_i|r_i\\
			    	    &= \lb |U_i|+M \rb \frac{\alpha r_i +\sum_{v\in V_b} c_v \rho_v}{\tilde \alpha} - \sum_{v\in V_b} \rho_v - |U_i|r_i\\
						&= \frac{|U_i|}{\tilde \alpha} \lb \alpha r_i + \sum_{v\in V_b} c_v \rho_v - \alpha r_i -r_i \sum_{v\in V_b} c_v \rb - \sum_{v\in V_b} \rho_v 
						    + \frac{M}{\tilde \alpha} \lb \alpha r_i + \sum_{v\in V_b} c_v \rho_v \rb \\
						&= |U_i| \frac{\sum_{v\in V_b} c_v (\rho_v - r_i)}{\tilde \alpha} + \frac{M}{\tilde \alpha} \alpha r_i - \sum_{v\in V_b} \rho_v 
						  + \frac{M}{\tilde \alpha} \sum_{v\in V_b} c_v \rho_v\\
						&= \frac{|U_i|}{\tilde \alpha} \sum_{v\in V_b}  c_v (\rho_v - r_i) + \frac{M}{\tilde \alpha} \alpha r_i - \frac{\alpha + \sum_{v\in V_b}  c_v}{\tilde \alpha}
						    \sum_{v\in V_b} \rho_v + \frac{M}{\tilde \alpha} \sum_{v\in V_b} c_v \rho_v\\
						&= \frac{|U_i|}{\tilde \alpha} \sum_{v\in V_b}  c_v (\rho_v - r_i) + \frac{M}{\tilde \alpha} \sum_{v\in V_b} c_v \rho_v 
						   + \frac{\alpha}{\tilde \alpha} \sum_{v\in V_b} (r_i-\rho_v) - \frac{1}{\tilde \alpha} \lb \sum_{v\in V_b} c_v \rb \lb \sum_{v\in V_b} \rho_v \rb \\
						&= \sum_{v\in V_b} \lb \alpha - |U_i|c_v \rb \frac{r_i-\rho_v}{\tilde \alpha} +         \frac{M-1}{\tilde \alpha} \sum_{v\in V_b}  c_v \rho_v - \frac{1}{\tilde \alpha}       \sum_{v\in V_b}c_v \lb \sum_{w\neq v} \rho_w \rb\\
						&= \frac{1}{\tilde \alpha} \sum_{v \in V_b} \lb \alpha + c_v \rb \pi^i_v + \frac{1}{\tilde \alpha} \sum_{v\in V_b} \lb c_v \lsb        (M-1) \rho_v - \sum_{w \neq v} \rho_w \rsb \rb,
		        \end{align*}
		        
		        where $\tilde \alpha =\displaystyle \sum_{u\in U_i} c_u + \sum_{v\in V_b}c_v$.
		\end{proof}
		\begin{proof}{[Proposition~\ref{prop:3}].}
	        In the case when a seller decides to convince the raters to update their ratings and also bribe new users to rate its product, the profit is given as 
	        
	        \underline{Case $n=m=1$}:
	            The ranking changes as:
	            \begin{equation*}
	                r_{\sigma^i} = \frac{\sum_{u\in U_i} c_u R_{ui} + c_u \rho_u + c_v \rho_v}{c_v + \sum_{u\in U_i} c_u}.
	            \end{equation*}
	            Using the following identity:
	            \begin{equation} \label{eq:identity_frac}
	                \frac{x}{y+z} = \frac{x}{y} - \frac{xz}{y(y+z)},
	            \end{equation}
	            the profit is given by
	            \begin{equation*}
	            \resizebox{.999\hsize}{!}{$
	                \begin{aligned}
	                    \pi_{\sigma_i} &= \lb |U_i|+1 \rb r_{\sigma^i} - \rho_u - \rho_v - |U_i| r_i\\
	                                   &= |U_i| \frac{\alpha r_i + c_u \rho_u + c_v \rho_v}{\alpha + c_v} + r_{\sigma^i} - \rho_u - \rho_v - |U_i|r_i \\
	                                   &= \lb \frac{|U_i|}{\alpha} - \frac{|U_i|c_v}{\alpha(\alpha+c_v)} \rb (\alpha r_i + c_u \rho_u + c_v \rho_v) + r_{\sigma^i} - \rho_u - \rho_v - |U_i| r_i\\
	                                   &= \frac{|U_i|}{\alpha} (\alpha r_i + c_u \rho_u) - \rho_u - |U_i| r_i + \frac{|U_i|}{\alpha} c_v \rho_v - \frac{|U_i| c_v}{\alpha(\alpha +c_v)} (\alpha r_i +c_u \rho_u + c_v \rho_v) + r_{\sigma^i} - \rho_v\\
	                                   &= \pi_{\sigma^i_u} 
	                                    + \frac{|U_i|}{\alpha} c_v \rho_v - \frac{c_v}{\alpha} |U_i| r_{\sigma^i} + r_{\sigma^i} -\rho_v \\
	                                   &= \pi_{\sigma^i_u} + \frac{|U_i|}{\alpha} c_v \rho_v + \lb 1- \frac{|U_i|}{\alpha}c_v \rb \frac{\alpha r_i +c_u \rho_u +c_v \rho_v}{\alpha + c_v} -\rho_v\\
	                                   &= \pi_{\sigma^i_u} + \lb 1 - \frac{c_v}{\bar c_{U_i}} \rb \frac{c_u \rho_u}{\alpha + c_v} + \frac{|U_i|}{\alpha} c_v \rho_v + \lb 1 - \frac{|U_i|}{\alpha} c_v \rb \frac{\alpha r_i + c_v \rho_v}{\alpha+c_v} - \rho_v \\
	                                   &= \pi_{\sigma^i_u} + \lb 1 - \frac{c_v}{\bar c_{U_i}} \rb \frac{c_u \rho_u}{\alpha + c_v} + \frac{|U_i|}{\alpha}c_v \rho_v + \frac{\alpha-|U_i|c_v}{\alpha} \frac{\alpha r_i +c_v \rho_v}{\alpha+c_v} - \rho_v \\
	                                   &= \pi_{\sigma^i_u} + \lb 1 - \frac{c_v}{\bar c_{U_i}} \rb \frac{c_u \rho_u}{\alpha + c_v} + \frac{|U_i|}{\alpha}c_v \rho_v + \frac{\alpha-|U_i|c_v}{\alpha+c_v} (r_i-\rho_v) + \frac{\alpha-|U_i|c_V}{\alpha+c_v} \rho_v - \rho_v \\ 
	                                   &\quad+  \frac{|U_i|}{\alpha}c_v \rho_v + \frac{\alpha-|U_i|c_v}{\alpha(\alpha+c_v)} c_v\rho_v\\
	                                   &= \pi_{\sigma^i_u} + \pi_{\sigma^i_v} + \lb 1- \frac{c_v}{\bar c_{U_i}} \rb \frac{c_u \rho_u}{\alpha+c_v} + \frac{\rho_v}{\alpha+c_v} \lb \alpha - |U_i|c_v - (\alpha+c_v) + \frac{|U_i|c_v(\alpha+c_v)}{\alpha} + \frac{\alpha-|U_i|c_v}{\alpha}c_v \rb \\
	                                   &= \pi_{\sigma^i_u} + \pi_{\sigma^i_v} + \lb 1- \frac{c_v}{\bar c_{U_i}} \rb \frac{c_u \rho_u}{\alpha+c_v},
 	                \end{aligned}
 	                $}
	            \end{equation*}
	            where $\alpha = \displaystyle\sum_{u \in U_i} c_u$.
	            
	            The above result tells us that the profit can be decomposed in the profit of two elementary strategies, $\pi_{\sigma^i_u}$ and $\pi_{\sigma^i_v}$, plus a extra term. 
	            
	        \underline{Case $n=N$ and $m=M$}: 
	            Let $U_b$ denote the set of $N$ bribed users that already rated item $i$, $V_b$ the set of $M$ bribed users that did not rate it, and
	            \[
	                \tilde \alpha = \alpha + \sum_{v\in V_b} c_v, \text{ where } \alpha = \sum_{u \in U_i} c_u.
	            \]
	            In this case the rating changes as:
	            \begin{equation*}
	                r_{\sigma^i} = \frac{1}{\tilde \alpha}\lb\sum_{u\in U_i} c_u R_{ui} + \sum_{u\in U_b} c_u \rho_u + \sum_{v\in V_b} c_v \rho_v\rb %{\sum_{u\in U_i} c_u + \sum_{v=1}^M c_v }
	            \end{equation*}
	            The profit for this combined strategy is 
	            \begin{equation*}
	                \begin{split}
	                    \pi_{\sigma^i} &= \lb |U_i| + |V_b| \rb r_{\sigma^i} - \sum_{u\in U_b} \rho_u - \sum_{v\in V_b} \rho_v - |U_i|r_i \\
	                        &=  \lb |U_i| + M \rb \frac{\alpha r_i + \sum_{v \in V_b} c_v \rho_v + \sum_{u\in U_b} c_u \rho_u}{\tilde \alpha} 
	                            - \sum_{v \in V_b} \rho_v - \sum_{u\in U_b} \rho_u - |U_i| r_i\\
	                        &= \lb |U_i| + M \rb  \frac{\alpha r_i + \sum_{v \in V_b} c_v \rho_v }{\tilde \alpha} - \sum_{v\in V_b} \rho_v  - |U_i|r_i
	                            + \lb |U_i|+M \rb \frac{\sum_{u\in U_b} c_u \rho_u}{\tilde \alpha} - \sum_{u\in U_b} \rho_u,
	                \end{split}
	            \end{equation*}
	            where we substituted the definition of $r_{\sigma^i}$ and re-arranged terms.
	            Substituting above the result of Proposition \ref{prop:2}, case $m=M$, we get
	            \begin{equation*}
	                \begin{split}
	                    \pi_{\sigma^i} = \frac{1}{\tilde \alpha} \sum_{v \in V_b} \lb \alpha +c_v \rb \pi_{v}^i 
	                                        + \frac{1}{\tilde \alpha} \sum_{v\in V_b} \lb c_v \lsb (M-1) \rho_v - \sum_{w \neq v} \rho_w \rsb \rb
	                                   + \frac{|U_i|}{\tilde \alpha} \sum_{i\in U_b} c_u \rho_u - \sum_{u \in U_b} \rho_u \\
	                                   + \frac{M}{\tilde \alpha} \sum_{u \in U_b} c_u \rho_u.
 	                \end{split}
	            \end{equation*}
	            Applying the identity, \eqref{eq:identity_frac}, to the third element of the right hand side in the above equation;
	            then, using the Proposition \ref{prop:1}, case $n=N$, we get
	            \begin{equation*}
	                \begin{split}
	                     \pi_{\sigma^i} &= \frac{1}{\tilde \alpha} \sum_{v \in V_b} \lb \alpha +c_v \rb \pi_{v}^i 
	                                        + \frac{|U_i|}{\alpha} \sum_{u\in U_b} c_u \rho_u - \sum_{u\in U_b} \rho_u  
	                                        - \frac{|U_i| \sum_{u\in U_b} c_u \rho_u}{\alpha \tilde\alpha} \sum_{v\in V_b} c_v\\
	                                        &+ \frac{1}{\tilde \alpha} \sum_{v\in V_b} \lb c_v \lsb (M-1) \rho_v - \sum_{w \neq v} \rho_w \rsb \rb
	                                        + \frac{M}{\tilde \alpha} \sum_{u \in U_b} c_u \rho_u \\
	                                    &= \frac{1}{\tilde \alpha} \sum_{v \in V_b} \lb \alpha +c_v \rb \pi_{v}^i + \sum_{u \in U_b} \pi_u^i + \frac{1}{\tilde \alpha} \sum_{v\in V_b} \lb c_v \lsb (M-1) \rho_v - \sum_{w \neq v} \rho_w \rsb \rb \\
	                                    &+ \frac{\sum_{u \in U_b} c_u \rho_u}{\tilde \alpha} \lb M- \frac{|U_i| \sum_{v]\in V_b}c_v}{\alpha} \rb \\
	                                    &= \sum_{u \in U_b} \pi_u^i + \frac{1}{\tilde \alpha} \sum_{v \in V_b} \lb \alpha +c_v \rb \pi_{v}^i 
	                                        + \frac{1}{\tilde \alpha} \sum_{v\in V_b} \lb c_v \lsb (M-1) \rho_v - \sum_{w \neq v} \rho_w \rsb \rb \\
	                                    &+ \frac{1}{\tilde \alpha} \lsb \sum_{v \in V_b} \lb 1 - \frac{c_v}{\bar c_{U_i}} \rb \sum_{u\in U_b}c_u \rho_u \rsb.
	                \end{split}
	            \end{equation*}
	    \end{proof}
	    \begin{proof}{[Proposition~\ref{prop:AllVoteCluster}].}
    		Following the same steps as in the proof of Proposition~\ref{prop:1} for one user, replacing $U_i$ by $U_i^{\Mm_s}$, we have that
    		\[
    			\bar \pi_{\sigma_v^i} = \bar J_{\sigma^i_v}-\bar J_i
    								  = \rho_v( c_v/\bar c_{U_i^{\Mm_s}}-1)>0.
    		\]
		\end{proof}
	    \begin{proof}{[Proposition~\ref{prop:7}].}
			Since $|U_i^{\Mm_s}|=0$, then
			\[
				\bar \pi_{\sigma^i_v} = \sum_{m\in\mathcal N_i} |U_i^{\Mm_m}| r_i^{\Mm_m} +  (|U_i^{\Mm_s}|+1) \frac{c_v\rho_v}{c_v} - \rho_v
				- \sum_{m\in\mathcal X_i} |U_i^{\Mm_m}| r_i^{\Mm_m} = 0.
			\]
		\end{proof}
		\begin{proof}{[Proposition~\ref{prop:8}].}
			By an adaptation of the proof of Proposition~\ref{prop:2}, the profit of $\sigma_v^i$ is 
			$$
				\bar\pi_{\sigma_v^i} = ( |U_i^{\Mm_s}|+1 ) r_{\sigma_v^i}^{\Mm_s}-\rho_v - |U_i^{\Mm_s}|r_i^{\Mm_s}
									 = ( \alpha -|U_i^{\Mm_s}| c_v ) \frac{r_i^{\Mm_s}-\rho_v}{\alpha+c_v},
			$$
			where $\alpha=\displaystyle\sum_{u\in U_i^{\Mm_s}}c_u$.
			It is profitable if $1)$ or $2)$ holds.		
		\end{proof}

		\begin{proof}{[Proposition~\ref{par:bribing_users_that_rated_item_i}].}
		There are two cases to explore: (i) both users are in the same cluster; (ii) each user is in a different cluster.\\
				(i) Suppose that $u,v\in\Mm_s$ are two users that already rated item $i$.
			By Proposition~\ref{prop:AllVoteCluster}, to have a positive profit $\bar \pi_{\sigma_u^i}$, when bribing user $u$, we need to have $c_u > \bar c_{U_i^{\Mm_s}}$.
			Thus, we do not consider strategies that bribe a user, $v$, s.t. $c_v<\bar c_{U_i^{\Mm_s}}$, because it  would not increase the wealth, $\bar J_i$.
			
			Let $c_u>c_v> c_{U_i^{\Mm_s}}$, we compute the profit per unit of invested resources, $\bar\pi_{\sigma_u^i}/\rho_u - \bar\pi_{\sigma_v^i}/\rho_v = (c_u-c_v)/\bar c_{U_i^{\Mm_s}}>0$.
			Thus, the profit per unit of invested wealth is larger for user $u$.
			Hence, as we obtained for the bipartite ranking systems, the optimal strategy is: to bribe users by decreasing reputation, investing all the wealth until either the lack of available profitable users ($c_u > \bar c_{U_i^{\Mm_s}}$) or the exhaustion of funds to bribe profitable users.\\
				(ii) When each reference user belongs to distinct clusters, $u\in \Mm_s$, $v\in \Mm_t$ and $s\neq t$, we have that if $|U_{i}^{\Mm_s}|\geq |U_i^{\Mm_t}|$, then the profit per unit of invested wealth ($\bar\pi_{\sigma_u^i}/\rho_u$ versus $\bar\pi_{\sigma_v^i}/\rho_v$) is larger for user $u$.
			If $|U_{i}^{\Mm_s}|< |U_i^{\Mm_t}|$ then the profit per unit of invested wealth is larger for user $u$ if $|U_i^{\Mm_s}|>(c_u-c_v)^{-1}$ and $|U_i^{\Mm_t}|<(| U_i^{\Mm_s}|c_u-1 )/c_v$, and larger for user $v$, otherwise.
			\end{proof}

\begin{proof}{[Proposition~\ref{par:bribing_users_that_did_not_rate_the_item_i}]}
Recalling Proposition~\ref{prop:7}, we only need to explore the case where the seller of item $i$ wants to bribe users belonging to clusters with users that already rated the item, clusters $m$ s.t. $i\in I^{\Mm_m}$, otherwise the profit is zero.
			Let users $u,v\in \Mm_s$ and $u,v\notin U_i$ be s.t. $c_u>c_v$, and let $$\alpha=\sum_{w\in U_i^{\Mm_s}}c_w, \,\gamma=\frac{\left|U_i^{\Mm_s}\right|c_u-\alpha}{c_u+\alpha}\text{ and }\delta = \frac{\left|U_i^{\Mm_s}\right|c_v-\alpha}{c_v+\alpha}.$$ 
			By Proposition~\ref{prop:8}, we have that the profits for bribing users $u$ and $v$ are
			$$
				\frac{\rho_u- r_i^{\Mm_s}}{c_u+\alpha}( |U_i^{\Mm_s}|c_u-\alpha ) \text{ and } \frac{\rho_v- r_i^{\Mm_s}}{c_v+\alpha}( |U_i^{\Mm_s}|c_v-\alpha ),
			$$
			respectively. 
			The difference of profits is $(\rho_u- r_i^{\Mm_s})\gamma - (\rho_v - r_i^{\Mm_s}) \delta$, hence, for the same amount of spent wealth, $\bar\pi_{\sigma_u^i}/(\rho_u-r_i) > \bar\pi_{\sigma_v^i}/(\rho_v-r_i^{\Mm_s})$, because $\gamma>\delta$.
			
			Again, the optimal strategy is to bribe users by decreasing order reputation, investing all the available wealth until either the exhaustion of profitable users ($c_u > \bar c_{U_i^{\Mm_s}}$) or funds.
\end{proof}

\begin{proof}{Proposition~\ref{par:general_case}}
        	We investigate when it is better to bribe a user $u\in U_i^{\Mm_s}$ or a non-rater user $v\notin U_i^{\Mm_s}$. 
			The result is the adaptation of the one for the general case in Section~\ref{sec:bipartite_ranking_systems}.
			We consider the profit change rate, which are $\displaystyle\frac{\bar\pi_{\sigma^i_u}}{\rho_u}=\delta$ and $\displaystyle\frac{\bar \pi_{\sigma_v^i}}{\rho_u-\bar r_i}=\gamma$, respectively.
			In the case, $c_u\geq c_v$ we always have $\delta \geq \gamma$.
			In the other case, $c_u<c_v$, we have $\gamma < \delta$ whenever either $\bar c_{U_i^{\Mm_s}}<1/|U_i^{\Mm_s}|$ and $c_u<\alpha$, or $\bar c_{U_i}\geq 1/|U_i^{\Mm_s}|$.
			Again, the optimal strategy is to order bribable users by decreasing reputation for each of the sets $U_i^{\Mm_s}$ and $U\setminus U_i^{\Mm_s}$, and start allocating wealth to $U_i^{\Mm_s}$ and, afterward, to $U\setminus U_i^{\Mm_s}$.	
\end{proof}

        \begin{proof}{[Theorem~\ref{prop:deutica}].}
			By definition, $\bar \pi_{\sigma^i}<\pi_{\sigma_i}$ is the same as 
			$$
				\lb \frac{\left|U_i^{\Mm_s}\right|c_v}{\sum_{u\in U_i^{\Mm_s}}c_u}-1\rb\rho_v < \lb \frac{\left|U_i\right|c_v}{\sum_{u\in U_i}c_u}-1\rb\rho_v,
			$$
			which is equivalent to $$|U_i^{\Mm_s}|\sum_{u\in U_i}c_u  <  |U_i|\sum_{u\in U_i^{\Mm_s}}c_u.$$
			Noticing that $U_i=U_i^{\Mm_s}\cup ( U_i\setminus U_i^{\Mm_s})$, we can rewrite it as
			$$
				|U_i^{\Mm_s}|(\sum_{u\in U_i^{\Mm_s}}c_u+\sum_{u\in U_i\setminus U_i^{\Mm_s}}c_u) < (|U_i^{\Mm_s}|+|U_i\setminus U_i^{\Mm_s}|)\sum_{u\in U_i^{\Mm_s}}c_u.	
			$$
			This is $\bar c_{\lb U_i\setminus U_i^{\Mm_s}\rb}  < \bar c_{U_i^{\Mm_s}}$.\qedhere
		\end{proof}

\end{document}